%% file: modification_v1_20250313.tex
\documentclass[twocolumn]{aastex631}
\usepackage{booktabs}
\usepackage{tablefootnote}

\begin{document}


\title{Search for the Optical Counterpart of Einstein Probe Discovered Fast X-ray Transients from Lulin Observatory }

\correspondingauthor{A.~Aryan}
\email{amar@astro.ncu.edu.tw}

\correspondingauthor{T.-W.~Chen}
\email{twchen@astro.ncu.edu.tw}

\correspondingauthor{S.~Yang}
\email{sheng.yang@hnas.ac.cn}

\author[0000-0002-9928-0369]{Amar~Aryan}
\affiliation{Graduate Institute of Astronomy, National Central University, 300 Jhongda Road, 32001 Jhongli, Taiwan}
\author[0000-0002-1066-6098]{Ting-Wan~Chen} 
\affiliation{Graduate Institute of Astronomy, National Central University, 300 Jhongda Road, 32001 Jhongli, Taiwan}
\author[0000-0002-2898-6532]{Sheng~Yang} 
\affiliation{Henan Academy of Sciences, Zhengzhou 450046, Henan, China}
%
%
%
%
\author[0000-0002-8094-6108]{James~H.~Gillanders}
\affiliation{Astrophysics sub-Department, Department of Physics, University of Oxford, Keble Road, Oxford, OX1 3RH, UK}
\author[0000-0002-5105-344X]{Albert~K.H.~Kong}
\affiliation{Institute of Astronomy, National Tsing Hua University, Hsinchu 300044, Taiwan}
\author[0000-0002-8229-1731]{S.~J.~Smartt}
\affiliation{Department of Physics, University of Oxford, Keble Road, Oxford, OX1 3RH, UK}
\affiliation{Astrophysics Research Centre, School of Mathematics and Physics, Queens University Belfast,
Belfast BT7 1NN, UK}
\author[0000-0002-0504-4323]{Heloise F. Stevance}
\affil{Astrophysics sub-Department, Department of Physics, University of Oxford, Keble Road, Oxford, OX1 3RH, UK}
\affil{Astrophysics Research Centre, School of Mathematics and Physics, Queens University Belfast,
Belfast BT7 1NN, UK}
\author[0000-0001-9108-573X]{Yi-Jung~Yang}
\affiliation{Graduate Institute of Astronomy, National Central University, 300 Jhongda Road, 32001 Jhongli, Taiwan}
%
%
%
%
\author[0000-0002-9085-8187]{Aysha~Aamer}
\affiliation{Astrophysics Research Centre, School of Mathematics and Physics, Queens University Belfast,
Belfast BT7 1NN, UK}
\author[0000-0003-4905-7801]{Rahul~Gupta}
\affiliation{Astrophysics Science Division, NASA Goddard Space Flight Center, Mail Code 661, Greenbelt, MD 20771, USA}
\affiliation{NASA Postdoctoral Program Fellow}
\author[0009-0007-4482-2101]{Lele~Fan}
\affiliation{Henan Academy of Sciences, Zhengzhou 450046, Henan, China}
\author{Wei-Jie~Hou}
\affiliation{Graduate Institute of Astronomy, National Central University, 300 Jhongda Road, 32001 Jhongli, Taiwan}
\author{Hsiang-Yao~Hsiao}
\affiliation{Graduate Institute of Astronomy, National Central University, 300 Jhongda Road, 32001 Jhongli, Taiwan}
\author[0000-0003-3164-8056]{Amit~Kumar}
\affiliation{Aryabhatta Research Institute of Observational Sciences (ARIES), Manora Peak, Nainital-263002, India}
\author{Cheng-Han~Lai}
\affiliation{Graduate Institute of Astronomy, National Central University, 300 Jhongda Road, 32001 Jhongli, Taiwan}
\author[0009-0003-0553-3340]{Meng-Han~Lee}
\affiliation{Graduate Institute of Astronomy, National Central University, 300 Jhongda Road, 32001 Jhongli, Taiwan}
\author[0009-0003-5139-9007]{Yu-Hsing~Lee}
\affiliation{Graduate Institute of Astronomy, National Central University, 300 Jhongda Road, 32001 Jhongli, Taiwan}
\author[0009-0008-4171-0152]{Hung-Chin~Lin}
\affiliation{Graduate Institute of Astronomy, National Central University, 300 Jhongda Road, 32001 Jhongli, Taiwan}
\author{Chi-Sheng~Lin}
\affiliation{Graduate Institute of Astronomy, National Central University, 300 Jhongda Road, 32001 Jhongli, Taiwan}

\author[0000-0001-8771-7554]{Chow-Choong~Ngeow}
\affiliation{Graduate Institute of Astronomy, National Central University, 300 Jhongda Road, 32001 Jhongli, Taiwan}
\author[0000-0002-2555-3192]{Matt~Nicholl}
\affiliation{Astrophysics Research Centre, School of Mathematics and Physics, Queens University Belfast,
Belfast BT7 1NN, UK}
\author[0000-0001-8415-6720]{Yen-Chen~Pan}
\affiliation{Graduate Institute of Astronomy, National Central University, 300 Jhongda Road, 32001 Jhongli, Taiwan}
\author[0000-0001-9309-7873]{Shashi~Bhushan~Pandey}
\affiliation{Aryabhatta Research Institute of Observational Sciences (ARIES), Manora Peak, Nainital, Uttarakhand, 263001, India}
\author[0009-0003-2609-3591]{Aiswarya~Sankar.K}
\affiliation{Graduate Institute of Astronomy, National Central University, 300 Jhongda Road, 32001 Jhongli, Taiwan}
\author[0000-0003-4524-6883]{Shubham~Srivastav}
\affiliation{Astrophysics sub-Department, Department of Physics, University of Oxford, Keble Road, Oxford, OX1 3RH, UK}
\author[0009-0009-7577-1516]{Guanghui~Sun}
\affiliation{Henan Academy of Sciences, Zhengzhou 450046, Henan, China}
\author{Ze-Ning~Wang}
\affiliation{Henan Academy of Sciences, Zhengzhou 450046, Henan, China}
%







\begin{abstract}

The launch of the Einstein Probe (EP) mission has revolutionized the detection and follow-up observations of fast X-ray transients (FXTs) by providing prompt and timely access to their precise localizations. In the first year of its operation, the EP-mission reports the discovery of 72 high signal-to-noise FXTs. Subjected to the visibility in the sky and weather conditions, we search for the optical counterparts of 42 EP-discovered FXTs from the Lulin Observatory. We successfully detect the optical counterparts of 12 FXTs, and five of those are first discovered by us from the Lulin Observatory. We find that the optical counterparts are generally faint ($r>20$\,mag) and decline rapidly ($>0.5$\,mag per day). 
We also find that 12 out of 42 FXTs show direct evidence of their association with Gamma-Ray Bursts (GRBs) through significant temporal and spatial overlapping. Furthermore, the luminosities and redshifts of FXTs with confirmed optical counterparts in our observations are fully consistent with the faintest end of the GRB population. However, the non-detection of any associated optical counterpart with a significant fraction of  FXTs suggests that EP FXTs are likely a subset of so-called `dark FXTs', similar to `dark GRBs'.
Additionally, the luminosities of {\bf two FXTs with confirmed redshifts} are also consistent with jetted tidal disruption events (TDEs). However, we find that the optical luminosities of FXTs differ significantly from typical supernova shock breakout or kilonova emissions.
Thus, we conclude that a significant fraction of EP-discovered FXTs are associated with events having relativistic jets; either a GRB or a jetted TDE.

\end{abstract}

\keywords{Transient sources (1851); High-energy astrophysics (739); X-ray
transient sources (1852); Optical identification (1167); Gamma-ray bursts (629).}


\section{Introduction} \label{sec:intro}

Fast X-ray transients (FXTs) are extragalactic bursts of soft X-rays (0.3--10 keV) lasting up to $\sim$\,10$^{1}$--10$^{4}$ seconds \citep[][]{2013ApJ...779...14J,2023ApJ...959...75P,2024ApJ...969L..14G, 2024arXiv240416350L}. The origin and nature of these FXTs, lasting from minutes to hours, still remain an intriguing mystery. However, numerous models have been proposed to unravel and quantify these enigmatic events. The supernova (SN) shock breakouts 
(SBO) \citep[][]{2008Natur.453..469S,2020ApJ...898...37N,2024MNRAS.52711823E}, binary neutron star (BNS) mergers \citep[][]{2019Natur.568..198X,2021ApJ...915L..11A,2022ApJ...927..211L,2024MNRAS.52711823E}, tidal disruption events (TDEs) of white dwarfs on intermediate-mass black holes \citep[][]{2013ApJ...779...14J,2015MNRAS.450.3765G, 2016Natur.538..356I, 2019ApJ...884L..34P, 2024ApJ...969L..14G}, and low-luminosity long gamma-ray bursts (GRBs) or short GRBs \citep[][]{2013ApJ...779...14J, 2015MNRAS.450.3765G, 2017MNRAS.467.4841B, 2024ApJ...969L..14G, 2024arXiv240416350L,2024arXiv240416425L} are among the most favorable scenarios for producing FXTs. These are among one of the newest class of transient objects first identified only about a decade ago in Chandra images of M86 \citep[][]{2013ApJ...779...14J}. Since then, intensive archival searches have resulted in the discovery of around 40 FXTs in Chandra, XMM-Newton images, and eROSITA \citep[e.g.,][]{2015MNRAS.450.3765G,2017MNRAS.467.4841B,2020ApJ...896...39A,2022A&A...663A.168Q,2023A&A...675A..44Q}. 

However, with the launch of the Einstein Probe (EP) mission \citep[][]{2018SPIE10699E..25Y,2022hxga.book...86Y} on 9 January 2024, the number of FXTs is significantly increasing. The EP-mission is an innovative mission dedicated to scanning the sky in the soft X-ray band with high precision. The Wide-field X-ray Telescope (EP-WXT) on-board EP with a large field of view ($60^\circ\times60^\circ$) systematically surveys and characterizes high-energy transients, while also monitoring variable objects with unprecedented sensitivity and frequency. Additionally, EP-mission also features a conventional narrow-field ($1^\circ\times1^\circ$) Follow-up X-ray Telescope (EP-FXT, \citealt[][]{2022hxga.book...86Y}) for follow-up observations and precise localization of newly discovered transients. 
During the commissioning phase (until 9 July 2024), EP-mission has already provided several useful communications regarding the discovery of new FXTs on various platforms such as NASA's General Coordinates Network (GCN) Circulars\footnote{\url{https://gcn.nasa.gov/circulars}}, GCN Notices\footnote{\url{https://gcn.nasa.gov/docs/schema/v4.1.0/gcn/notices/einstein_probe}} and the Astronomer's Telegram (ATel)\footnote{\url{https://astronomerstelegram.org}}. The commissioning phase has been completed on 10$^{\rm th}$ of July 2024, now it is promptly providing publicly available alerts for transient objects, ensuring timely access to new discoveries. 

\begin{figure*}
\centering
    \includegraphics[width=\textwidth,angle=0]{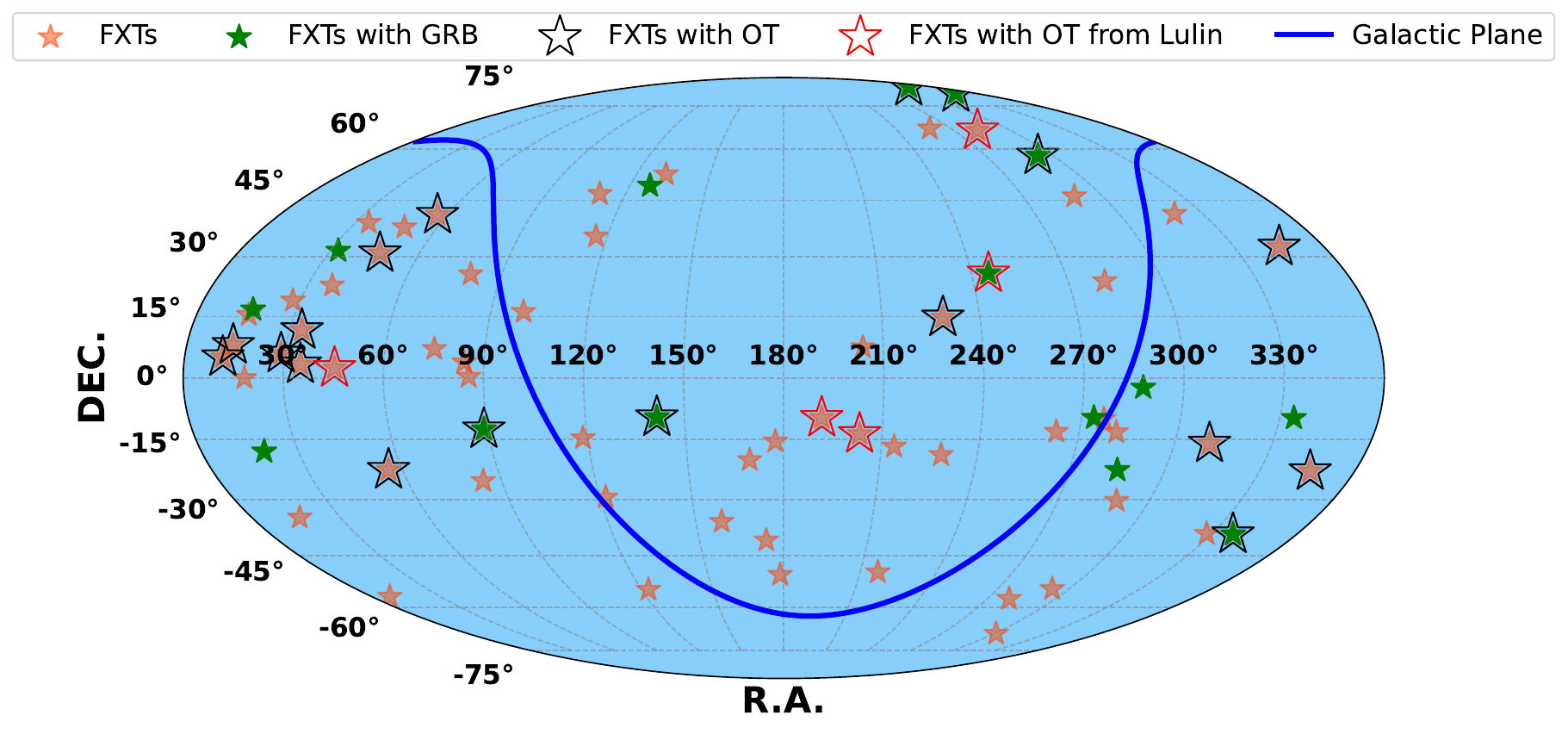}
    \includegraphics[width=0.65\textwidth,angle=0]{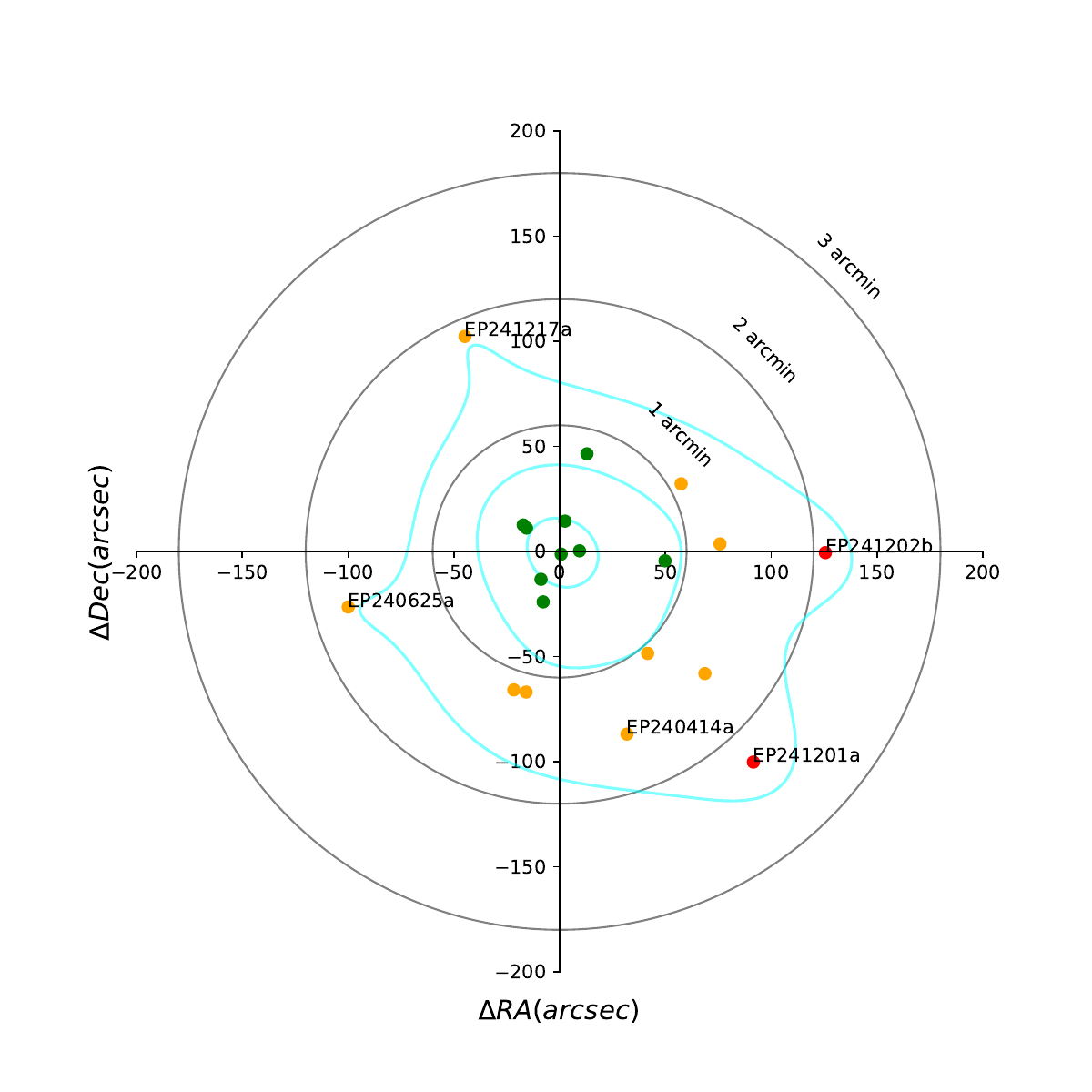}
   \caption{$Top$: The sky localizations of 72 high SNR FXTs discovered by the EP-mission in the first year of its operation. Within the context of the limited number of FXTs detected, the occurrences of FXTs seem to be isotropic in nature, similar to GRBs. The FXTs with associated GRBs, FXTs with corresponding optical transients (OTs), and FXTs with respective OTs discovered from the Lulin Observatory are highlighted. $Bottom$: Angular offset distribution between the EP-WXT candidates and their associated optical counterparts, plotted as $\Delta$RA versus $\Delta$Dec in arcseconds. A point located at the center indicates perfect positional agreement between the EP-WXT detection and the corresponding OT. Black concentric circles mark angular separation radii of 1, 2, and 3 arcminutes, corresponding to the typical positional uncertainty of the EP-WXT ($\sim$3$\arcmin$). Each point represents a candidate, color-coded by event group, with several labeled for clarity. Cyan contours denote the 1, 2, and 3 $\sigma$ levels of the event population density, illustrating the statistical spread in positional accuracy. This visualization highlights both the average localization precision of EP-WXT and the reliability of optical counterpart identification. The references for localizations are credited in the Appendix.}
    \label{fig:sky_distribution}
\end{figure*}

In about one year of its operation (January 9$^{\rm th}$ 2024 to January 9$^{\rm th}$ 2025), the EP-mission has reported the discovery of 72 high signal-to-noise ratio (SNR) FXTs. Fig.\,\ref{fig:sky_distribution}, top panel shows the sky localization of those 72 FXTs. Similar to the GRBs, the occurrences of EP-discovered FXTs seem isotropic in nature, although the sample size currently is not very large. 
Currently, the EP-WXT provides the localization with an error circle of radius $\lesssim$\,3$\arcmin$ for most of the detected FXTs. The localization improves significantly with EP-FXT, which provides more precise localization with error circles having radii between 10 and 30$\arcsec$ only. Such precise localization is ideal for enabling many narrow-field, large-aperture, follow-up telescopes to search for the multi-wavelength counterpart of any new FXT detection by the EP-mission. Fig.~\ref{fig:sky_distribution}, bottom panel shows the distribution of angular offsets between the EP-WXT locations and the associated optical counterpart candidates. All identified counterparts fall within the EP-WXT error circles, indicating good positional consistency. Throughout this paper, we use the term FXTs specifically to refer to events detected by EP.

The groundbreaking first detection of multi-wavelength counterparts (X-ray, optical, and radio) of EP240315a has unlocked exciting new pathways for exploring the intriguing origins and nature of FXTs. Following \citet[][]{2024ApJ...969L..14G,2024arXiv240416350L,2025NatAs.tmp...34L}, the role of the optical counterpart in deciphering the origin and nature of FXTs has proved essential. Additionally, early optical follow-up (within several hours post-discovery) helps us to probe the early evolution of the underlying FXT in its rest frame. The role of such follow-up is further amplified for high-redshift discoveries, where the observed-frame optical evolution over a few days probes nearly the initial 24 hrs of the FXT's rest-frame evolution. Similar to high-redshift GRBs, with detection up to $z$\,$\gtrsim$\,5, FXTs demonstrate great potential for probing the early universe (e.g., EP240315a; $z=4.859$). Based on the multi-wavelength observations, several studies \citep[][]{2024ApJ...969L..14G,2024arXiv240416350L,2025NatAs.tmp...34L} infer EP240315a to originate from a relativistic event with two possible scenarios: a long GRB or a jetted TDE, although the measured optical and radio luminosity in their analysis lie consistently with long GRBs. Recently, \citet[][]{2024arXiv240718311R} present the long-term radio monitoring of EP240315a and report evidence of a relativistic jet.  

About a month after EP240315a, the EP-mission reports the discovery of FXT EP240414a \citep[][]{2024GCN.36091....1L}. Later, EP240414a is also discovered with multi-wavelength counterparts \citep[][]{2024arXiv241002315S,2024arXiv240919056V,2024arXiv240919055B,2025ApJ...978L..21S}. Owing to the extremely blue spectrum in the early phase, \citet[][]{2024arXiv240919056V} link the EP240414a to luminous fast blue optical transients (LFBOTs). However, due to the non-thermal origin of the early phase followed by the decaying phase nicely explained by a power law, \citet[][]{2025ApJ...978L..21S} 
find evidence to link EP240414a to GRBs. {\bf The prompt and afterglow phase investigations by \citet[][]{2025ApJ...986L...4H} place EP240414a between long GRBs and low-luminosity GRBs.} Further, the late-time photometric and spectral evolutions of EP240414a match with Type Ic-BL SNe as presented in 
\citet[][]{2024arXiv241002315S} and also in \citet[][]{2024arXiv240919056V}. Type Ic-BL SNe are occasionally found associated with several long-GRBs \citep{modjaz2016}.  {\bf Another study by \citet[][]{2025ApJ...985...21Z} proposes that EP240414a could be the result of an off-axis view of a jet-cocoon system from an expanded helium progenitor star.} Recently, elaborative analyses of EP241021a displaying exceptional rebrightening also suggest a likely link between EP-mission discovered FXTs and low-luminosity GRBs \citep[][]{2025arXiv250314588B}. {\bf Detailed X-ray, optical, and radio analyses by \citet[][]{2025arXiv250507665X} indicate the launch of relativistic jet for the case of EP241021a, later, \citet[][]{2025arXiv250508781Y} confirm this relativistic jet launch scenario through radio observations. Similar to EP240414a, \citet[][]{2025arXiv250505444G} propose an off-axis jet and cocoon scenario for EP241021a. Another detailed X-ray, optical, and radio analyses of EP241021a by \citet[][]{2025arXiv250512491W} suggest it to be an explosion-type event accompanied by a moderately relativistic jet.} In early January 2025, the EP-mission reports the discovery of an interesting FXT EP250108a \citep[][]{2025GCN.38861....1L}. Several studies report the Type Ic-BL SN~2025kg to be associated with EP250108a as its optical counterpart \citep[][]{2025arXiv250408886E,2025arXiv250408889R,2025arXiv250417516S}. Rigorous modeling of well-sampled multi-wavelength light curves suggests a mildly relativistic outflow as the origin of this event \citep[][]{2025arXiv250417034L}. The detailed analysis by \citet[][]{2025arXiv250408886E} indicates that the observed X-ray and radio properties are consistent with a collapsar-powered low-energy jet that fails to break out of the dense material surrounding it. Additionally, they further suggest that the optical emissions possibly originate from a shocked cocoon resulting from the trapped jet. Further, a related work by \citet[][]{2025arXiv250408889R} indicates similar results with the broadband data being consistent with a trapped or low-energy jet-driven explosion from a collapsar with a zero-age main sequence (ZAMS) mass of 15-30\,M$_{\odot}$. Finally, detailed analyses by \citet[][]{2025arXiv250417516S} strengthen the link of EP250108a with low-luminosity GRBs. 

Due to the limited number of available FXT detections, their collective origin and nature are poorly constrained. Moreover, the localization by EP-WXT allowed 
the rapid detection of optical and radio counterparts for EP240315a, opening up new avenues to investigate these rather new transient events. Specifically, the early detection of the optical counterpart enables us to probe the early evolution of FXTs and put constraints on the origin of these enigmatic events. 

In this paper, we report the results of our extensive search for the optical counterparts of EP-mission discovered FXTs in its first year of operation, utilizing the Lulin Observatory\footnote{\url{https://www.lulin.ncu.edu.tw}} telescopes. In Sec.~\ref{sec:Lulin_obs}, we present the design and implementation of our follow-up observation campaign at the Lulin Observatory. In Sec.~\ref{sec:results}, we present our observational results from the Lulin Observatory and emphasize the important role that meter- and sub-meter-class telescopes can play in the follow-up of sources discovered by the EP-mission. {\bf Furthermore, in Sec.~\ref{sec:discussions}, we discuss the constraints of possible aspects on the nature and possible origin of FXTs following from our observations.} Finally, in Sec.~\ref{sec:conclusions}, we  summarize our findings and provide concluding remarks based on the current study. 
Throughout the paper, we assume a $\Lambda$CDM cosmology using \citet[][]{2016A&A...594A..13P} with a Hubble constant of $H_0$ = 67.7 km\,s$^{-1}$ Mpc$^{-1}$, $\Omega_{\rm M}$ = 0.309 and $\Omega_{\Lambda}$ = 0.691. All magnitudes are reported in the AB system, and time is in Coordinated Universal Time (UTC).

\begin{figure*}
\centering
    \includegraphics[width=\columnwidth,height=3.5cm,angle=0]{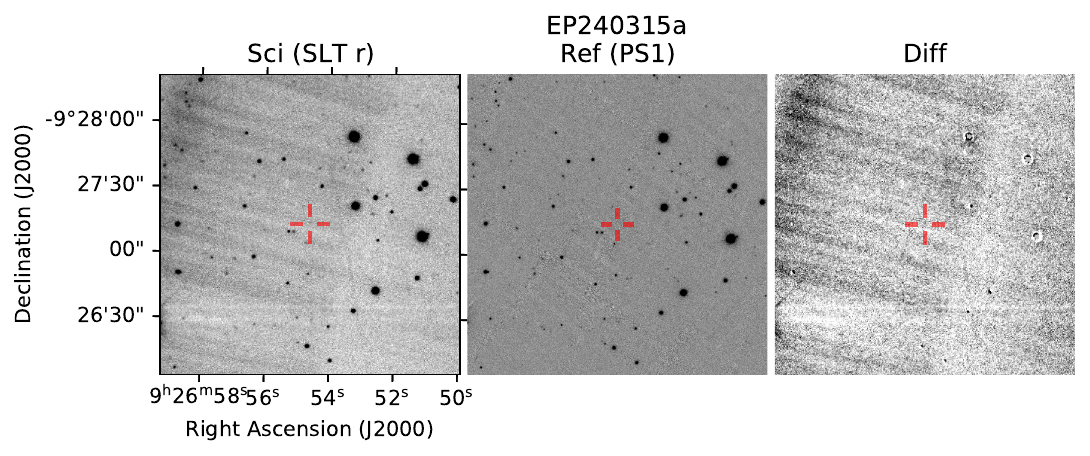}
    \includegraphics[width=\columnwidth,height=3.5cm,angle=0]{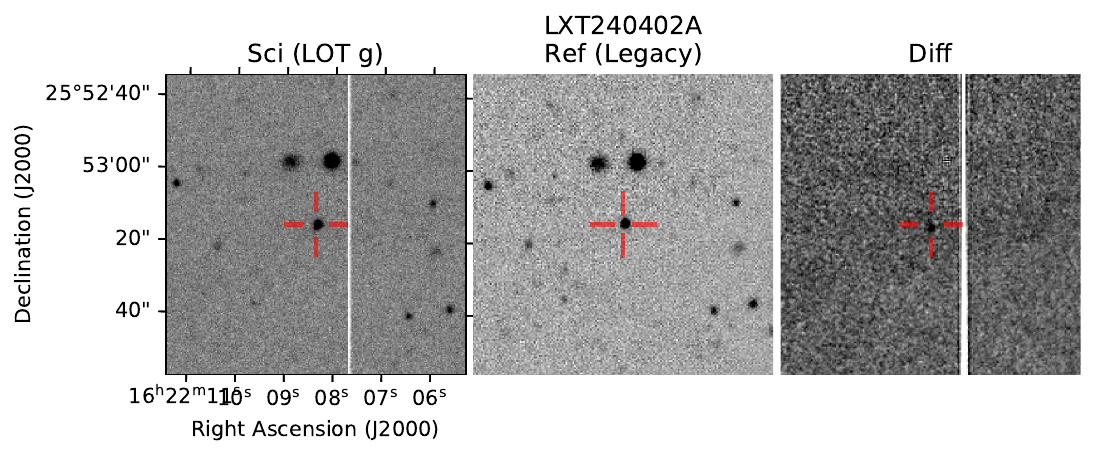}
    \includegraphics[width=\columnwidth,height=3.5cm,angle=0]{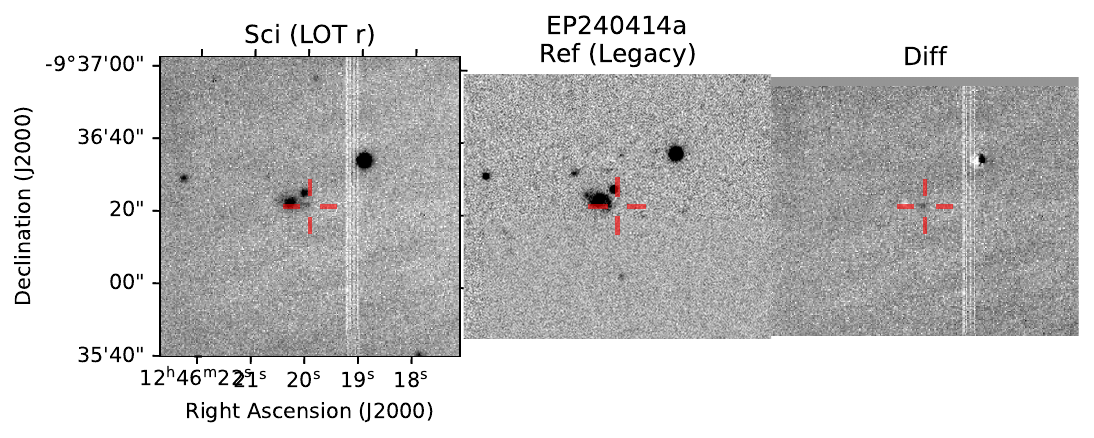}
    \includegraphics[width=\columnwidth,height=3.5cm,angle=0]{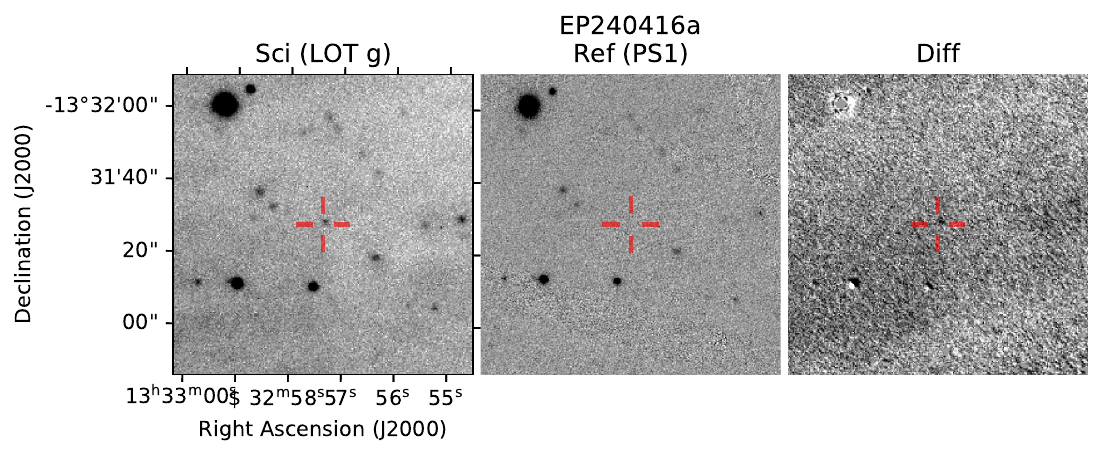}
    \includegraphics[width=\columnwidth,height=3.5cm,angle=0]{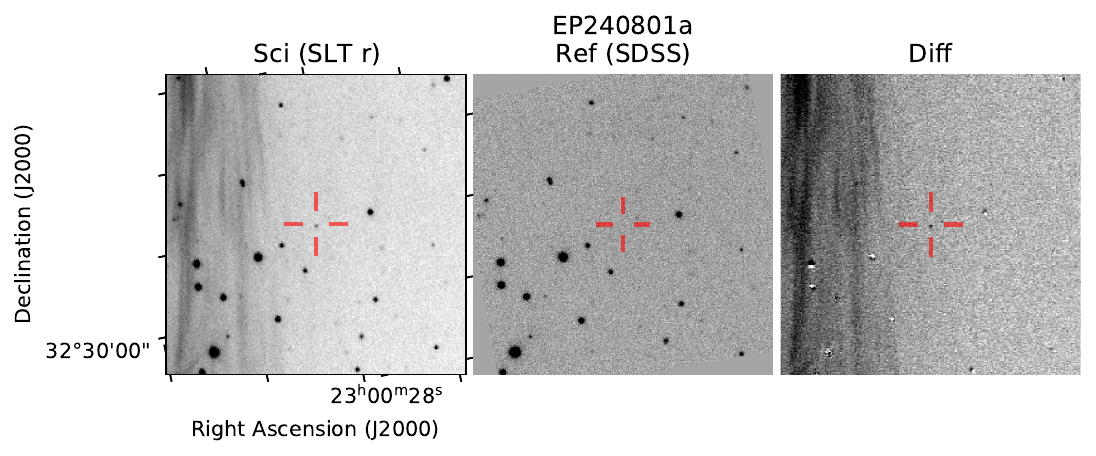}
    \includegraphics[width=\columnwidth,height=3.5cm,angle=0]{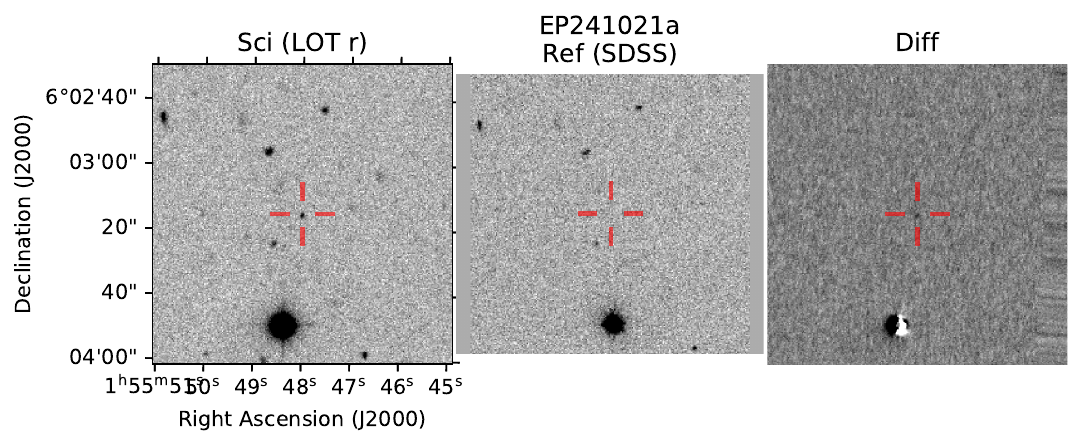}
    \includegraphics[width=\columnwidth,height=3.5cm,angle=0]{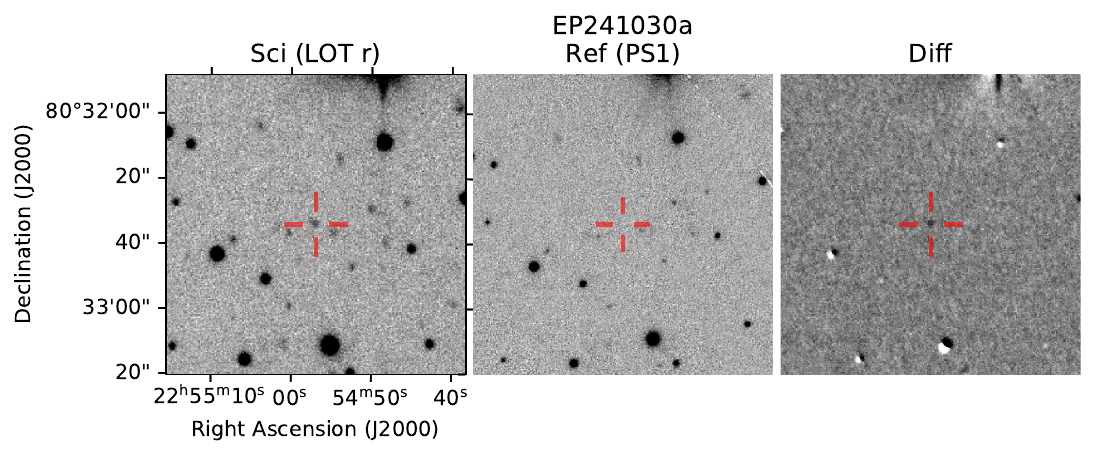}
    \includegraphics[width=\columnwidth,height=3.5cm,angle=0]{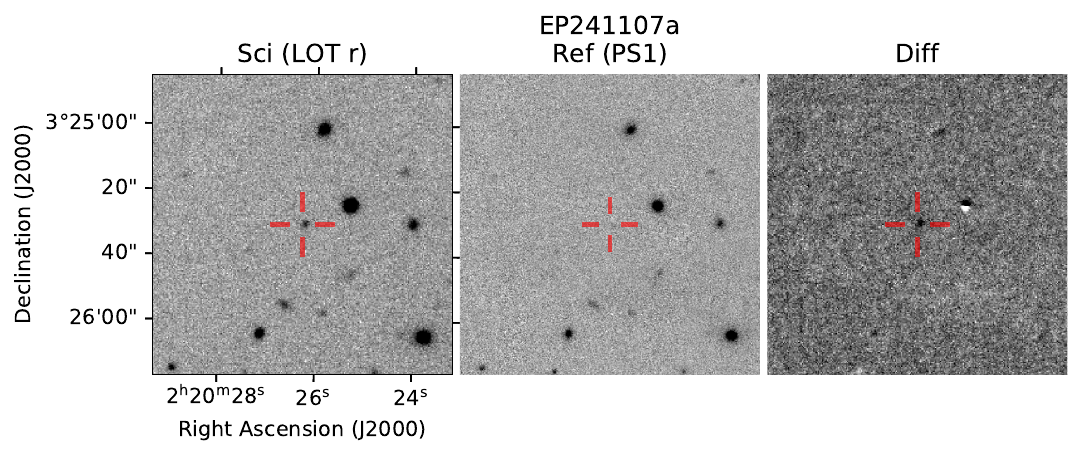}
    \includegraphics[width=\columnwidth,height=3.5cm,angle=0]{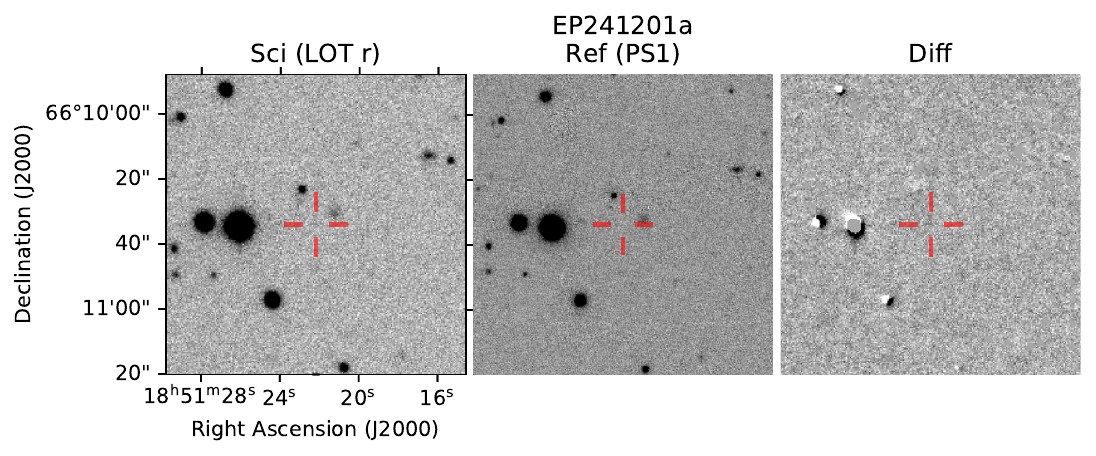}
    \includegraphics[width=\columnwidth,height=3.5cm,angle=0]{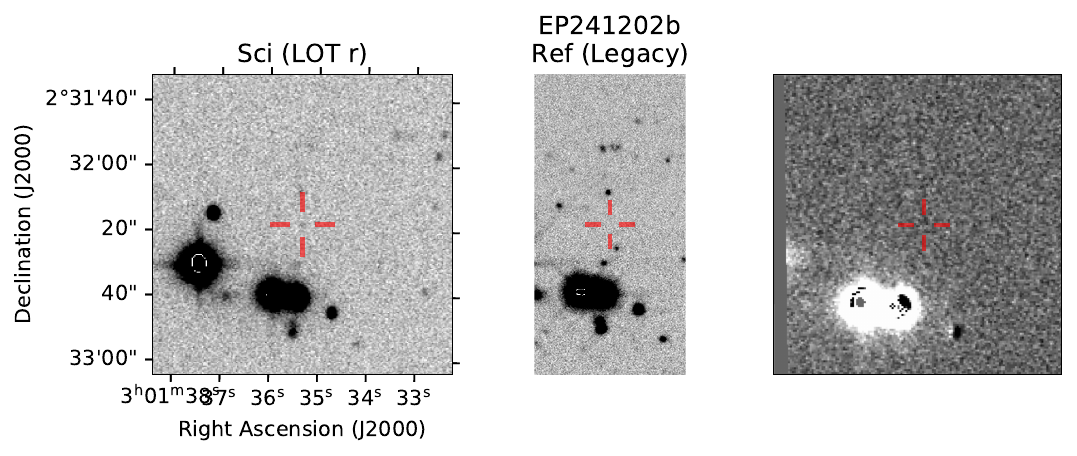}
    \includegraphics[width=\columnwidth,height=3.5cm,angle=0]{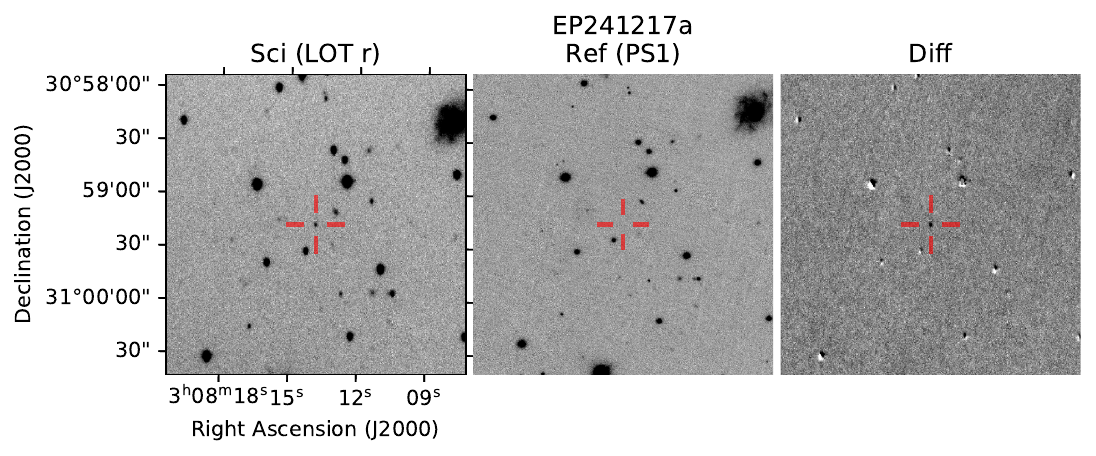}
    \includegraphics[width=\columnwidth,height=3.5cm,angle=0]{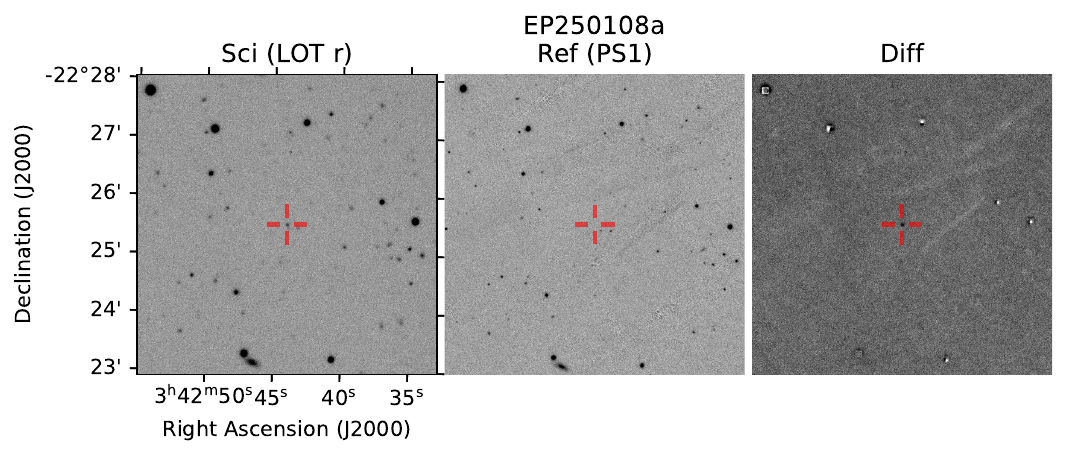}
    
   \caption{{\tt Kinder} observations with Lulin SLT and LOT were carried out for EP FXT follow-up. For 12 EP events, we detected optical counterparts, and for five of these, we were the first to report the discovery. In each panel, the EP event is labeled at the top. The left column shows stacked science frames obtained with the corresponding telescope and filter; the middle column presents reference images from various all-sky surveys; and the right column displays the difference frames produced by subtracting the reference images from the science frames.}
    \label{fig:kinder_12detections}
\end{figure*}

\section{Lulin Observatory Search for the EP-mission discovered FXTs} \label{sec:Lulin_obs}

In about first year of its operation (up to January 9$^{\rm th}$ 2025), the EP-mission reported the discovery of 72 high SNR FXTs. Subject to their visibility from the Lulin Observatory and weather conditions, we triggered our telescopes to search for the optical counterparts of 42 out of 72 high SNR FXTs reported by the EP-mission. {\bf Tables~\ref{tab:detection} and \ref{tab:Coordinates} list the coordinates and error circle radii for EP-WXT and EP-FXT for 42 FXTs, followed from the Lulin Observatory.} 

The optical follow-up Program of FXTs from the Lulin Observatory is a part of the kilonova finder (\citealp[{\tt Kinder};][]{2025ApJ...983...86C}) project. It primarily focuses on searching for kilonovae without relying on gravitational wave triggers. Leveraging the advantageous geographic location of the Lulin Observatory, Taiwan 
(\(23^\circ 28' 10.0''\,\text{N},\quad 120^\circ 52' 21.5''\,\text{E}\))
and scheduling flexibility on the small telescopes, we are able to significantly contribute to capturing short-lived events and quickly respond to unique astronomical events. For example, {\tt Kinder} observed the early light curve and color evolution of the nearby type II SN~2024ggi, right when ATLAS first detected its explosion \citep{2025ApJ...983...86C}. 

We use both the 40cm SLT telescope and the Lulin One-meter Telescope (LOT) for imaging follow-ups. The SLT is an RC Optical Systems Carbon 16-inch f/8.4 telescope with Ascension 200HR mount. It is equipped with an EMCCD camera Andor 934 with BEX2-DD chip (blue sensitive) with gain 1.088 electrons/ADU, readout noise 8.831 electrons RMS, and pixel scale of 0\arcsec.76/pixel. The LOT is a Cassegrain f/8 telescope on an ASA DDM mount, equipped with SOPHIA 2048B CCD camera, having gain 0.92 electrons/ADU, readout noise 7.27 electrons RMS, and pixel scale of 0\arcsec.34/pixel. Both SLT and LOT have the same Astrodon photometrics Sloan filters, providing excellent multi-band observational capabilities (see Fig.\,\ref{fig:filters} for the filter transmission curves\footnote{The SDSS, Pan-STARRS, and SkyMapper filters' response curves are obtained from \url{http://svo2.cab.inta-csic.es/theory/fps/}, while the DESI Legacy Survey filter response curves are obtained from its DR10 website, \url{https://www.legacysurvey.org/dr10/description/}}). 

The {\tt Kinder} workflow and observation steps, from receiving EP triggers to completing observations, analyzing the data, and reporting, are outlined as follows: Initially, we notice new EP triggers through the EP internal Slack channel by collaborating with EP members, where they are discussed before the night begins at Lulin, prompting us to initiate follow-up observations. Primarily, our team relies on GCN Circulars from the EP team as the main source for receiving these triggers. 
Furthermore, we utilize the Kafka client provided by GCN Notices to automatically retrieve alert notifications from EP. These alerts contain essential information such as trigger time (in UTC), localization, count rate, and significance.
During nighttime, we automatically convert these alerts into a format compatible with the ACP Observatory Control Software used at the Lulin Observatory\footnote{\url{http://acp.dc3.com/index2.html}}. Each target is then marked with the highest priority (`First Priority'), and visibility plots are generated. These steps allow the telescope operators in our control room to efficiently point the telescope directly toward the designated targets.
During the daytime, we compile these alert notifications and generate visibility plots, which are then sent to our dedicated Slack channel for further discussions, enabling the team to decide if the target will be observed during the upcoming night. Additionally, the alerts and related information are automatically emailed to all members of our team, ensuring everyone stays promptly informed.

\begin{figure*}
\centering
    \includegraphics[width=0.67\textwidth,angle=0]{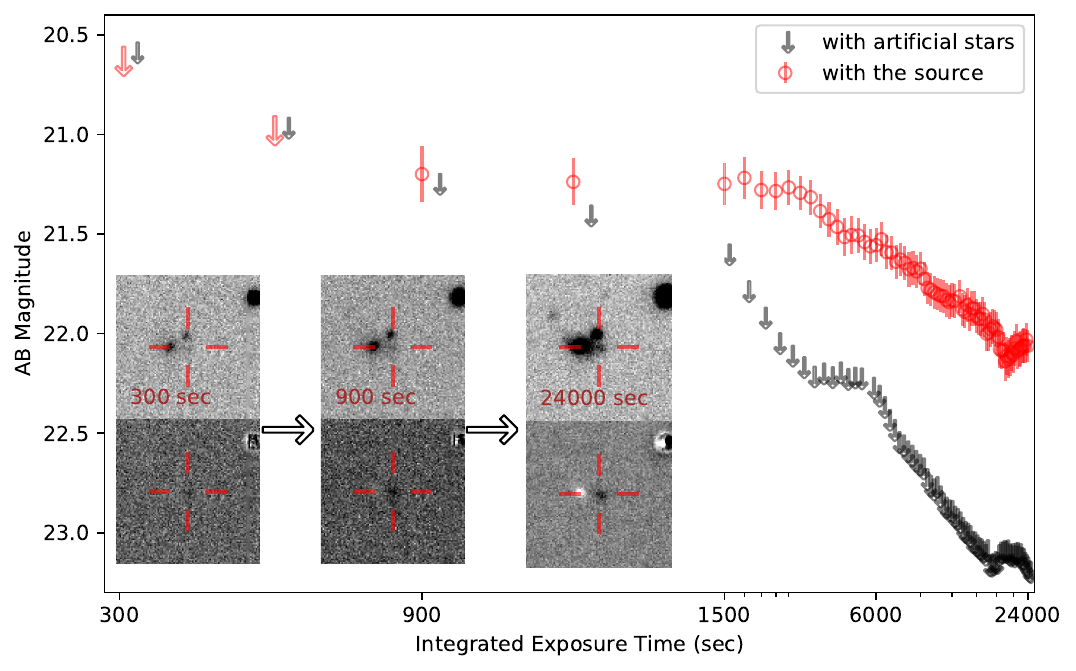}
    \includegraphics[width=0.65\textwidth,angle=0]{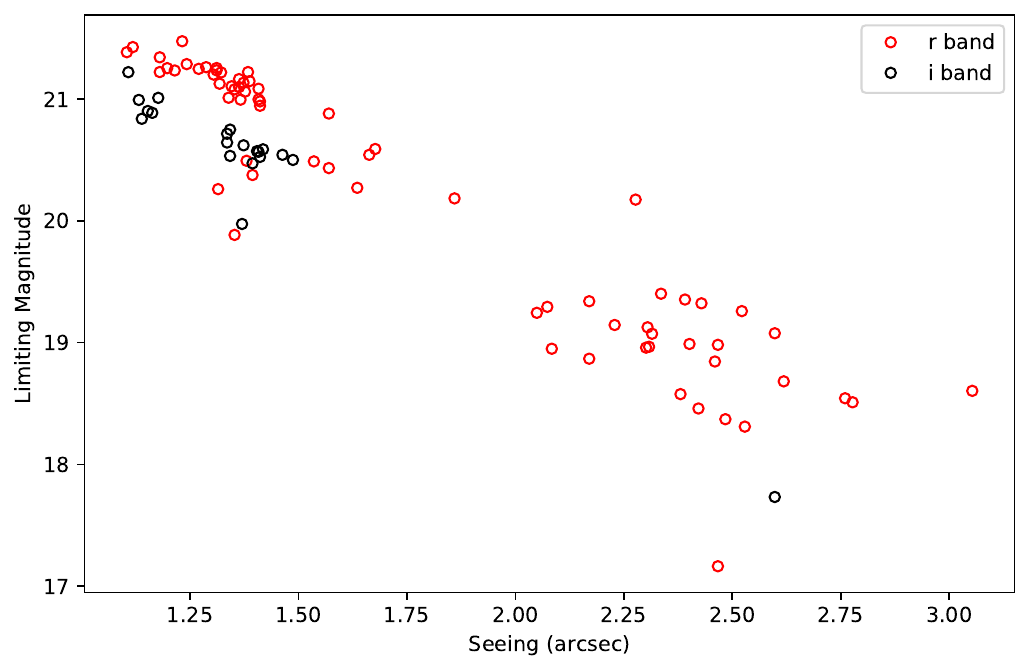}
   \caption{\textit{Top:} Integrated exposure time {\bf enhances} the detectable magnitude depth. In the lower left panel, the top row displays images with increasing integrated exposure time, while the bottom row shows the corresponding template-subtracted results. The marker at the center indicates the location of the optical counterpart, AT~2024gsa, of EP240414a. For a target at 21.3 mag, an integrated exposure time of 900 seconds is required to achieve a detection above 3-sigma. With a stacked exposure time of 1800 seconds, the detection depth reaches 22 mag, can even reach 23 mag in some cases with better conditions. Typically, we use long exposures (30 minutes) for detections. 
   \textit{Bottom:} Limiting magnitude as a function of seeing. }
    \label{fig:limit_seeing}
\end{figure*} 

On-site observers or telescope operators then conduct follow-up observations, typically with long exposures (30 minutes for LOT and 2 hours for SLT), and upload the images in real time to Google Drive. After downloading the data to our lab studio, we process it using the custom-built pipeline\footnote{\url{https://hdl.handle.net/11296/98q6x4}}, which follows the standard procedure of bias, dark subtraction, and flat-field correction. We have deliberately disabled the cosmic ray removal function to preserve faint and weak detections of optical counterparts.

We employ three methods to search for optical counterparts of EP events:
First, we perform a catalog search by plotting the observed combined image using ds9 and overlaying known catalog sources from SDSS, USNO, and Gaia with region markers; any source that is not marked may be considered a candidate optical counterpart. Second, we visually compare the template image with the observed combined image using blink frames in ds9, a method that successfully led to the identification of five optical counterparts. Third, we apply template subtraction by subtracting the observed images from baseline images obtained from major survey projects (e.g., Pan-STARRS1, \citealt[][]{2016arXiv161205560C}; SDSS, \citealt[][]{2000AJ....120.1579Y}; DESI Legacy Survey, \citealt[][]{2019AJ....157..168D}; and SkyMapper, \citealt[][]{2018PASA...35...10W}) to reveal optical counterparts that may be obscured by the host galaxy's background. For template subtraction, we employ either the {\tt Kinder} pipeline \citep[][]{2021A&A...646A..22Y} using the ``hotpants" algorithm \citep{2015ascl.soft04004B} or the Python-based package AutoPhOT \citep[][]{2022A&A...667A..62B} with the ``hotpants", ``sfft" \citep{2022ApJ...936..157H}, or ``Zogy" \citep[][]{2016ApJ...830...27Z} algorithms; any prominent residual in the difference image may be considered a candidate optical counterpart. In this paper, however, all subtractions are conducted using the {\tt Kinder} pipeline to ensure a uniform analysis. Fig.\,\ref{fig:kinder_12detections} displays the {\tt Kinder}-observed science images for the EP-FXT follow-up, along with the corresponding reference images and the resulting difference images by subtracting the reference from the science frames.

We provide a general overview of our {\tt Kinder} pipeline below. The {\tt Kinder} pipeline was specifically developed for the {\tt Kinder} project and is based on the SNOoPY pipeline\footnote{SNOoPy is a package for SN photometry using PSF fitting and/or template subtraction developed by E. Cappellaro. A package description can be found at \url{http://sngroup.oapd.inaf.it/ecsnoopy.html}}. Similar to SNOoPY, the {\tt Kinder} pipeline is a collection of Python scripts that interface with standard IRAF tasks via PyRAF, along with additional specialized analysis tools. These include ``SExtractor'' \citep[][]{sextractor} for source extraction and star/galaxy classification, ``DAOPHOT'' \citep[][]{daophot}  for PSF-fitting photometry, and ``Hotpants'' for image subtraction with PSF matching. Photometry is performed using PSF-fitting, with the sky background first subtracted using a low-order polynomial fit to the surrounding regions. The PSF is constructed by averaging the profiles of isolated field stars automatically selected from the frame. After fitting and removing the {\bf OT} from the original frame, a new local background is estimated and the fitting procedure is iterated. Residuals from the PSF fit are visually inspected to validate the results. Photometric errors are estimated using artificial star experiments. A fake star, with brightness similar to the optical transient ({\bf OT}), is inserted near the {\bf OT} position in the residual image. The PSF-fitting procedure is repeated multiple times at slightly different positions, and the dispersion in these measurements provides an estimate of the instrumental error. This is then combined in quadrature with the formal ``DAOPHOT'' error. If a suitable template image is available, the pipeline also allows for template subtraction using ``Hotpants'' after PSF matching. In these cases, the {\bf OT} magnitude is again measured via PSF fitting, which has been shown to be less sensitive to noise in the difference image. Once subtraction is completed, sources with Gaussian-like profiles are likely to be real, whereas non-Gaussian profiles often indicate artifacts caused by detector effects or poor subtraction. Although automatic real-bogus classification using image stamps is under development, the pipeline currently relies primarily on visual inspection for candidate validation.

\subsection{Detection limits for LOT and SLT}
The Fig.\,\ref{fig:limit_seeing}, top panel, helps one to assess the detectability of the optical counterpart AT~2024gsa associated with EP240414a. We analyze LOT observations obtained from 2024 April 14 to 17, where each individual frame has an exposure time of 300 seconds. To investigate the effect of the integrated exposure time on the depth of detection, we stack images with increasing total exposure times. For a target with magnitude 21.3, we find that a minimum integration of 900 seconds is required to achieve a detection above the $3\sigma$ level. With a total exposure time of 1800 seconds, the detection depth improves to approximately 23 mag. This demonstrates that even when the transient fades to fainter than 22 mag, it remains clearly detectable with LOT under similar observing conditions.

We also investigate the dependence of limiting magnitude on seeing conditions using the same dataset in both $r$ and $i$ bands, see Fig.\,\ref{fig:limit_seeing}, bottom panel. Adopting the method outlined in \citet{2018MNRAS.474..411B}, we first measure the average full width at half maximum (FWHM) of point sources in each frame and generate point spread functions (PSFs) accordingly. Artificial stars with a range of magnitudes are randomly injected into the images. After performing image subtraction, we assess the recovery rate of these injected sources. The detection efficiency is defined as the ratio of recovered to injected stars, and the limiting magnitude is determined at the 50\% recovery threshold. This analysis allows us to quantify the impact of seeing on our detection sensitivity in a systematic way.

\begin{figure*}
\centering
    \includegraphics[width=0.6\textwidth,angle=0]{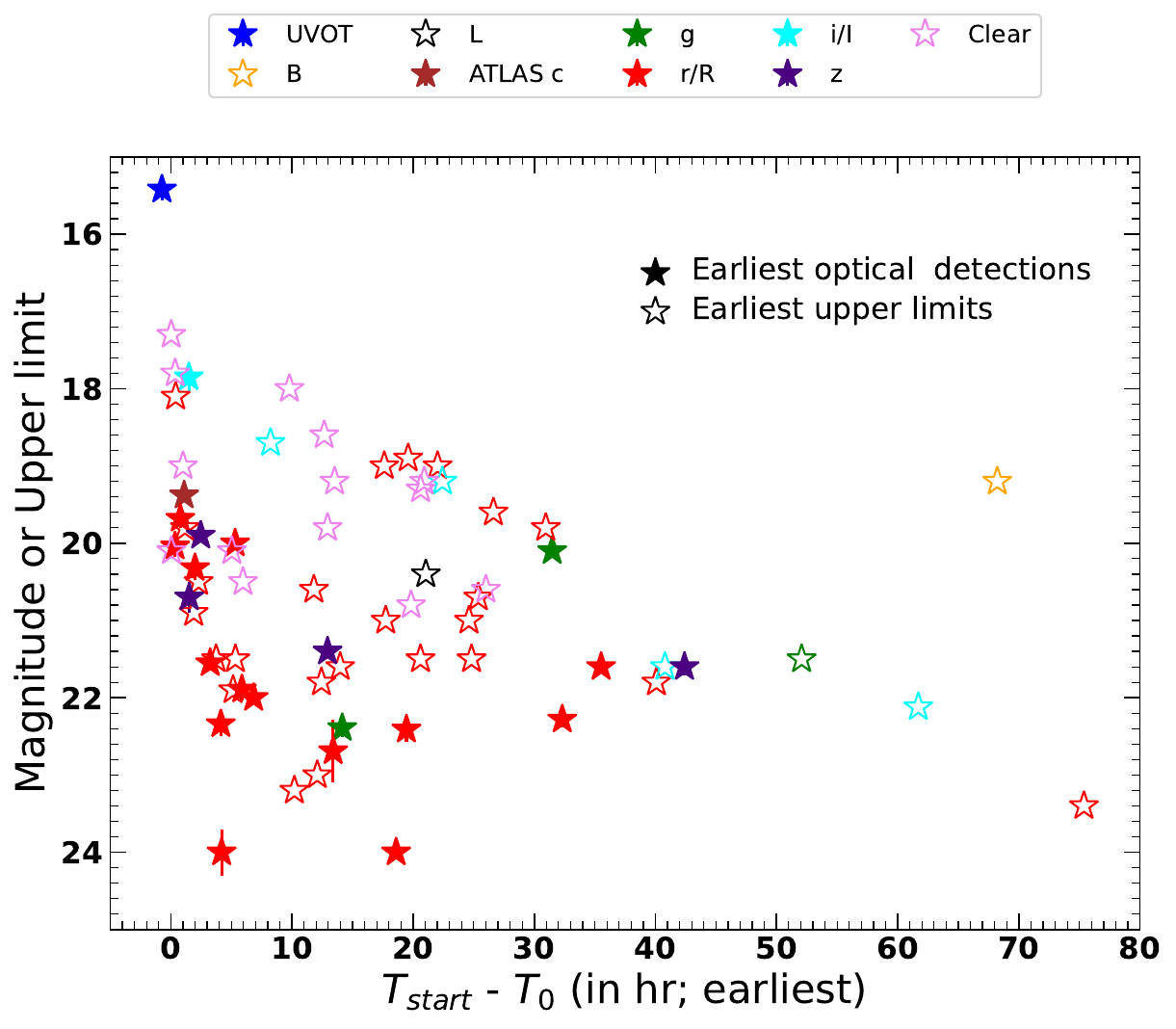}
    \includegraphics[width=0.6\textwidth,angle=0]{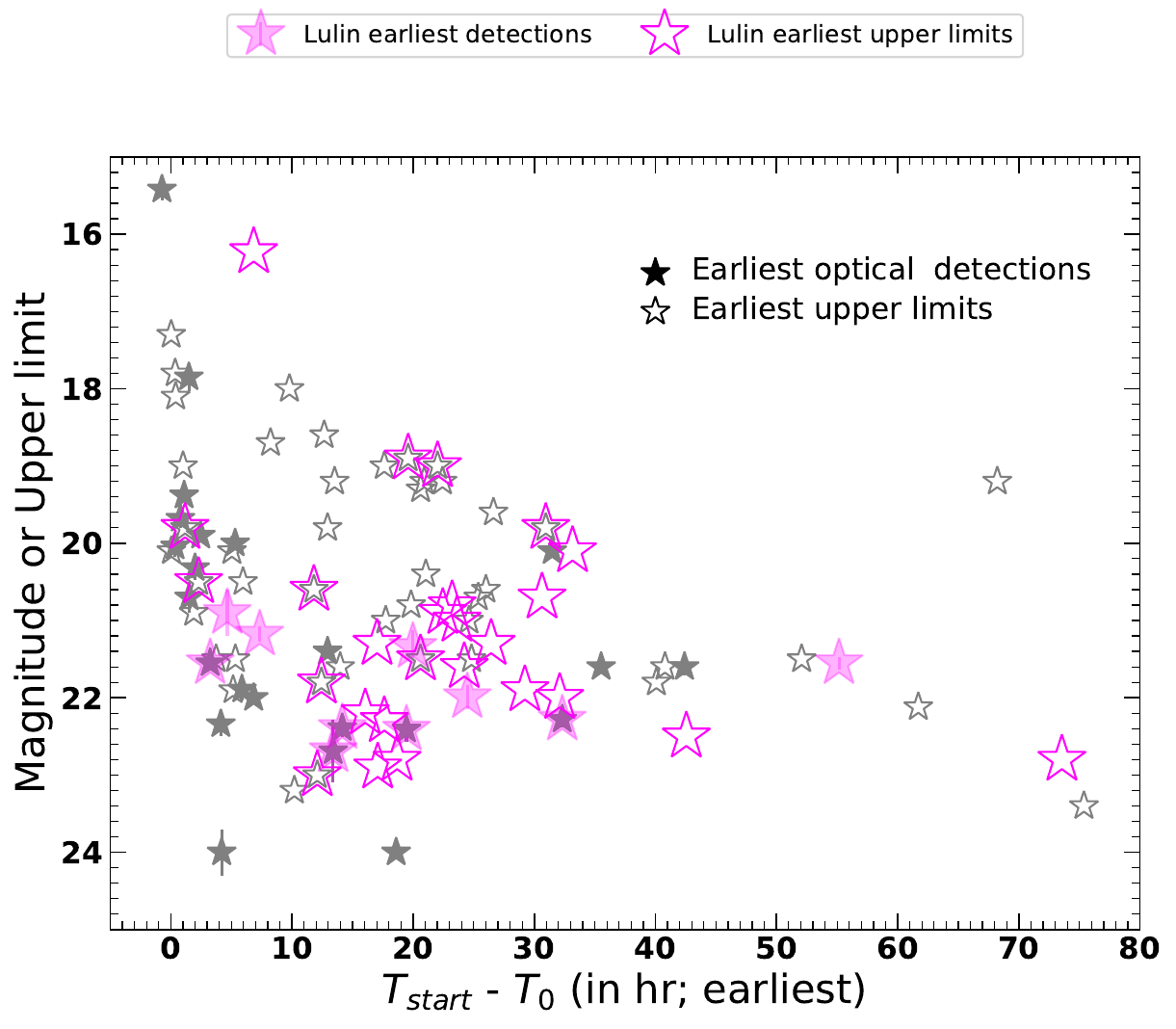}
   \caption{$Top$: Earliest optical detections (or upper limits in case of non-detection of any optical counterpart) for the entire FXTs' sample reported in GCN Circulars. Out of 72 FXTs discovered by the EP, 23 showed corresponding optical counterparts. All reported times are in the observer frame, where $T_{\rm start}$ indicates the beginning of the observation, and $T_{\rm 0}$ refers to the trigger time of the EP-WXT detection. 
   $Bottom$: Comparison of $T_{\rm start}-T_{\rm 0}$ of {\tt Kinder}/Lulin observations with other observations reported in GCNs. 12 out of 23 FXTs with optical counterparts were also detected from the Lulin Observatory. Further, 5 out of those 12 were first discovered by us. The {\tt Kinder} optical detections for EP241021a and EP250108a were delayed; thus not shown here. We had only obtained optical upper limits in their respective first epoch of observations; those upper limits have been included in this plot. The pre-EP discovery Swift-UVOT optical counterpart detection was for EP241030a, which was associated with GRB 241030A \citep[][]{2024GCN.37956....1K}.  The references for first detections or upper limits are credited in the Appendix.}
    \label{fig:gcn_summary}
\end{figure*}

\section{Results} \label{sec:results}

\subsection{Kinder's observational contributions to EP follow-up}

In this paper, we present a statistical study of the FXT sources detected by the EP-mission and followed up by the Lulin Observatory during the first year of the EP-mission, up to 9$^{\rm th}$  January 2025.
Out of the 72 high SNR FXT detections by EP-mission, we have conducted follow-up observations for 42. Out of the 42 sources followed from the Lulin Observatory, 12 sources ({\bf namely, EP250315a, LXT240402A, EP240414a, EP240416a, EP240801a, EP241021a, EP241030a, EP241107a, EP241201a, EP241202b, EP241217a, EP250108a}) have confirmed optical detections through our observations (see Fig.\,\ref{fig:kinder_12detections}), and for five of these, we were the first to report the discovery, while for the remaining 30, we have provided corresponding 3-sigma upper limits. If we include the optical follow-up observations throughout the world, then a total of 23 FXTs out of 72 ({\bf namely, EP250315a, LXT240402A, EP240414a, EP240416a, EP240420a, EP240625a, EP240708a, EP240801a, EP240804a, EP240806a, EP240908a, EP241021a, EP241025a, EP241026a, EP241026b, EP241030a, EP241107a, EP241126a, EP241201a, EP241202b, EP241217a, EP250108a, EP250109a}) have shown confirmed optical counterparts. Thus, some fraction of FXTs also show optical counterparts.

Followed by their X-ray detection, most of the optical counterparts are discovered early; only four among those 23 are discovered 24 hours post their EP-WXT detection. The top panel of Fig.\,\ref{fig:gcn_summary} shows the earliest optical detections or upper limits reported publicly by the community. This figure indicates that the optical counterparts are often bright ($\sim$15.5 to 21.5 mag) if discovered early within a few hours (exception, EP240708a discovered with $r$-band of 24.0$\pm$0.3 mag as reported in \citealt[][]{2024GCN.36842....1J}), however, the late discoveries (beyond $\sim$ 10 hours) are expected to be faint. {\bf As presented above, a total of 23 sources are discovered in optical bands, and 12 of them have also been detected from the Lulin Observatory. Thus, among the remaining 11 (out of 23) sources, four, namely, EP240625a (late follow-up by us), EP240708a (rapidly decaying faint optical counterpart), EP240908a (late follow-up by us, faint OT with $r\sim$24 mag, \citealt[][]{2024GCN.37438....1Q}), EP241026b (late follow-up by us), EP241126a (rapidly decaying faint optical counterpart), were followed from the Lulin Observatory, but the optical counterparts were not detected in observations; although the remaining six (namely, EP240420a, EP240804a, EP240806a, EP241025a, EP241026a, and EP250109a) were visible from the Lulin Observatory, they were missed due to unfavorable weather conditions.}
 
The bottom panel of Fig.\,\ref{fig:gcn_summary} shows our observations plotted over the publicly reported observations. In total, 41 GCN Circulars and 1 ATel have been issued for the EP follow-up by the {\tt Kinder} project; three papers using {\tt Kinder} data have been published: EP240315a \citep{2024ApJ...969L..14G}, EP240414a/AT~2024gsa \citep{2025ApJ...978L..21S}, EP240801a \citep{2025arXiv250304306J}, and one under review: EP250108a/SN~2025kg 
\citep[the `kangaroo'; ][]{2025arXiv250408889R}. Furthermore, two works are in preparation for the events EP250226a (Lee et al. in prep.), and EP250304a (Cotter et al. in prep.).

Tables\,\ref{tab:detection} and \ref{tab:Coordinates} list the details of the Lulin observations for the FXTs, and Tables\,\ref{tab:Photometry1}, \ref{tab:Photometry2}, \ref{tab:Photometry8}, and \ref{tab:Photometry9} list all photometry measurements.

\subsection{Optical counterparts discovered by Kinder}
In this Sec., we list five FXTs discovered by the EP-mission, for which the optical counterparts were discovered from the Lulin Observatory as part of {\tt Kinder} observations. 

\subsubsection{LXT240402A}
\label{sec:lxt240402a}
LXT240402A was first detected by LEIA \citep[][]{2022ApJ...941L...2Z,2023RAA....23i5007L} at 2024-04-02T08:47:41 with sky coordinates of R.A. = 245.438 deg and DEC. = 25.800 deg with an uncertainty of 1.5${\arcmin}$ in radius \citep[][]{2024GCN.36016....1X}. About 35 hours post LEIA detection at 2024-04-03T19:45:00, EP-FXT also performed follow-up observations and detected the X-ray afterglow at sky coordinates of R.A. = 245.451 deg and DEC. = 25.763 deg with an uncertainty of only 10$\arcsec$ in radius \citep[][]{2024GCN.36022....1J}. Followed by the detection of an X-ray counterpart, we searched for any associated optical counterpart and discovered a clear evidence of an uncataloged source at R.A.= 16$^{h}$21$^{m}$48$^{s}$.225 and DEC = +25$^{\circ}$45$\arcmin$47$\farcs$57 \citep[][]{2024GCN.36027....1Y} with $r$-band magnitude of $22.28\pm0.10$ and $g$-band magnitude of $22.86\pm0.09$ \citep[][]{2024GCN.36027....1Y}. We independently discovered this optical counterpart candidate, although while drafting the GCN Circular, we noticed that it was also reported in \citet[][]{2024GCN.36025....1L}. 

The Gravitational wave high-energy Electromagnetic Counterpart All-sky Monitor\footnote{\url{https://gecam.nssdc.ac.cn/}} (GECAM)-C observations of the FXT field provided crucial insights into the nature of the transient event LXT240402A. The time-averaged spectrum of the GECAM-C instrument revealed characteristics that were strongly indicative of a long {\bf GRB 240402B} \citep[][]{2024GCN.36017....1X}. {\bf The GECAM-C localization of GRB 240402B was found to be consistent with the LEIA localization of LXT240402A within the error \citep[][]{2024GCN.36017....1X}. The associated burst was also detected by Konus-Wind \citep[][]{2024GCN.36028....1R}. Further, \citet[][]{2024GCN.36253....1Y} reported the detection of an X-ray source by Chandra with high significance at the location of the optical counterpart. The Glowbug gamma-ray telescope \citep[][]{2020grbg.conf...57G,2022SPIE12181E..1OW}, operating on the International Space Station, also reported the detection of LXT 240402A/GRB 240402B.} The spectroscopic observations by VLT X-shooter detected a weak continuum throughout the spectrum and claimed to have detected multiple emission lines due to [O~II], H-$\beta$, [O~III], H-$\alpha$, and Ly-$\alpha$ at a common redshift of $z=1.551$ \citep[][]{2024GCN.36385....1T}. The authors proposed these emission lines to be associated with the host. Thus, in our work, we adopted a redshift of 1.551 for LXT240402A.

\subsubsection{EP240414a/AT~2024gsa}
\label{sec:ep240414a}
EP240414a was discovered by EP-WXT at the sky coordinates of R.A. = 191.498 deg and DEC. = -9.695 deg with an uncertainty of 3${\arcmin}$ in radius \citep[][]{2024GCN.36091....1L}. The X-ray transient triggered the EP-WXT  processing unit at 2024-04-14T09:50:12. Followed by the EP-WXT discovery of the X-ray transient, search for the optical counterpart of EP240414a began. About two hours post EP-WXT trigger at 2024-04-14T11:50:01, the EP-FXT also performed follow-up observations. The follow-up EP-FXT observations unambiguously detected an X-ray source at sky coordinates of R.A. = 191.509 deg and DEC. = -9.718 deg with an uncertainty of 10${\arcsec}$ in radius \citep[][]{2024GCN.36129....1G}. 

Followed by the EP-WXT discovery and EP-FXT follow-up observations, several telescopes were quickly triggered to search for the optical counterpart. Our {\tt Kinder} observations from the LOT in $r$ band started about 3.13 hours post EP-WXT trigger. We discovered an optical counterpart candidate at R.A.= 12$^{h}$46$^{m}$01$^{s}$.72 and DEC = -09$^{\circ}$43$\arcmin$08$\farcs$87 \citep[][]{2024GCN.36094....1A} with an $r$-band magnitude of $21.52\pm0.12$. We registered it in the Transient Name Server with an IAU name AT~2024gsa \citep{2024TNSTR1096....1A}. The optical counterpart candidate was associated with the galaxy SDSS~J124601.99-094309.3, having an $r$-band magnitude of 19.04 mag in SDSS DR15 \citep[][]{2019ApJS..240...23A}. The spectroscopic observations of SDSS~J124601.99-094309.3 showed unambiguous hydrogen and oxygen features at $z=0.41$ \citep[][]{2024GCN.36110....1J}. Post the optical counterpart discovery, several optical/NIR follow-up observations also detected the optical counterpart candidate \citep[][]{2024GCN.36105....1X,2024GCN.36150....1S,2024GCN.36154....1L,2024GCN.36171....1W,2024GCN.36189....1K,2024GCN.36355....1L}. Later, \citet[][]{2024GCN.36362....1B} reported the radio detection from MeerKAT. Thus, EP240414a was just the second FXT detected with X-ray, optical, and radio counterparts within a month after EP240315a. 

EP240414a displays complex temporal evolution across X-ray, optical, and radio bands, with a red, non-thermal optical rebrightening and late-time SN emergence \citep{2025ApJ...978L..21S,
2025ApJ...981...48B,2024arXiv241002315S}. Several theoretical interpretations have been proposed, including a weak relativistic jet associated with a broad-lined Type Ic SN \citep{2024arXiv241002315S}, afterglow emission from a mildly relativistic, possibly off-axis or choked jet \citep{2025ApJ...981...48B}, jet breakout through an extended circumstellar medium \citep{2025arXiv250316243H}, cocoon-dominated emission viewed off-axis \citep{2025arXiv250324266Z}, and shock interactions analogous to luminous fast blue optical transients \citep{2025ApJ...982L..47V}.

\subsubsection{EP240416a}
\label{sec:ep240416a}
EP240416a was discovered by EP-WXT at the sky coordinates of R.A. = 203.150 deg and DEC. = -13.612 deg with an uncertainty of 3$\arcmin$ in radius while performing calibration observation \citep[][]{2024GCN.36138....1C}. The EP-WXT detected this FXT at 2024-04-16T02:42:13. 
Followed by the EP-WXT discovery, several telescopes were triggered to search for the optical counterpart. Our observations from LOT in $g$ band started about 14.15 hours post EP-WXT discovery. We discovered a plausible optical counterpart candidate at R.A.= 13$^{h}$32$^{m}$34$^{s}$.51 and DEC = -13$^{\circ}$37$\arcmin$49$\farcs$05 \citep[][]{2024GCN.36139....1C} with a $g$-band magnitude of $22.39\pm0.12$. Followed by the discovery of optical counterpart candidate, several telescope performed optical/NIR follow-up observations, but none of those observations reported any further detection \citep[][]{2024GCN.36142....1A,2024GCN.36155....1X,2024GCN.36157....1P,2024GCN.36158....1B,2024GCN.36172....1M}. The reported upper limits were all later and also shallower than ours. Probably, due to these reasons, none of them detected the proposed optical counterpart in follow-up observations.

\subsubsection{EP241201a}
\label{sec:ep241201a}
The EP-WXT discovered EP241201a at 2024-12-01T21:01:22. It was detected at sky coordinates of R.A. = 282.596 deg, DEC = 66.081 deg with an uncertainty of 2$\arcmin$.343 in radius \citep[][]{2024GCN.38415....1C}. About 12 hr post EP-WXT discovery, a follow-up observation was performed using EP-FXT, which detected an uncataloged X-ray source at R.A. = 282.4865 deg, DEC. = 66.0693 deg with an uncertainty of 10$\arcsec$ in radius \citep[][]{2024GCN.38430....1C}.
Followed by the EP-WXT discovery and EP-FXT follow-up observations, a few telescopes were triggered to search for the optical counterpart. The observations by \citet[][]{2024GCN.38416....1L} provided a shallow limit of about 19.0 -- 19.5 mag in their four pointings. Our observations from LOT in $r$ band started about 13.40 hours post EP-WXT discovery. We proposed a plausible optical counterpart at  R.A. = 18$^{h}$50$^{m}$38$^{s}$.079
DEC. = +66$^{\circ}$03$\arcmin$11$\farcs$36 with $r$-band magnitude of $22.69\pm0.41$\,mag \citep[][]{2024GCN.38418....1L}. The further optical follow-up observations from other facilities were either delayed or had shallower limits than us. None of them found any evidence of the proposed optical counterpart \citep[][]{2024GCN.38425....1L,2024GCN.38427....1P,2024GCN.38446....1D}.

\subsubsection{EP241202b}
\label{sec:ep241202b}
EP241202b was first detected by EP-WXT at 2024-12-02T15:12:55 with the sky coordinates of R.A.= 45.302 deg, DEC. = 2.441 deg with an uncertainty of 2$\arcmin$.6 in radius \citep[][]{2024GCN.38426....1Z}. 
Followed by EP-WXT discovery, our observations utilizing LOT in the $r$ band started about 19.45 hours later. We proposed a plausible optical counterpart candidate at R.A. = 03$^{h}$01$^{m}$20$^{s}$.862
DEC. = +02$^{\circ}$26$\arcmin$27$\farcs$04 with $r$-band magnitude of $22.41\pm0.16$\,mag \citep[][]{2024GCN.38433....1N}. Only a few telescopes searched further to confirm the proposed optical counterpart. None of them detected the proposed optical counterpart as the limits were shallow \citep[$>$19.5 mag in $clear$ band at $\sim$19.2 hours in][]{2024GCN.38444....1Z} and \citep[$>$21.5 mag in $clear$ band at $\sim$14.4 hours in][]{2024GCN.38428....1L}.

\section{Discussions} \label{sec:discussions}

\subsection{Coincidences of FXTs with GRBs or their detection in gamma-ray bands}

{\bf We noticed that a few FXTs in our sample were associated with GRBs. EP240315a was found coincident with GRB 240315C (see Sec.~\ref{sec:ep240315a}), LXT240402A was associated with GRB 240402B through significant coincidence (see Sec.~\ref{sec:lxt240402a}), EP240801a was also discovered by Fermi-GBM and its multi-wavelength properties resembled GRB 221009A (BOAT, \citealt[][]{2023ApJ...946L..31B}) under the assumption of two-component jet model (see Sec.~\ref{sec:ep240801a}), and EP241030a was coincident with GRB 241030A (see Sec.~\ref{sec:ep241030a}). Thus, among the sources having confirmed optical detections in our observations, four of those showed their direct association with GRBs either through coincidence or through direct detection in gamma-ray bands.

Among the source with upper limits in our observations, eight FXTs, namely, EP240617a (a weak gamma-ray transient was discovered within EP-WXT localization, see Sec.~\ref{sec:ep240617a}), EP240703a (coincident with GRB 240703A, see Sec.~\ref{sec:ep240703a}), EP240802a (coincident with GRB 240802A, see Sec.~\ref{sec:ep240802a}), EP240913a (coincident with GRB 240913C, see Sec.~\ref{sec:ep240913a}), EP240919a (coincident with GRB 240919A, see Sec.~\ref{sec:ep240919a}), EP241104a (coincident with GRB 241104A, see Sec.~\ref{sec:ep241104a}), EP241115a (coincident with GRB 241115D, see Sec.~\ref{sec:ep241115a}), and EP241208a (coincident with a long and soft transient discovered by SVOM-ECLAIRS, see Sec.~\ref{sec:ep241208a})  provided strong evidence of their association with GRBs through significant spatial and temporal coincidences.}

Thus, in our sample of 42 FXTs followed from the Lulin Observatory, {\bf twelve} of them showed strong evidence of their association with GRBs either through spatial and temporal coincidences {\bf or through direct detection in gamma-ray bands.}

\subsection{FXTs with no/faint optical counterparts and the context of `dark FXTs'}
{\bf For a subset of FXTs, specifically EP240413a, EP240918b, EP240918c, EP241104a, EP241115a, EP241125a, EP241206a, EP241201a, EP241202b, and EP241208a, the optical follow-up was limited, with only one to three telescopes triggering post EP-WXT detection. This limited response was primarily either due to significant delays in the release of EP-detection GCN alerts (exceeding 10 hours post-detection), or the timing of the alerts coinciding with weekends. Except EP241201a and EP241202b (with reported optical counterpart fainter than $\sim$22.4 mag), no others were detected in optical bands. 

We notice that the majority (49 out of a total of 72) of FXTs were not detected with any associated optical counterpart. In our sample of 42 FXTs, 30 were not detected in optical bands. As presented in Table~\ref{tab:Photometry8} and Table~\ref{tab:Photometry8}, 11 FXTs (namely, EP240408a, EP240413a, EP240506a, EP240617a, EP240625a, EP240626a, EP240802a, EP240908a, EP240913a, EP241026b, EP241115a) were followed $\gtrsim$ 24\,hr from the Lulin Observatory, post their detection in X-ray by EP-WXT. Among these, EP240625a (a weak and ambiguous candidate reported with $z$ = 20.7$\pm$0.2 at 1.52 hr, \citealt[][]{2024GCN.36761....1F}), EP240908a (very faint optical counterpart with an $r$-band magnitude of $\sim$24 mag at $\sim$18.6 hr, \citealt[][]{2024GCN.37438....1Q}), and EP241026b (late discovery of an optical counterpart with $r$ = 21.6\,mag discovered at 35.52 hr although galactic origin of the source could not be completely ruled out, \citealt[][]{2024GCN.37938....1R}) had optical counterpart, but missed by us either due to their faintness or late trigger of telescopes. However, we notice that despite early follow-up from other telescopes, no credible optical counterpart candidates in the following sources were found: EP240413 (GOTO $L$-band limit of $>$20.4 mag at $\sim$21 hr, \citealt[][]{2024GCN.36098....1P}), EP240626a (KAIT optical upper limit of $>$19.0 mag in $clear$-band at 1.01 hr, \citealt[][]{2024GCN.36773....1Z}), EP240802a (KAIT optical upper limit of $>$21.0 mag in $clear$-band at $\sim$17.8 hr, \citealt[][]{2024GCN.37094....1Z}), EP240913a (JinShan optical upper limits of $>$21.9 mag in $r$-band at 5.1 hr, \citealt[][]{2024GCN.37500....1Z}), EP241115a (Global MASTER-Net upper limit of 18.6\,mag in $clear$-band at $\sim$12.7 hr, \citealt[][]{2024GCN.38242....1L}). Thus, a) delayed trigger of telescopes to search for the optical counterpart, b) intrinsically faint nature of the associated optical counterpart, and/or c) optical counterpart being intrinsically absent, could be the possible reasons for their non-detection.

The intrinsically faint nature of the associated optical counterpart and/or optical counterpart being intrinsically absent despite their detection in X-ray bands suggest that a major fraction of FXTs could be `dark FXTs' similar to `dark GRBs' \citep[][]{2001A&A...369..373F,2004ApJ...617L..21J, 2011A&A...526A..30G,2022JApA...43...11G, 2024A&A...683A..55C}. Originally, the term `dark GRB' was coined for the GRBs that were detected in X-rays but without an optical counterpart \citep[][]{2001A&A...369..373F}. Following this analogy, the FXTs with no optical counterparts should be classified as `dark FXTs'.  Later, the nomenclature of `dark GRBs' was made more specific by adding a time and brightness limit, e.g., $R>$23 mag at 1-2 days of the burst \citep[][]{2001ApJ...562..654D}. These constraints on the time and brightness limit will still make several FXTs even with optical counterparts from categorizing them as `dark FXTs' since they are usually fainter than 23\,mag in optical bands (Fig.~\ref{fig:gcn_summary}) at 1-2 days post their discovery in X-ray bands. A more advanced and robust condition of optical-to-X-ray spectral index $\beta_{\mathrm {OX}} < 0.5$ was proposed by \citet[][]{2004ApJ...617L..21J}. The estimation of $\beta_{\mathrm {OX}}$ for the sources with optical detection is beyond the scope of this paper, although we quote the $\beta_{\mathrm {OX}}$ for a few FXTs available from the literature. While performing the multi-wavelength afterglow fitting for EP240315a, \citet[][]{2024arXiv240416425L} reported a spectral index of -0.93$^{+0.04}_{-0.04}$, consistent with EP250315a falling under the category of `dark FXTs'. For EP240414a, \citet[][]{2025ApJ...978L..21S} fit a single power law to the optical and X-ray data points, and measure a spectral index of -0.76$\pm$0.05. With $\beta < 0.5$ for the optical and X-ray data, EP240414a also fall under the category of `dark FXTs'. Further, for EP241021a, \citet[][]{2025arXiv250314588B} estimated the optical-to-IR spectral index ($\beta_{\mathrm {OIR}}$) of -1.09$^{+0.05}_{-0.06}$, and the extrapolation of the OIR spectral index to X-ray data was consistent. Thus, EP241021a fell under the category of `dark FXTs'. 
With these results, we conclude that the majority of FXTs are `dark FXTs' due to the non-detection of any associated optical counterpart. Although several FXTs have optical counterpart detection, those can still fall under the category of `dark FXTs' due to $\beta_{\mathrm {OX}} < 0.5$.}

\begin{figure*}
\centering
    \includegraphics[width=\textwidth,angle=0]{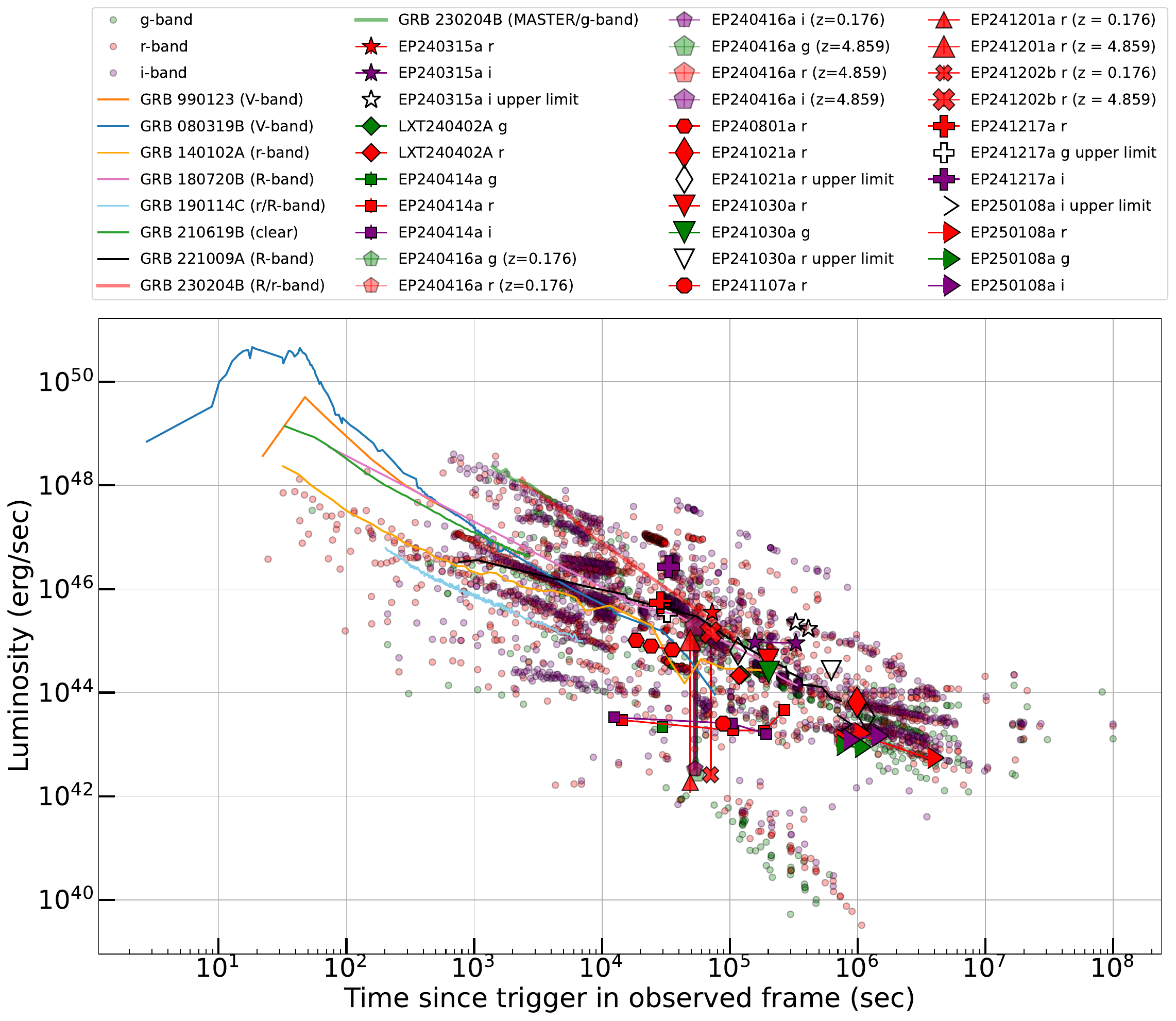}
  
   \caption{The comparison of the optical luminosity of FXTs in our study with a sample of 535 GRBs from \citet[][]{2024MNRAS.533.4023D}  in $g$-, $r$-, and $i$-bands (round markers). A few peculiar and bright GRBs are also shown and highlighted for comparison. The optical luminosities of FXTs are consistent with GRBs. The redshifts of EP240416a, EP241201a, and EP241202b are unknown; however, the vertical lines joining the smaller and larger markers show the range of luminosities had they occurred within the known range of redshift for FXTs in our sample (i.e., $z = 0.176$ to 4.859). The sources of optical light curve of highlighted GRBs are as follow: GRB~990123 from \citet[][]{1999Natur.398..400A}, GRB~080319B from \citet[][]{2008Natur.455..183R}, GRB~140102A from \citet[][]{2021MNRAS.505.4086G}, GRB~180720B from \citet[][]{2021RMxAC..53..113G}, GRB~190114C from \citet[][]{2020ApJ...892...97J}, GRB~210619B from \citet[][]{2023NatAs...7..843O}, GRB~221009A from \citet[][]{2023ApJ...942...34R}, and GRB~230204B from \citet[][]{2024arXiv241218152G}.}
    \label{fig:comp_GRB}
\end{figure*} 

\begin{figure*}
\centering
    \includegraphics[width=\textwidth,angle=0]{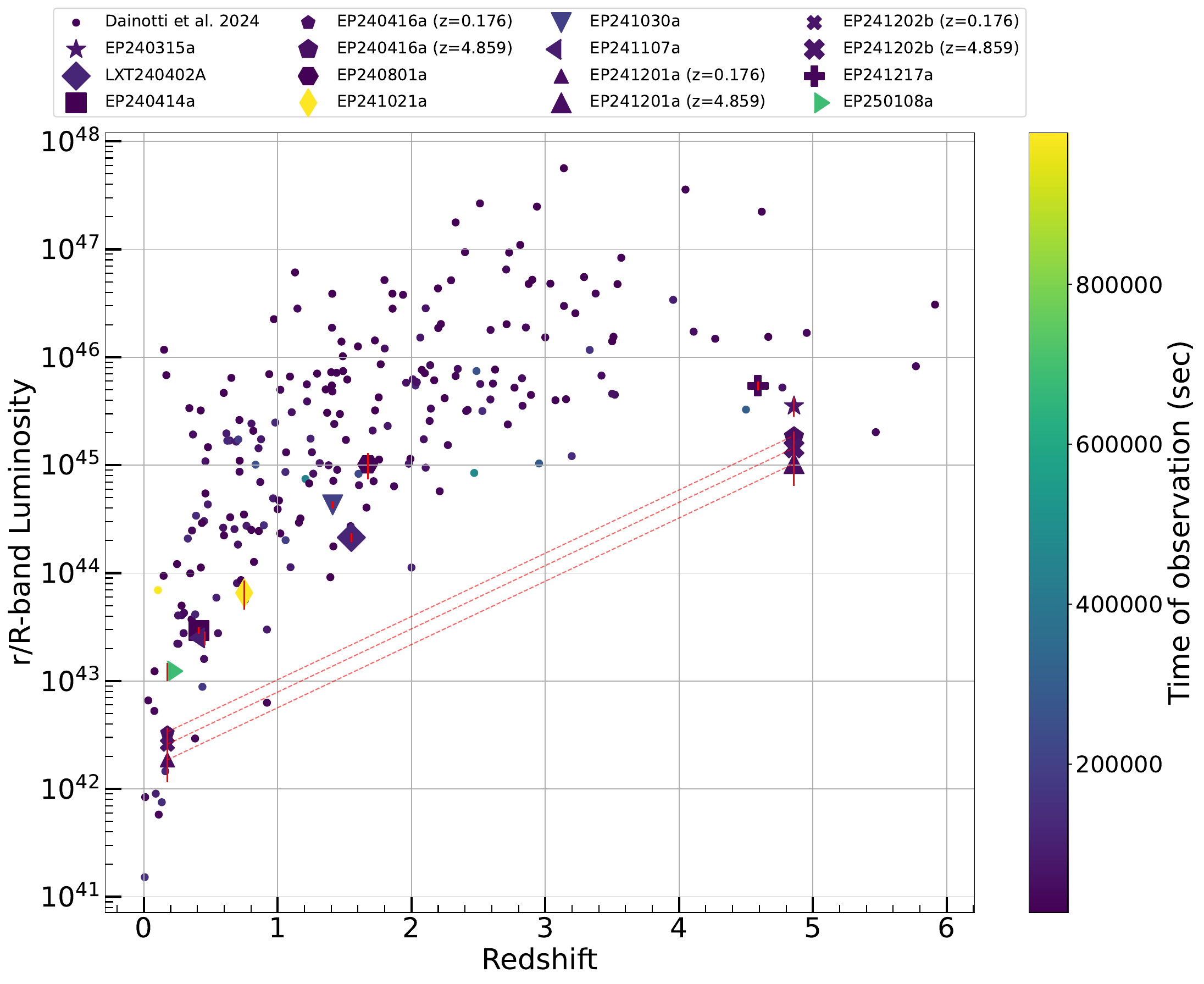}
  
   \caption{The redshift of the occurrence of the FXTs in our sample and their earliest $r/R$-band luminosity compared with a sample of GRBs from \citet[][]{2024MNRAS.533.4023D}. Once again, the redshift and luminosity of FXTs are consistent with GRBs. The dashed lines show the range of luminosities of EP240416a, EP241201a, and EP241202b, had they occurred within the observed redshift range of FXTs in our sample (i.e., $z=0.176$ to 4.859). The luminosities of EP240416a, EP241201a, and EP241202b lie towards the fainter limit had they occurred at high redshifts. They are more consistent with low-redshift GRBs. The redshifts of GRBs are compiled in \citet[][]{2024arXiv241218152G}.}
    \label{fig:GRB_z}
\end{figure*}

\subsection{Comparison with GRBs, TDEs, FBOTs, SBO and kilonova}
\label{subsec:comparison}

First, we compare the optical luminosity of FXTs with GRBs. 
To estimate the optical luminosity from the observed magnitudes, we proceed as follows. The apparent magnitude $m$ is first converted to absolute magnitude $M$ using the formula:
\[
M = m - 5\log_{10}\left(\frac{D_{\mathrm{L}}}{10~\mathrm{pc}}\right) + 2.5\log_{10}(1+z) - A_\lambda,
\]
where $D_{\mathrm{L}}$ is the luminosity distance derived from the redshift $z$ assuming a flat $\Lambda$CDM cosmology, and $A_\lambda$ represents the Galactic extinction. {\bf We ignored the host galaxy extinction as the FXTs appear mostly hostless or far from the host.} The term $2.5\log_{10}(1+z)$ serves as an approximate $k$-correction to account for the redshifted spectral energy distribution. The absolute magnitude is then converted to monochromatic luminosity (in erg\,s$^{-1}$\,Hz$^{-1}$) via:
\[
L = 3.0128 \times 10^{35} \times 10^{-0.4M},
\]
where the zero-point $3.0128 \times 10^{35}$ erg\,s$^{-1}$\,Hz$^{-1}$ corresponds to $M = 0$. This approach yields an estimate of the rest-frame luminosity that incorporates both distance and an approximate $k$-correction. 

\begin{figure*}
\centering
    \includegraphics[width=\textwidth,angle=0]{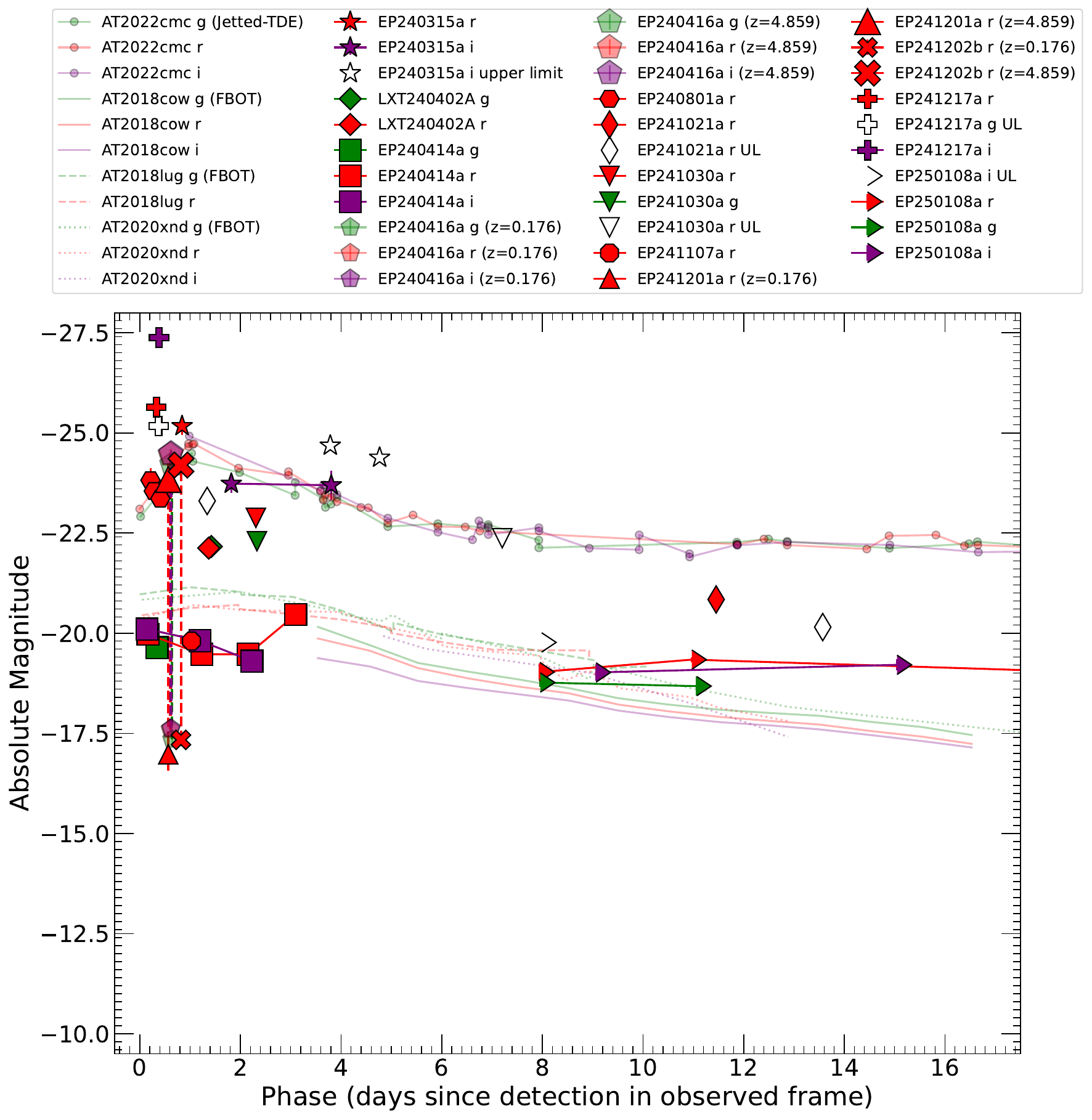}
  
   \caption{The comparison of the absolute optical magnitudes of the FXTs in our sample with the only confirmed jetted TDE, AT~2022cmc.  The optical data for AT~2022cmc are obtained from \citet[][]{2022Natur.612..430A}. Besides matching the GRBs, FXTs EP240315a and EP240801a are also consistent with jetted TDE. It also shows the comparison with the FBOT events AT~2018cow, AT~2018lug, and AT~2020xnd. The dashed, vertical lines show the range of optical absolute magnitudes had EP240416a, EP241201a, and EP241202b occurred within the range of known redshift for FXTs in our sample ($z=0.176$ to 4.859). The sources of optical data are: AT~2018cow \citep[][]{2018ApJ...865L...3P}; AT~2018lug \citep[][]{2020ApJ...895...49H}; and AT~2020xnd \citep[][]{2021MNRAS.508.5138P}. The dashed lines show the range of optical absolute magnitudes if EP240416a, EP241201a, and EP241202b had occurred within the range of known redshift for FXTs in our sample ($z=0.176$ to 4.859).}
    \label{fig:comp_TDE_FBOT}
\end{figure*} 

{\bf The top panel of Fig.~\ref{fig:gcn_summary} shows the earliest optical detections (upper limits in case of non-detections) for the sample of FXTs reported in GCN. 23 out of 72 FXTs are detected with optical counterparts. The sources with confirmed optical detections have a mean (median) optical magnitude of 20.89 mag (21.48 mag), with a standard deviation of 1.81 mag. Further, the mean (median) time of optical detection post X-ray discovery is 10.92 hr (4.76 hr), with a standard deviation of 12.74 hr. The earliest mean (median) $r/R$-band magnitudes of GRBs estimated from \citet[][]{2024MNRAS.533.4023D} is 19.10 mag (19.23 mag) with a standard deviation of 2.50 mag. Additionally, the mean (median) time of first $r/R$-band observations is 9.70 hr (1.79 hr) with a standard deviation of 26.54 hr. The earliest mean (median) $r/R$-band magnitude of GRBs appears brighter than FXTs because a large number of GRBs are discovered in optical bands much earlier in time when compared to the discovery of FXTs, as shown in Fig.~\ref{fig:comp_GRB}. For the sources with upper limits, the mean (median) upper limits in optical bands is 20.32 mag (20.5 mag), respectively with a standard deviation of 1.44 mag. Additionally, the mean (median) time of first upper limit is 18.87 hr (13.98 hr) since detection, respectively, with a standard deviation of 17.63 hr.}

Fig.~\ref{fig:comp_GRB} shows the comparison of the optical luminosities of FXTs in our analysis with a sample of 535 GRBs from \citet[][]{2024MNRAS.533.4023D}. 
A few peculiar and bright GRBs are also highlighted. The optical luminosities of FXTs are very well consistent with GRBs. The redshifts of EP240416a, EP241201a, and EP241202b are unknown; however, the vertical lines joining the smaller and larger markers show the range of luminosities, had they occurred within the known range of redshift for FXTs in our sample (i.e., $z=0.176$ to 4.859). The optical luminosities of all the FXTs in our sample with confirmed optical counterparts are consistent with the luminosities of GRBs, no matter how early or late after the X-ray detections their optical counterparts have been discovered. 

  

Fig.~\ref{fig:GRB_z} shows the redshifts of the occurrences of the FXTs with confirmed optical counterparts in our sample and their luminosity beyond 14,300 seconds since detection compared with a sample of GRBs from \citet[][]{2024MNRAS.533.4023D}. The choice of a cutoff of 14,300 seconds is chosen since our earliest $r$-band detection is at 14,292 seconds for EP240414a. For the rest of the FXTs in our sample, their detections are beyond 14,292 seconds. The dashed lines show the range of luminosities of EP240416a, EP241201a, and EP241202b, had they occurred within the observed redshift range of FXTs (i.e., $z = 0.176$ to 4.859) in our sample. Once again, we find that the redshifts and luminosities of FXTs are consistent with GRBs. In particular, {\bf for a given redshift, the FXTs seem to lie towards the faint end of GRB populations.}

Fig.~\ref{fig:comp_TDE_FBOT} shows the comparison of the absolute magnitudes of the FXTs having confirmed optical counterparts in our sample with the only confirmed jetted TDE event AT~2022cmc. The optical data for AT~2022cmc are obtained from \citet[][]{2022Natur.612..430A}. The dashed lines show the range of optical absolute magnitudes if EP240416a, EP241201a, and EP241202b had occurred within the range of known redshifts for FXTs in our sample ($z = 0.176$ to 4.859). Besides being consistent with GRBs, the optical luminosities of  FXTs EP240315a and EP240801a are also consistent with the jetted TDE event. However, EP240315a is confirmed to be associated with GRB~240315C \citep{2025NatAs.tmp...34L}, which diminishes the probability of EP240315a having a jetted TDE origin.  {\bf Recently, \citet[][]{2025arXiv250304306J} reported the detection of EP240801a by Fermi-GBM. Further, the detailed multi-wavelength analysis by \citet[][]{2025arXiv250304306J} showed the resemblance of EP240801a with GRB 221009A (BOAT, \citealt[][]{2023ApJ...946L..31B}) under the assumption of two component jet models. Thus, EP240801a was probably another off-axis jet or intrinsically weak GRB.} Among the sources with unknown redshifts, the optical luminosities of EP240416a, EP241201a, and EP241202b are also consistent with jetted TDE if we consider them to occur at the highest known redshift for FXTs. None of the other FXTs produce luminosities consistent with the jetted TDE event.

{\bf In the case of EP240414a, the entire optical evolution was divided into three phases \citep[][]{2024arXiv241002315S,2024arXiv240919056V}. Followed by the first blue peak in optical bands, the source luminosity reduced slowly with time. It then reached the second peak with a fast rise rate, and the light curve subsequently faded rapidly thereafter. Such fast optical evolution resembled the evolution of luminous FBOTs. The observations by \citet[][]{2024arXiv240919056V}  associated the progenitors of typical long GRBs to some FXTs even in the absence of a detected GRB, and find a possible connection between FXTs and LFBOTs.} Additionally, several FXTs have detection with corresponding radio counterparts. Thus, we find it reasonable to compare the luminosities of FXTs with FBOTs. Fig.~\ref{fig:comp_TDE_FBOT} also shows the comparison of the absolute magnitudes in optical bands of the FXTs in our sample with luminous and radio-loud FBOT events \citep{2023ApJ...949..120H}, including AT~2018cow \citep[][]{2018ApJ...865L...3P}, AT~2018lug \citep[][]{2020ApJ...895...49H}, and AT~2020xnd \citep[][]{2021MNRAS.508.5138P}. The dashed lines show the range of optical absolute magnitudes had EP240416a, EP241201a, and EP241202b occurred within the range of known redshift for FXTs in our sample ($z=0.176$ to 4.859). Besides EP240314a in its early phases and EP250108a in its late stages, none of the other FXTs have luminosities comparable to these FBOTs.   

\begin{figure*}
\centering
    \includegraphics[width=\textwidth,angle=0]{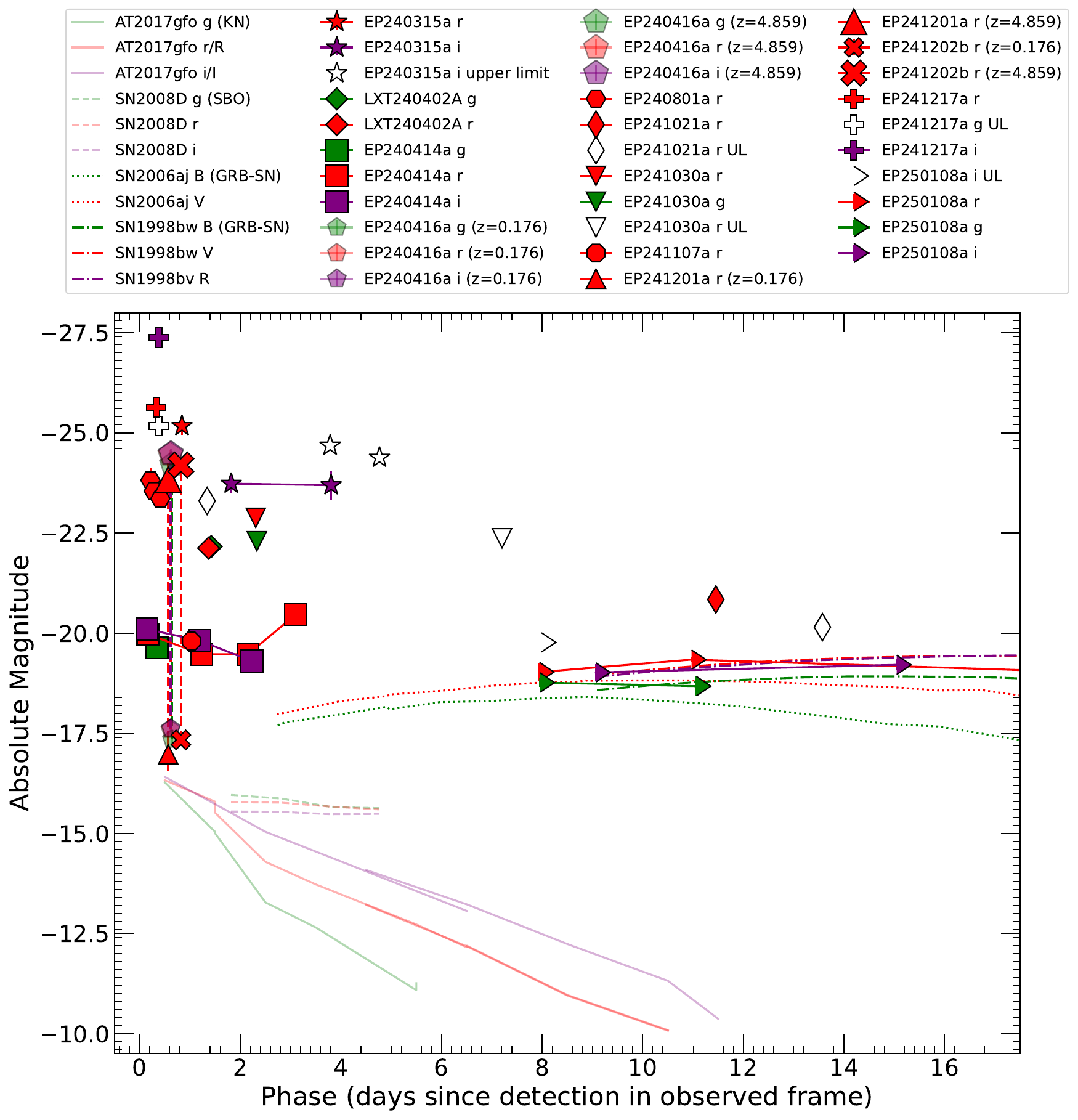}
  
   \caption{The comparison of the absolute magnitudes in optical bands of the FXTs in our sample with the Kilonova event AT~2017gfo. The absolute magnitudes are also compared with the early SBO component of SN2008D, a weak x-ray flash (XRF) event followed by an early optical detection \citep[][]{2008Sci...321.1185M}. SN 2008D proves to be a possible link between X-ray transients and the SBO phenomena. {\bf Additionally, the comparison with GRB-SNe 1998bw (data from \citealt[][]{2011AJ....141..163C}) and 2006aj (data from \citealt[][]{2006A&A...457..857F}) are also shown}. The dashed, vertical lines show the range of optical absolute magnitudes if EP240416a, EP241201a, and EP241202b had occurred within the range of known redshift for FXTs in our sample ($z=0.176$ to 4.859).}
    \label{fig:comp_KNe_SBO}
\end{figure*}  

To compare the optical bands' absolute magnitudes of FXTs in our sample with NS-NS merger events in Fig.~\ref{fig:comp_KNe_SBO}, we use the only confirmed kilonova source AT~2017gfo. {\bf The kilonovae (KNe) are the robust indicator of binary merger (NS-NS or NS-BH) events \citep[][]{2010MNRAS.406.2650M} and they are among the rapidly evolving transients.} The $g$, $r$/$R$, $i$/$I$-band data are obtained from \citet[][]{2017Natur.551...67P,2017Natur.551...75S}. The light curves are corrected for galactic extinction using the NED. Additionally, following \citet[][]{2017ApJ...848L..31H}, we use a redshift of 0.009783 for AT~2017gfo. The detection of SN~2008D, preceded by the first discovery in the X-ray band, suggested a possible link between X-ray transients and SN shock-breakout (SBO) events. Fig.~\ref{fig:comp_KNe_SBO} also shows the comparison of the absolute magnitudes in optical bands of the FXTs in our sample with SN~2008D SBO. The $g$, $r$, and $i$-band data are obtained from \citet[][]{2008Natur.453..469S}. Following \citet[][]{2008Sci...321.1185M}, we correct the light curves of SN~2008D for a total extinction, E(B-V) of 0.65 mag. Further, following \citet[][]{2008MNRAS.388..603L}, we adopt a redshift of 0.006494 for SN~2008D while estimating the absolute magnitudes. As displayed in this Figure, the FXTs' absolute magnitudes seem much brighter than the SBO and kilonova events. {\bf The emergence of SN component in the last evolutionary stages of EP240414a \citep[][]{2025ApJ...978L..21S,2024arXiv241002315S} and EP250108a \citep[][]{2025arXiv250408889R,2025arXiv250417516S,2025arXiv250417034L} is well established, thus, we also show the comparison of the FXTs in our sample with the GRB-SNe events, SN~1998bw and SN~2006aj. SN~1998bw was the first GRB-SN event \citep[][]{1998Natur.395..670G}. The intrinsic $BVR$ data for SN~1998bw are taken from \citet[][]{2011AJ....141..163C}. The total extinction corrected $B$ and $V$-band data (subjected to host-flux correction too) for SN~2006aj are obtained from \citet[][]{2006A&A...457..857F}. The FXT EP250108a appears similarly luminous as GRB-SN events.}

\begin{figure*}
\centering
    \includegraphics[height=6cm]{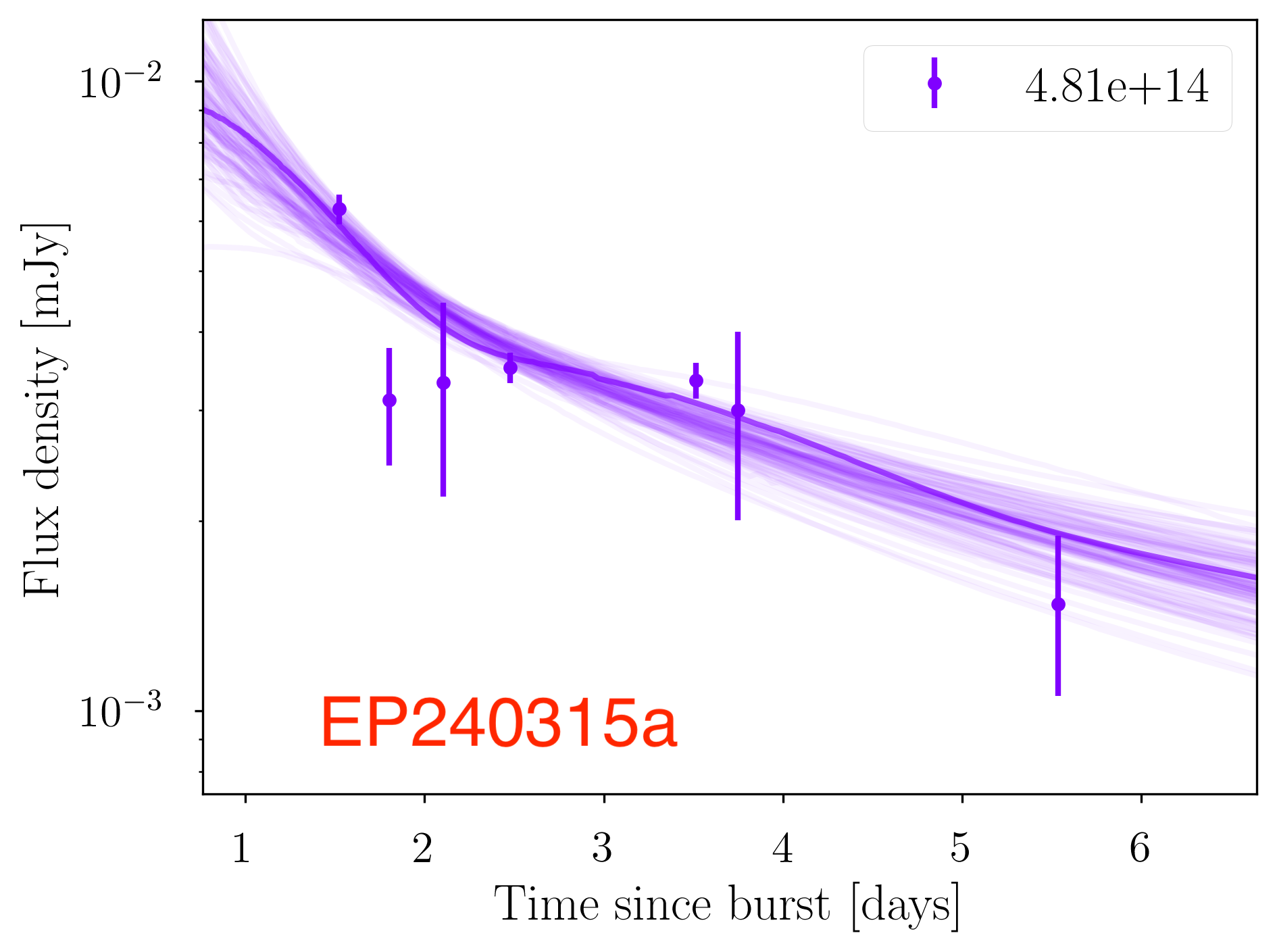}
    \includegraphics[height=6cm]{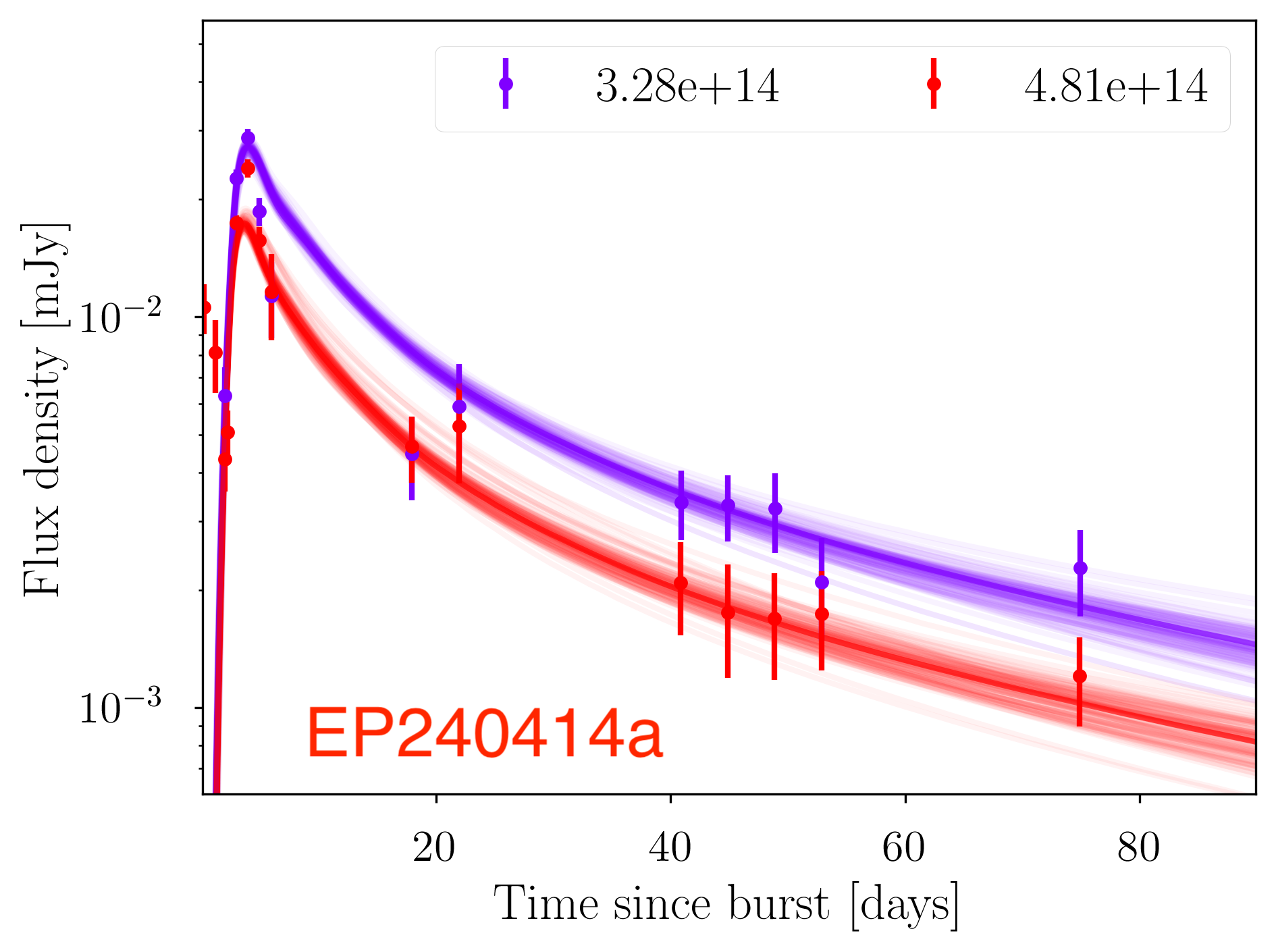}
    \includegraphics[height=6cm]{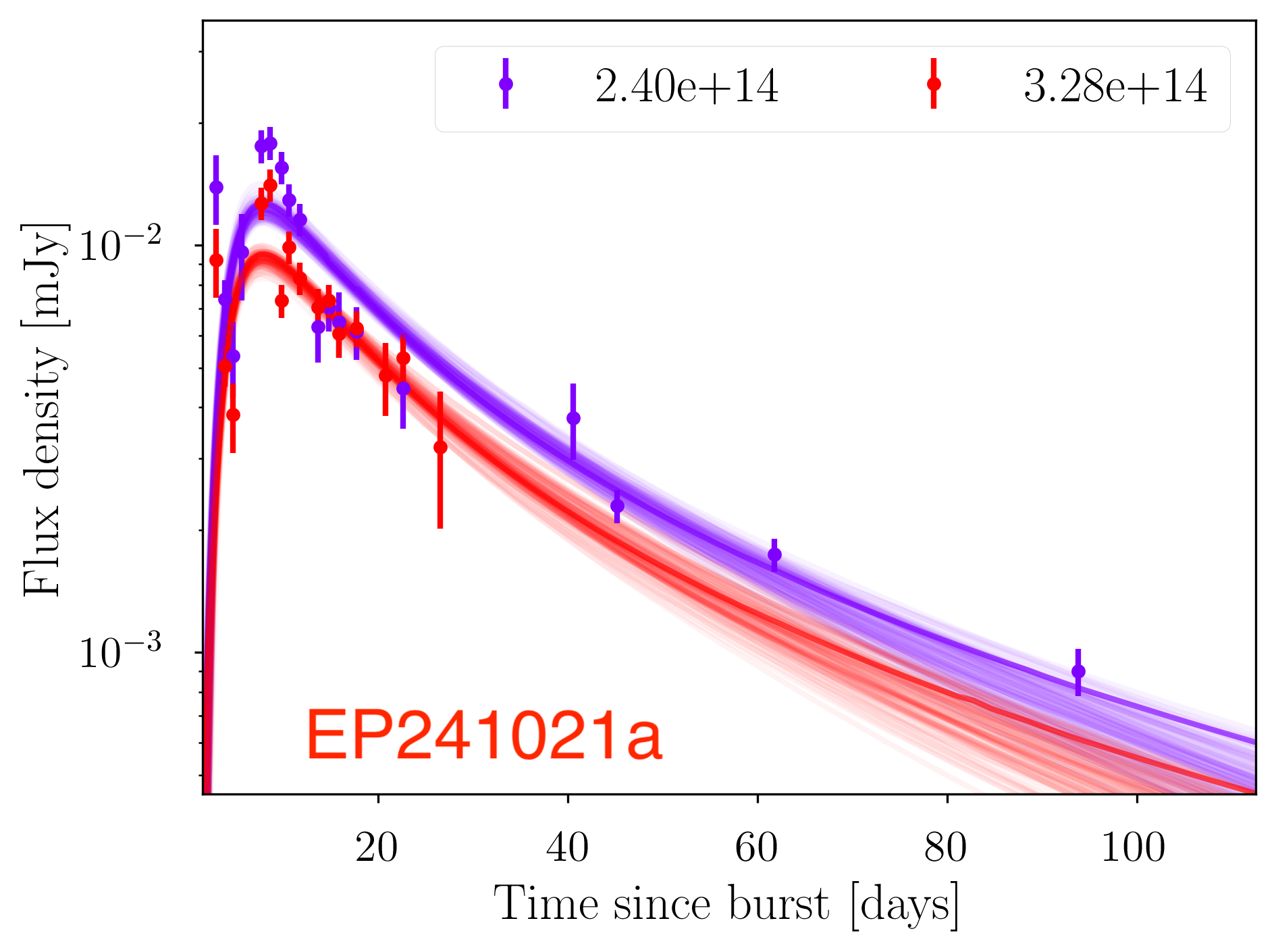}
  
   \caption{Outcomes of the broadband afterglow (assuming the ``gaussiancore" model) fitting using {\tt REDBACK} \citep[][]{sarin_redback} to the $i$-(4.81e+14 Hz), $z$-(3.28e+14 Hz) and $J$-band (2.40e+14 Hz) light curves of well-studied FXTs. The top-left panel shows the fitting to the $i$-band light curve of EP240315a; the top-right panel shows the fittings to the $z$- and $i$-band light curves of EP240414a, and the bottom panel shows the fittings to the $J$-band and $z$-band light curves of EP241021a. Sources of light curves: EP240315a \citep[][]{2024ApJ...969L..14G}, EP240414a \citep[][]{2025ApJ...978L..21S}, and EP241021a \citep[][]{2025arXiv250314588B}. To avoid the problem of Lyman-${\alpha}$ leaking, only bands redder than $i$ bands are used. The best-fit parameters are listed in Table~\ref{tab:redback}.}
    \label{fig:redback}
\end{figure*}

\begin{figure*}
    \centering
    \includegraphics[width=\textwidth]{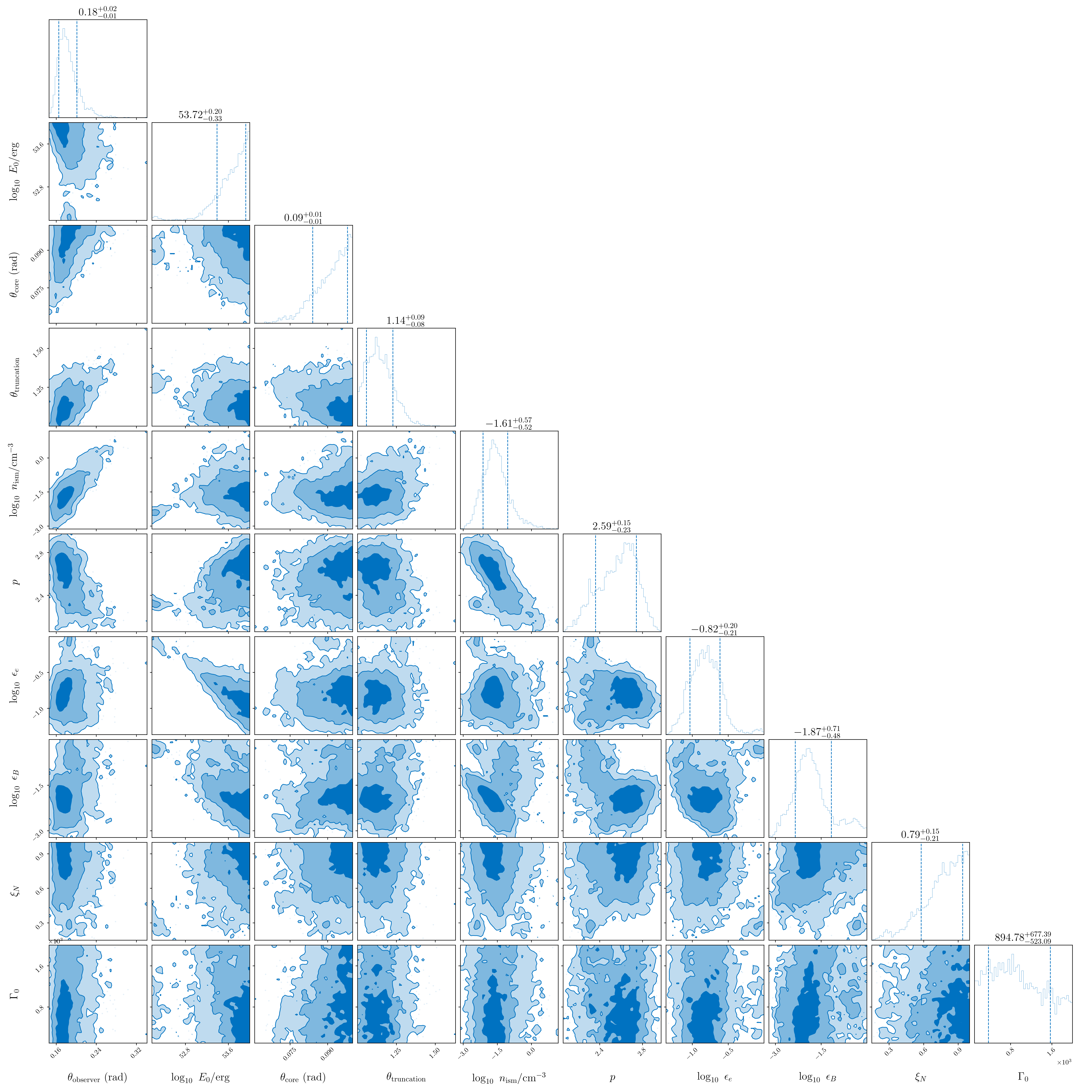}
    \caption{The corner plot of the posteriors for the ``gaussiancore'' afterglow model using {\tt REDBACK} for EP241021a.}
    \label{fig:redback_corner}
\end{figure*}

Overall, we find that the properties of FXTs with optical counterparts in our sample align well with low luminosity GRBs. Thus, we utilize {\tt REDBACK} \citep[][]{sarin_redback} to model the broadband afterglow, incorporating the ``gaussiancore" structured jet model to fit the optical light curves of three best studied and best followed FXTs in literature, i.e., EP240315a, EP240414a, and EP241021a. {\bf This model has been successful in explaining the GRB afterglows} \citep[e.g.,][]{2024arXiv241218152G}. To avoid the problem of Lyman-${\alpha}$ leaking, only bands $\ge$ $i$-bands are used in fitting. The ``gaussiancore" structured jet model is parameterized by a normalization ${E_0}$, a width $\theta_{core}$, and a truncation angle $\theta_{truncation}$ outside of which the energy is initially zero. $\theta_{observer}$ is the viewing angle, $\epsilon_{B}$ is the fraction of thermal energy in the magnetic field, and a fraction $\xi_{N}$ of the electron population is shock-accelerated electrons with a fraction $\epsilon_{e}$ of thermal energy. The model is explicitly discussed in \citet[][]{2020ApJ...896..166R}. We used the default range of parameters specified in {\tt REDBACK}. Fig.~\ref{fig:redback} shows that barring the initial decline after the first detection, the rest of the luminosities could be nearly produced by the afterglow model (for examples on GRB afterglows, see \citealt[][]{1998ApJ...497L..17S, 1999ApJ...519L..17S, 2023arXiv231216265G}). The details of the posterior distribution for each of the sources are mentioned in Table~\ref{tab:redback}. The `corner' plot displaying the posteriors from fitting for EP240414a is shown in Fig.~\ref{fig:redback_corner}. These results further favor the FXT-GRB link. 

\section{Conclusions} \label{sec:conclusions}

Following the detection of the optical counterpart of EP240315a, a growing number of telescopes have initiated prompt and systematic follow-up observations whenever the EP-mission reports a new FXT. This increased attention from the astronomical community has led to a surge in the follow-up observations of FXTs, aiming to capture and analyze these transient events across different wavelengths. In this study, we have conducted an extensive search for the optical counterparts of FXTs discovered by the EP-mission during its first year of operation, utilizing the 40cm SLT and 1m LOT telescopes at the Lulin Observatory as a part of the {\tt Kinder} project. We were able to observe the fields of 42 FXTs. Out of these, 12 FXTs were clearly detected in our observations. For five out of these 12, we discovered the optical counterpart candidates. There were a total of 30 remaining FXTs for which we did not detect any optical counterpart in our observations from the Lulin Observatory. We reported upper limits across various optical bands for such FXTs. Additionally, we utilized the valuable communications available on the GCN and ATel platforms to derive significant insights and conclusions: 

\begin{enumerate}

    \item {EP sources with detected optical counterparts are typically identified within seven hours after the EP-WXT discovery. These counterparts often exhibit relatively bright early-time magnitudes (16-18~mag), followed by a rapid fading trend typically declining to between 22 and 24 mag, with most counterparts clustering around 20 mag. In cases where the counterpart is discovered more than ten hours post-trigger, the observed brightness is generally fainter, around 22 mag.  {\bf We find that the sources with confirmed optical detections have a mean (median) optical magnitude of 20.89 mag (21.48 mag), with a standard deviation of 1.81 mag. Further, the mean (median) time of optical detection post X-ray discovery is 10.92 hr (4.76 hr), with a standard deviation of 12.74 hr. When compared with GRBs, the earliest mean (median) $r/R$-band magnitude of GRBs appears brighter than FXTs because a large number of GRBs are discovered in optical bands much earlier (and hence brighter than FXTs) in time, when compared to FXTs.}}

    \item The EP sources with non-detection in the optical bands are reported with a mean (median) limiting magnitude of 20.32~mag (20.50~mag), with a standard deviation of 1.44~mag. The average (median) delay in reporting the first upper limits is 18.87~hr (13.98~hr), with a standard deviation of 17.63~hr. Hence, we conclude that the following three causes for the non-detection of the optical counterparts for a major fraction of FXTs: a) delayed trigger of the telescopes to search for optical counterparts, b) the optical counterparts are intrinsically very faint, and c) there are no optical counterparts at all, similar to many GRBs.

    \item {The redshift and luminosity distributions of FXTs observed in our study closely match with GRBs, especially at the faint end. {\bf For a given redshift, the FXTs lie towards the faint end of the GRB populations.} Additionally, only 23 out of the 72 high SNR FXTs discovered by the EP-mission show optical afterglows, and the remaining 49 are not detected in optical wavelengths. The non-detection in optical wavelengths supports the interpretation that a major fraction of EP-mission discovered FXTs represents a population of `dark FXTs', similar to `dark GRBs'}

    \item {The multi-band optical luminosities of EP240315a and EP240801a were also consistent with jetted TDE events. However, the direct association of EP240315a with GRB 240315A ruled out the possibility of it being linked to the jetted TDE event. {\bf The detection of EP240801a by Fermi-GBM ruled out the possibility of it being a TDE.} Additionally, EP240416a, EP241201a, and EP241202b were consistent with the jetted TDE event had they been high redshift ($z=4.859$) events.}


    \item {We notice that for a few FXTs, namely, EP240413a, EP240918b, EP240918c, EP241104a, EP241115a, EP241125a, EP241206a, EP241201a, EP241202b, EP241208a, only one, two or three telescopes were triggered to search for their optical counterparts. For these sources, either the EP-detection GCNs are delayed ($>$10 hr since detection) or the GCNs are on weekends. Hence, automated follow-up observations and a wide, time-zone-distributed observational network are crucial for improving the detection rate of optical counterparts.}

\end{enumerate}
These findings highlight the critical role of prompt optical follow-up in unveiling the nature of FXTs. Their high-energy origins are likely associated with GRBs, particularly at the faint end of the luminosity distribution. The large fraction of EP-mission discovered FXT events are detected without optical counterparts, which supports the interpretation that many EP FXTs constitute a population of `dark FXTs', analogous to the so-called `dark GRBs'. While TDEs remain a possible origin for some events, our sample disfavors SN SBOs and kilonovae as dominant contributors. Continued rapid-response observations will be essential for building a statistically significant sample and advancing our understanding of these energetic transients. Future efforts combining multi-wavelength and spectroscopic follow-up will be crucial to further constrain their physical properties and explosion mechanisms.

\section*{Acknowledgments}
{\bf We thank the anonymous referee for providing constructive comments that further improved the manuscript significantly.} We acknowledge the Einstein Probe team for their great work and useful communication. AA and T.-W.C thank Ed Hsing-Wen Lin for useful discussions.

AA and T.-W.C 
acknowledge the Yushan Young Fellow Program by the Ministry of Education, Taiwan for the financial support (MOE-111-YSFMS-0008-001-P1). 
T.-W.C acknowledges funding from the National Science and Technology Council, Taiwan (NSTC grant 114-2112-M-008-021-MY3) to support GREAT Lab. S.Y. acknowledges the funding from the National Natural Science Foundation of China under grant No. 12303046, and the Startup Research Fund of Henan Academy of Sciences No.242041217. SJS acknowledges funding from STFC Grants ST/T000198/1, ST/Y001605/1, a Royal Society Research Professorship and the Hintze Charitable Foundation. 
HFS is supported by Schmidt Sciences.
AAa is supported by the European Research Council (ERC) under the European Union's Horizon 2020 research and innovation programme (grant agreement No. 948381).
RG was sponsored by the National Aeronautics and Space Administration (NASA) through a contract with ORAU. The views and conclusions contained in this document are those of the authors and should not be interpreted as representing the official policies, either expressed or implied, of the National Aeronautics and Space Administration (NASA) or the U.S. Government. The U.S. Government is authorized to reproduce and distribute reprints for Government purposes notwithstanding any copyright notation herein. 
MN is supported by the European Research Council (ERC) under the European Union's Horizon 2020 research and innovation programme (grant agreement No.~948381).
YCP is supported by the National Science and Technology Council (NSTC grant 112-2112-M-008-026-MY3).

This publication has made extensive use of data collected at the Lulin Observatory, partly supported by the TAOVA grants NSTC 112-2740-M-008-002 and NSTC 113-2740-M-008-005. We are thankful to the Lulin Observatory observing staffs for prompt conduct of our ToO observations. We are also thankful to the PIs of the accepted proposals for LOT to facilitate smooth ToO observations from the Lulin Observatory.


%

\vspace{5mm}
\facilities{1m LOT, 40cm SLT, Einstein Probe (EP-WXT and EP-FXT), Swift(XRT and UVOT)}


\software{IRAF \citep[][]{1986SPIE..627..733T,1993ASPC...52..173T}, astropy \citep[][]{2013A&A...558A..33A,2018AJ....156..123A}, source extractor \citep[][]{1996A&AS..117..393B}, hotpants \citep[][]{2015ascl.soft04004B}, scipy \citep[][]{2020NatMe..17..261V}, numpy \citep[][]{2020Natur.585..357H}, matplotlib \citep[][]{2007CSE.....9...90H}, ChatGPT\footnote{ChatGPT serves as grammar checker and paraphrasing tool; OpenAI. (2022). Introducing ChatGPT. \url{https://openai.com/blog/chatgpt}}}



\bibliography{modification_v1_20250313}{}

\input{Tables/Detection_list}

\input{Tables/FXT_coordinates}

\input{Tables/Photometry1}

\input{Tables/Photometry2}

\clearpage
\appendix
\setcounter{figure}{0}
\counterwithin{figure}{section}
\renewcommand{\thefigure}{A\arabic{figure}}

\section{Follow-up description}
\label{sec:followup_summary}

\subsection{Summary: Sources with Confirmed Detection}
{\bf In this Sec., we discuss the observational details of the remaining FXTs whose optical counterparts were also detected from the Lulin Observatory, but we were not the first to discover these.}

\subsubsection{EP240315a}
\label{sec:ep240315a}
EP240315a was discovered at 2024-03-15T20:10:44 by the EP-WXT while performing a calibration observation. In the J2000 epoch, the reported X-ray transient had a Right Ascension (R.A.) = +141.644 deg and corresponding Declination (DEC.) = -9.547 deg with an uncertainty of 3${\arcmin}$ in radius \citep[][]{2024GCN.35931....1Z}. Soon after the EP-WXT discovery, \citet[][]{2024GCN.35932....1S} noticed that the normal survey mode observations from ATLAS\footnote{\url{https://atlas.fallingstar.com}} had serendipitously visited the EP-WXT localization of this FXT multiple times. Thus, a careful examination of the field resulted in the detection of an optical counterpart candidate AT 2024eju within the EP-WXT localization error circle. This candidate was detected just 1.1 hours after the EP-WXT discovery of EP240315a, at coordinates R.A. = 141.64763 deg, DEC. = -9.53401 deg (09$^{h}$26$^{m}$35$^{s}$.43,-09$^{\circ}$32$\arcmin$02$\farcs$4) with ATLAS $cyan$-band magnitude of 19.38$\pm$0.08 \citep[][]{2024ApJ...969L..14G}. Post the serendipitous optical counterpart detection by ATLAS, we were the first to perform the follow-up observations using the 40cm SLT in $r$-band. The details of our photometry are presented in Table~\ref{tab:Photometry1}. Later, \citet[][]{2024ApJ...969L..14G} reported the first-ever discovery of optical and radio counterparts for an FXT. Thus, EP240315a was the first FXT with detection in X-ray, optical, and radio bands.%

Several studies were undertaken to explore the nature and origin of EP240315a. A consistent pattern emerged across all these studies, leading to the compelling conclusion that EP240315a had a GRB origin \citep[][]{2024ApJ...969L..14G,2024arXiv240416350L,2024arXiv240416425L,2024arXiv240718311R}. For instance, the timing and spatial localization of EP240315a, when cross-referenced with GRB 240315C, revealed a strong correlation indicating its GRB origin \citep[][]{2024arXiv240718311R,2024arXiv240416425L}. Additionally, models simulating the evolution of EP240315a under the assumption of a GRB origin have successfully reproduced key observational features \citep[][]{2024arXiv240718311R}. Moreover, the temporal and spectral characteristics of EP240315a closely matched those typically associated with GRBs \citep[][]{2024arXiv240416350L,2024arXiv240416425L}.

\subsubsection{EP240801a}
\label{sec:ep240801a}
EP240801a was first detected by EP-WXT at the sky coordinates of R.A. = 345.140 deg and DEC. = +32.610 deg with an uncertainty of 2$\arcmin$.4 in radius \citep[][]{2024GCN.36997....1Z}. The EP-WXT trigger began at 2024-08-01T09:06:03. The EP-FXT trigger started autonomously 180s post EP-WXT detection. The follow-up observations from EP-FXT detected an uncataloged X-ray source at the sky coordinates of R.A. = 345.1630 deg and DEC. = +32.5927 deg with an uncertainty of 10$\arcsec$ in radius \citep[][]{2024GCN.36997....1Z}. {\bf Recently, \citet[][]{2025arXiv250304306J} reported its detection by Fermi-GBM, thus indicating EP240801a also had GRB link.}

Followed by the EP-WXT discovery and EP-FXT confirmation, several telescopes searched for the optical counterpart candidate associated with EP240801. The optical observations by \citet[][]{2024GCN.36998....1F} proposed the detection of an optical counterpart at R.A. = 23$^{h}$00$^{m}$39$^{s}$.019
DEC. = +32$^{\circ}$35$\arcmin$37$\farcs$20 with an $R$-band magnitude of 20.32$\pm$0.16. The proposed candidate optical counterpart was further confirmed by multiple observations \citep[][]{2024GCN.36999....1L,2024GCN.37000....1Z,2024GCN.37002....1A,2024GCN.37004....1A,2024GCN.37007....1P,2024GCN.37008....1M,2024GCN.37010....1Z,2024GCN.37012....1M,2024GCN.37014....1M,2024GCN.37015....1T,2024GCN.37016....1P,2024GCN.37017....1M,2024GCN.37040....1P,2024GCN.37049....1M,2024GCN.37146....1P}. Our observations in $r$-band using SLT started about 4.66 hrs post EP-WXT discovery \citep[][]{2024GCN.37002....1A}. The details of our photometry are presented in Table~\ref{tab:Photometry1}. GTC spectroscopy of the proposed optical counterpart candidate detected a well-defined continuum over a wavelength range of 5100--10000 \text{\AA} \citep[][]{2024GCN.37013....1Q}. The presence of several absorption lines in the GTC spectrum was interpreted to originate from Mg~II and Fe~II ions, indicating a common redshift of 1.673 for EP240801a. Later, the Keck/LRIS spectroscopic observations by \citet[][]{2024GCN.37228....1Z} also confirmed the GTC proposed redshift.{\bf The detailed multi-wavelength analysis by \citet[][]{2025arXiv250304306J} showed the resemblance of EP240801a with GRB 221009A (BOAT, \citealt[][]{2023ApJ...946L..31B}) under the assumption of two component jet models. Thus, EP240801a was probably another off-axis jet or intrinsically weak GRB.}

\subsubsection{EP241021a}
\label{sec:ep241021a}
The EP-WXT discovered EP241021a at 2024-10-21T05:07:56. It was detected at sky coordinates of R.A. = 28.852 deg, DEC. = 5.957 deg with an uncertainty of 2$\arcmin$.4 in radius \citep[][]{2024GCN.37834....1H}. A follow-up observation using EP-FXT was also performed about 36.5 hr post EP-WXT discovery. The EP-FXT follow-up observation detected an uncataloged X-ray source at R.A. = 28.8483 deg, DEC = 5.9395 deg with an uncertainty of 10$\arcsec$ in radius \citep[][]{2024GCN.37848....1W}.

Followed by EP-WXT discovery and EP-FXT follow-up observations, several telescopes were triggered to search for the optical counterpart. Some early observations with shallow limits did not detect any optical counterpart \citep[][]{2024GCN.37835....1G,2024GCN.37839....1L}. However, about 1.77 days post EP-WXT discovery, \citet[][]{2024GCN.37840....1F} proposed a plausible candidate at R.A. = 01$^{h}$55$^{m}$23$^{s}$.41
DEC. = +05$^{\circ}$56$\arcmin$18$\farcs$01 with $z$-band magnitude of 21.60$\pm$0.11. The proposed optical counterpart candidate was rigorously followed in several observations \citep[][Freeburn et al., GCN38062; Schneider et al., GCN38071]{2024GCN.37842....1F,2024GCN.37844....1L,2024GCN.37845....1R,2024GCN.37846....1L,2024GCN.37849....1Z,2024GCN.37850....1M,2024GCN.37869....1B,2024GCN.37875....1K,2024GCN.37877....1B,2024GCN.37878....1B,2024GCN.37892....1J,2024GCN.37910....1M,2024GCN.37911....1F}.

About 6.8 days post EP-WXT discovery, \citet[][]{2024GCN.37930....1Q} reported the re-brightening of the optical counterpart. The optical re-brightening was further confirmed by several other observations \citep[][]{2024GCN.37942....1F,2024GCN.37951....1M,2024GCN.37968....1P,2024GCN.37990....1K,2024GCN.38022....1S,2024GCN.38030....1B}.

The optical spectroscopy using VLT/FORS2 detected the presence of a strong emission line of [O~II] at 3728 \text{\AA} and Mg~II doublet absorption feature at 4897 \text{\AA} around a common redshift of z = 0.75 \citep[][]{2024GCN.37852....1P}. This redshift was also confirmed by \citet[][]{2024GCN.37858....1P} and \citep[][]{2024GCN.38294....1Z}. Thus, we adopt a redshift z = 0.75 in the current work.

The first epoch of observations from 40cm SLT in $r$-band started about 1.3 days post EP-WXT discovery; however, we did not detect the proposed optical counterpart up to a limit of 19.6 mag \citep[][]{2024GCN.37843....1Y}. Although during the re-brightening episode of the FXT, we detected the optical counterpart. Our observations during the re-brightening episode using the 1m LOT in $r$-band started about 11.44 days post EP-WXT trigger (Aryan et al., GCN38042). The details of our photometry are presented in Table~\ref{tab:Photometry1}.

About 3 days post EP-WXT trigger, e-MERLIN radio observations did not reveal any significant radio source \citep[][]{2024GCN.37906....1G}. Interestingly, a significant radio counterpart was detected in ATCA radio observation about 8.4 days post EP-WXT trigger \citep[][]{2024GCN.37949....1R}. Thus, EP241021a was the third FXT in our sample with X-ray, optical, and radio counterparts. Later, the radio counterpart was also detected by \citet[][]{2024GCN.38014....1C} and \citet[][]{2024GCN.38640....1S}. About 65 days post EP-WXT trigger, SMA radio observations at 235 GHz did not reveal any significant source up to an upper limit of 0.8 mJy \citep[][]{2025GCN.38924....1A}. 

{\bf Elaborative analysis of EP241021a displaying exceptional rebrightening suggested a likely link between EP-discovered FXTs and low-luminosity GRBs \citep[][]{2025arXiv250314588B}. Detailed multi-wavelength analyses by \citet[][]{2025arXiv250507665X} indicated the launch of a relativistic jet, which was further confirmed by \citet[][]{2025arXiv250508781Y} through radio observations. The study by \citet[][]{2025arXiv250505444G} proposed an off-axis jet and cocoon scenario. Another multi-wavelength analysis of EP241021a by \citet[][]{2025arXiv250512491W} suggested it to be an explosion-type event accompanied by a moderately relativistic jet.}

\subsubsection{EP241030a}
\label{sec:ep241030a}
EP241030a was discovered by EP-WXT at the sky coordinates at 2024-10-30T06:33:18 with R.A.= 343.013 deg, DEC. = 80.449 deg having an uncertainty of 2$\arcmin$.4 in radius \citep[][]{2024GCN.37997....1W} and was found coincident with GRB 241030A \citep[][]{2024GCN.37955....1F}. About 2.8 days post EP-WXT discovery, EP-FXT follow-up observations also detected the afterglow in X-ray (Liang et al., GCN38040).

Soon after the EP-WXT discovery, several telescopes were triggered to search for the optical counterpart. The optical counterpart of GRB 241030a/EP241030a was first discovered by \citet[][]{2024GCN.37956....1K} at R.A. = 22$^{h}$52$^{m}$33$^{s}$.57
DEC. = +80$^{\circ}$26$\arcmin$59$\farcs$9 in Swift-UVOT with an estimated magnitude of 15.42$\pm$0.14 (quoted error was 1-$\sigma$). The optical counterpart was confirmed in several other observations \citep[][Moskvitin et al., GCN38032; Yan et al., GCN38035; Adami et al., GCN38041; Reguitti et al., GCN38107]{2024GCN.38019....1B}.

The spectroscopic observations of the optical counterpart showed the presence of numerous narrow absorption lines, including Mg~II doublets at redshifts of 0.456, 0.862, and 1.411 \citep[][]{2024GCN.37959....1Z}.
Further spectroscopic follow-up observations by \citet[][]{2024GCN.38027....1L} revealed a strong absorption line of Al~II at 1671 \text{\AA}, along with weaker absorption lines of Fe II at 2374 \text{\AA}, 2383 \text{\AA}, 2587 \text{\AA}, and 2600 \text{\AA}, Mg II at 2800 \text{\AA}, and Mg I at 2852 \text{\AA}. All these metal features indicated a common redshift of around 1.4, consistent with \citet[][]{2024GCN.37959....1Z}. Thus, we adopt z = 1.411 for EP241030a in the present work. Our observations from 1m LOT in $r$-band started about 2.30 days post EP-WXT discovery (Aryan et al., GCN38042). The details of our photometry are presented in Table~\ref{tab:Photometry1}.

\subsubsection{EP241107a}
\label{sec:ep241107a}
EP241107a was discovered by EP-WXT at 2024-11-07T14:10:23. About 5 minutes later, an autonomous observation by EP-FXT also discovered an X-Ray source at R.A. = 35.0085 deg, DEC. = 3.3329 deg with an uncertainty of 10$\arcsec$ in radius (Zhou et al. GCN38112).

Followed by EP-WXT and EP-FXT observations, several telescopes were quickly triggered to search for the optical counterpart. Only about 90 minutes post EP-WXT trigger, Odeh et al., GCN38115 reported the discovery of a bright optical counterpart candidate at R.A. = 02$^{h}$20$^{m}$02$^{s}$.45
DEC. = +03$^{\circ}$20$\arcmin$02$\farcs$2 in $Ic$-band with 17.85$\pm$0.18 mag. The proposed optical counterpart was confirmed in several optical/NIR follow-up observations \citep[][SVOM Team, GCN38116; Mohan et al., GCN38118; Bussman et al., GCN38120; Adami et al., GCN38122;]{2024GCN.38127....1L,2024GCN.38128....1O,2024GCN.38136....1Z}. Our observations from 1m LOT in $r$-band started about 1.02 days post EP-WXT trigger \citep[][]{2024GCN.38131....1K}. The details of our photometry are presented in Table~\ref{tab:Photometry1}.

A radio follow-up observation using VLA revealed a point radio source at a location consistent with the optical counterpart \citep[][]{2024GCN.38584....1B}. Thus, EP241107a is the fourth source in our sample having X-ray, optical, and radio counterparts.

\subsubsection{EP241217a}
\label{sec:ep241217a}
The EP-WXT discovered EP241201a at 2024-12-17T05:36:03. It was discovered at sky coordinates of R.A. = 46.957 deg, Dec. = 30.901 deg with an uncertainty of 2$\arcmin$.8 in radius \citep[][]{2024GCN.38586....1Z}. Followed by EP-WXT discovery, an EP-FXT follow-up observation was also performed, which detected an uncataloged X-ray source at R.A. = 46.9398, Dec. =  30.9299 deg with an uncertainty of about 20$\arcsec$ in radius \citep[][]{2024GCN.38586....1Z}. The Fermi-GBM observations in a time interval of -50s -- +500s from the EP-WXT trigger did not reveal any GRB-like detection coinciding with the EP241217a spatially and temporally. However, between a time interval of T$_{0}$+24.2 ks to T$_{0}$+25.7 ks, the Swift-XRT detected an uncatalogued source consistent with the EP-FXT localization \citep[][]{2024GCN.38596....1W}.

Quickly after the EP-WXT discovery, several telescopes were triggered to search for any associated optical counterpart. Approximately 2.5 hours post EP-WXT discovery, observations by \citet[][]{2024GCN.38587....1L} discovered a candidate optical counterpart at R.A. = 03$^{h}$07$^{m}$46$^{s}$.20
DEC. = +30$^{\circ}$55$\arcmin$45$\farcs$9 with $z$-band magnitude of $\sim$19.9. The candidate optical counterpart was confirmed in several other follow-up observations \citep[][]{2024GCN.38588....1I,2024GCN.38607....1J,2024GCN.38612....1M,2024GCN.38613....1Z,2024GCN.38615....1B,2024GCN.38636....1L}. 

The spectroscopic observations of the optical counterpart by \citet[][]{2024GCN.38593....1L} detected a bright continuum with a strong break at $\sim$6820~\text{\AA}, along with several narrow absorption lines of Si~II, C~II, weak C~IV, Fe~II, Al~II, all at a common redshift of z = 4.59. Thus, in the present work, we adopted a redshift of 4.59 for EP241217a. Our observations from 1m LOT in $r$-band started about 7.34 hours post EP-WXT discovery. We clearly detected the optical counterpart candidate \citep[][]{2024GCN.38592....1F}. The details of our photometry are presented in Table~\ref{tab:Photometry2}.

The field of EP241217a was also scanned by a few radio telescopes. About 2.45 days post EP-WXT discovery, the e-MERLIN radio observations at 5 GHz did not reveal any significant radio source down to 96~$\mu$Jy \citep[][]{2024GCN.38652....1R}. However, about 3.81 days post EP-WXT discovery, the VLA observations detected a point radio source consistent with EP241217a having 20$\pm$6.6~$\mu$Jy at 3 GHz, 58$\pm$4~$\mu$Jy at 6 GHz, and 99.3$\pm$4.1~$\mu$Jy at 10 GHz \citep[][]{2025GCN.38749....1A}. Thus, EP241217a is the fifth source in our sample having X-ray, optical, and radio counterparts.

\subsubsection{EP250108a}
\label{sec:ep250108a}
EP241201a was discovered by EP-WXT at 2024-12-17T05:36:03 with the sky coordinates of R.A.= 55.623 deg, DEC. = -22.509 deg with an uncertainty of 2$\arcmin$.2 in radius \citep[][]{2025GCN.38861....1L}. Followed by EP-WXT discovery, an EP-FXT follow-up observation was also performed about 22.2 hours later, but no x-ray counterpart was detected \citep[][]{2025GCN.38888....1L}. 

Followed by the EP-WXT discovery and EP-FXT follow-up observations, several telescopes were triggered to search for any associated optical counterpart candidate. About 31.5 hours post EP-WXT discovery, \citet[][]{2025GCN.38878....1E} proposed an optical counterpart candidate at R.A. = 03$^{h}$42$^{m}$28$^{s}$.38
DEC. = -22$^{\circ}$30$\arcmin$21$\farcs$1 with $g'$-band magnitude of 20.10$\pm$0.06. The proposed optical counterpart candidate was confirmed by several follow-up observations \citep[][]{2025GCN.38885....1Z,2025GCN.38902....1M,2025GCN.38907....1K,2025GCN.38909....1L,2025GCN.38912....1I,2025GCN.38914....1Z,2025GCN.38925....1M,2025GCN.38972....1S}. The spectroscopic observations by \citet[][]{2025GCN.38908....1Z} detected a bright and blue continuum. Emission lines from H$_{\alpha}$, [O~II] 3727/29~\text{\AA}, faint [O~III] 5007~\text{\AA}, (narrow) absorption from Ca~II H and K were detected  at a common redshift z = 0.176. Thus, we adopted a redshift of 0.176 in the present work for EP250108a.

About 10.3 days post EP-WXT trigger, \citet[][]{2025GCN.38983....1E} claimed to observe the re-brightening episode, which was confirmed in further follow-up observations \cite[][]{2025GCN.38984....1X,2025GCN.38987....1L,2025GCN.39002....1R}. Additionally, the spectroscopic follow-up observations during the re-brightening episode revealed consistent features matching Type Ic-BL SNe (\citealt[][]{2025GCN.38984....1X,2025GCN.38987....1L}). SNe Type Ic-BL are typically associated with collapsar origin GRBs. Our observations from 1m LOT in $r$-band started about 7.95 days post EP-WXT discovery. We clearly detected the optical counterpart candidate. The details of our photometry are presented in Table~\ref{tab:Photometry2}. No significant radio counterpart was detected \citep[][]{2025GCN.38958....1C,2025GCN.38970....1S,2025GCN.38998....1A}.

{\bf The optical counterpart of the X-ray transient EP250108a was identified as SN 2025kg, a broad-lined Type Ic SN \citep[][]{2025arXiv250408886E,2025arXiv250408889R,2025arXiv250417516S}. Analysis of the well-sampled light curves by \citet[][]{2025arXiv250417034L} pointed toward a mildly relativistic outflow as the likely origin of the event. \citet[][]{2025arXiv250408886E} found that the observed X-ray and radio properties were consistent with a low-energy jet, likely produced by a collapsar, that probably failed to penetrate the surrounding dense material. They further proposed that the optical emission might originate from a cocoon of shocked material surrounding the trapped jet. Supporting this scenario, \citet[][]{2025arXiv250408889R} indicated similar results with the broadband data being consistent with a trapped or low-energy jet-driven explosion from a collapsar with a ZAMS mass of 15-30\,M$_{\odot}$. Additionally, \citet[][]{2025arXiv250417516S} further reinforced the connection of EP250108a with low-luminosity GRBs.}

\subsection{summary: Sources with Upper Limits}
{\bf This Sec. presents a summary of FXTs followed from the Lulin Observatory that do not have confirmed optical detections in our observations, and we only provide optical upper limits.} 

\subsubsection{EP240331a}
EP240331a was discovered by EP-WXT  at 2024-03-31T22:07:17. It was detected at sky coordinates of R.A. = 169.414 deg and DEC. = -20.042 deg with an uncertainty of 20$\arcmin$ in radius \citep[][]{2024ATel16564....1P}. The large uncertainty in localization was attributed to the incomplete calibration of the detector.
Followed by the EP-WXT discovery of EP240331a, a few telescopes started the search for the optical counterpart of EP240331a within a day. None of these observations detected any optical counterpart \citep[][]{2024GCN.36007....1G,2024GCN.36009....1L,2024GCN.36010....1F}. Our observations from 1m LOT in $r$-band started about  17.62 hours post the EP-WXT trigger. We also utilized the 40cm SLT for $g$-band observations. For our observations, we used four different pointings in each band to significantly cover the entire error circle of localization. In $r$-band, each pointing had a total exposure time of 4$\times$300s, while in $g$-band, three pointings had an exposure time of 4$\times$300s while one had 3$\times$300s. We find no evidence of a new source in any of the four pointings for each band. The upper limits in different bands are presented in Table~\ref{tab:Photometry8}. We provided the deepest upper limits that further confirmed the non-detection of any significant optical counterpart associated with EP240331a. Our preliminary photometry was reported in \citet[][]{2024GCN.36011....1C}.

\subsubsection{EP240408a}
The EP-WXT detected EP240408a at 2024-04-08T17:56:30. It was detected at the sky coordinates of R.A. = 158.840 deg and DEC. = -35.749 deg with an uncertainty of 3$\arcmin$ in radius \citep[][]{2024GCN.36053....1H}.
Besides EP-mission, Swift-XRT follow-up observations also confirmed the new FXT EP240408a, about 33 hours post EP-WXT discovery \citep[][]{2024GCN.36057....1H}. None of the follow-up observations in optical bands detected any optical counterpart within the specified EP-WXT error circle \citep[][]{2024GCN.36056....1P,2024GCN.36060....1A,2024GCN.36079....1L}.
Although \citet[][]{2024GCN.36059....1R} suggested a potential counterpart detection in the NIR J and H bands, they noted that the identified source was also there in the $z$-band data from the Legacy survey DR10\footnote{\url{https://www.legacysurvey.org/}}. The 10$^{th}$ Legacy survey data was released in September 2023 (much earlier than EP240408a detected); thus, it was difficult to constrain whether the detected NIR counterpart candidate was actually associated with EP240408a. About 42.38 hours post the EP-WXT discovery, our observations from 1m LOT started in $r$-band for a total exposure of 26$\times$180s. Also, the $i$-band follow-up observations using 40cm SLT began at 42.74 hours post EP-WXT discovery for a total exposure of 15$\times$300s. The upper limits in different bands for our follow-up observations are presented in Table~\ref{tab:Photometry8}.

\subsubsection{EP240413a}
The EP team reported the detection of EP240413a by EP-WXT at 2024-04-13T14:39:37. The sky coordinates of this source were R.A. = 228.794 deg and DEC. = -18.800 deg with an uncertainty of 20$\arcmin$ in radius \citep[][]{2024GCN.36086....1L}. The big uncertainty in the localization, compared to the usual value, was due to the incomplete calibration of the EP-WXT module. Followed by the EP-WXT discovery, EP-FXT  also performed the follow-up observations. The EP-FXT unambiguously detected an X-ray source at R.A. = 228.815 deg and DEC. = -18.503 deg with an uncertainty of 30$\arcsec$ in radius \citep[][]{2024GCN.36092....1L}.
Followed by the  EP-WXT discovery and EP-FXT follow-up observations of EP240413a, a few telescopes searched for the possible optical counterpart associated with the FXT. None of them found any optical counterpart \citep[][]{2024GCN.36098....1P,2024GCN.36135....1X}. Our observations from 1m LOT started about 73.55 hours after the EP-WXT trigger for a total exposure of 6$\times$300s in $r$-band. The upper limit in $r$-band is presented in Table~\ref{tab:Photometry8}.  

\subsubsection{EP240506a}
EP240506a was discovered by EP-WXT  at 2024-05-06T05:01:39 while performing the calibration tests.  The sky coordinates of the EP source were R.A. = 213.978 deg and DEC. = -16.715 deg with an uncertainty of 3$\arcmin$ in radius \citep[][]{2024GCN.36405....1L}.
Several telescopes were triggered to search for the possible optical counterpart associated followed by the EP-WXT discovery. However, none of them found any detection within reported EP-WXT localization \citep[][]{2024GCN.36412....1P,2024GCN.36413....1T, 2024GCN.36417....1P, 2024GCN.36419....1J}. Our observations from 40cm SLT started about 32.10 hours after the EP-WXT trigger with a total exposure of 30$\times$300s in $r$-band. The upper limits in different bands are presented in Table~\ref{tab:Photometry8}. Our preliminary was reported in \citet[][]{2024GCN.36408....1A}. Further, \citet[][]{2024GCN.36414....1C} did extensive archival radio searching. They found one radio detection in the Rapid ASKAP Continum Survey \citep[RACS;][]{2021PASA...38...58H} data within the EP-WXT localization error circle. The radio fluxes of the specified source showed a power-law spectral index of -1.47. This radio source was found to be 2$\arcmin$ away from a variable star SDSS J141557.82-164317.2. On inspecting the ALLWISE\footnote{\url{https://wise2.ipac.caltech.edu/docs/release/allwise/}} source catalog further, they found WISEA J141556.28-164124.4 to be only 2$\arcsec$ away from the RACS source, ultimately concluding that the detected radio source in archival radio images could be a background galaxy.

\subsubsection{EP240617a}
\label{sec:ep240617a}
EP240617a was discovered by EP-WXT  at 2024-06-17T12:19:13. It was detected at sky coordinates of R.A. = 285.030 deg and DEC. = -22.561 deg with an uncertainty of 3$\arcmin$ in radius \citep[][]{2024GCN.36691....1Z}.
Followed by the EP-WXT discovery of EP240617a, \citet[][]{2024GCN.36692....1Y} reported the discovery of a weak, untriggered gamma-ray transient in Fermi-GBM\footnote{\url{https://fermi.gsfc.nasa.gov/science/instruments/gbm.html}} data within the EP240617a occurrence time interval. The location of the Fermi-GBM detected transient aligned with the EP-WXT localization. Based on the spatial and temporal coincidence of EP240617a, \citet[][]{2024GCN.36692....1Y} concluded it to be a GRB event. About 77 hours post the EP-WXT discovery, \citet[][]{2024GCN.36722....1S} reported a weak X-ray signal from Swift-XRT consistent with the EP-WXT localization. Several telescopes also triggered to search for the optical counterpart, but none of them detected any candidate \citep[][]{2024GCN.36693....1P,2024GCN.36707....1S}. Our observations from 40cm SLT started about 30.95 hours post the EP-WXT trigger with a total exposure of 6$\times$300s in $r$-band. The upper limits in different bands are presented in Table~\ref{tab:Photometry8}. 

\subsubsection{EP240618a}
The EP team reported the EP240618a detection by EP-WXT discovered at 2024-06-18T05:43:43. The sky coordinates of this source were R.A. = 281.627 deg and DEC. = +23.820 deg with an uncertainty of 3$\arcmin$ in radius \citep[][]{2024GCN.36690....1S}. About an hour post the EP-WXT discovery, EP-FXT also began follow-up observations,  and unambiguously detected an X-ray source at R.A. = 281.648 deg and DEC. = +23.833 deg with an uncertainty of 30$\arcsec$ in radius \citep[][]{2024GCN.36690....1S}.
About 58 hours the EP-WXT discovery of EP240618a, Swift-XRT also observed the EP-FXT localization. However, no significant X-ray counterpart was detected within the EP-FXT localization error circle with a radius of 30$\arcsec$ \citep[][]{2024GCN.36723....1C}. Additionally, exploring the Fermi-GBM data, \citet[][]{2024GCN.36725....1R} also reported the non-detection of any significant gamma-ray source. Several optical/NIR  follow-up observations were also triggered after the EP-WXT and EP-FXT detections of EP240618a. However, none of those observations found any strong evidence of a counterpart candidate \citep[][]{2024GCN.36701....1W,2024GCN.36702....1A,2024GCN.36710....1T,2024GCN.36712....1A,2024GCN.36714....1P,2024GCN.36731....1L,2024GCN.36736....1L,2024GCN.36741....1L}. Our observations from 40cm SLT started about 11.81 hours after the EP-WXT trigger for a total exposure of 10$\times$300s in $r$-band. We did multi-epoch follow-up observations but did not detect any optical counterpart. The upper limits in different bands are presented in Table~\ref{tab:Photometry8}.

\subsubsection{EP240625a}
EP240625a was discovered by EP-WXT at 22024-06-25T01:48:23. The sky coordinates of this source were R.A. = 310.760 deg and DEC. = -15.966 deg with an uncertainty of 2$\arcmin$ in radius \citep[][]{2024GCN.36757....1P}. Following the EP-WXT discovery, EP-FXT performed follow-up observations about an hour post EP-WXT discovery. The EP-FXT clearly detected a slowly fading X-ray source at R.A. = 310.7308 deg and DEC. = -15.9760 deg with an uncertainty of 10$\arcsec$ in radius \citep[][]{2024GCN.36760....1P}.
Followed by the EP-WXT and EP-FXT detection, a few telescopes were triggered to search for the possible optical counterpart. \citet[][]{2024GCN.36761....1F} proposed a marginally detected source within the EP-FXT localization error circle as the optical counterpart candidate for EP240625a; however, they could not completely rule out the possibility of their detection being a background fluctuation. The observations reported by \citet[][]{2024GCN.36771....1A} also did not find any optical counterpart candidate. We began our observations from 40cm SLT about 40.99 hours after the EP-WXT trigger with a total exposure of 10$\times$300s in $i$-band. The upper limits in different bands are presented in Table~\ref{tab:Photometry8}.

\subsubsection{EP240626a}
The EP team reported the EP240626a detection by EP-WXT at 2024-06-26T06:28:28. The sky coordinates of this source were R.A. = 263.023 deg and DEC. = -13.051 deg with an uncertainty of 2$\arcmin$ in radius \citep[][]{2024GCN.36766....1W}.  Following the EP-WXT discovery, EP-FXT also performed follow-up observations about an hour post EP-WXT discovery. The EP-FXT clearly detected a faint X-ray source at R.A. = 263.0171 deg and DEC. = -13.0490 deg with an uncertainty of 30$\arcsec$ in radius \citep[][]{2024GCN.36770....1W}.
Several telescopes were triggered to search for the optical counterpart of EP240626a. None of those observations claimed to have optical counterpart detection \citep[][]{2024GCN.36767....1P,2024GCN.36773....1Z,2024GCN.36774....1L}.  Our observations from 40cm SLT started about 33.16 hours after the EP-WXT trigger with a total exposure of 6$\times$300s in $r$-band. The corresponding upper limit is presented in Table~\ref{tab:Photometry8}.

\subsubsection{EP240702a}
EP240702a was discovered by EP-WXT  at 2024-07-02T00:50:05. It was detected at sky coordinates of R.A. = 328.203 deg and DEC. = -38.980 deg with an uncertainty of 3$\arcmin$ in radius \citep[][]{2024GCN.36801....1C}. 
Followed by the EP-WXT discovery of EP240702a, a few telescopes were triggered to search for the optical counterpart. None of them found any traces of a new uncatalogued source in their observations \citep[][]{2024GCN.36806....1L,2024GCN.36831....1T}. Our observations from 40cm SLT started about 16.05 hours post the EP-WXT trigger in $r$-band with a total exposure of 35$\times$300s. The upper limit in $r$-band is presented in Table~\ref{tab:Photometry8}. Additionally, Swift-XRT follow-up observations about 8 hours post EP-WXT discovery did not reveal any significant X-ray source within the EP-WXT localization error circle.

\subsubsection{EP240703a}
\label{sec:ep240703a}
The EP team reported the detection of EP240703a by EP-WX at 2024-07-03T00:38:40. It was detected at sky coordinates of R.A. = 273.803 deg and DEC. = -9.681 deg with an uncertainty of 3$\arcmin$ in radius \citep[][]{2024GCN.36807....1W}.
Only $\sim$\,2s before the EP-WXT discovery of EP240703a, Konus-Wind\footnote{http://www.ioffe.ru/LEA/kw/index.html} detected the long GRB 240703A. Owing to the temporal and spatial coincidence of EP240703a and GRB 240703A, \citet[][]{2024GCN.36809....1F} claimed EP240703a to be the GRB 240703A counterpart. Followed by the EP-WXT and Konus-Wind detection, Swift-XRT also triggered for follow-up observations; however, \citet[][]{2024GCN.36827....1S} claimed to have no significant X-ray source detection within the EP-WXT localization of EP240703a. Further, several telescopes were also triggered to search for the optical counterpart, but none of those claimed to have any significant detection \citep[][]{2024GCN.36820....1A,2024GCN.36821....1B,2024GCN.36822....1F,2024GCN.36824....1V,2024GCN.36825....1Z,2024GCN.36832....1T}.  Our observations from 1m LOT started about 12.43 hours after the EP-WXT trigger in $r$-band with a total exposure of 3$\times$300s. The upper limits in different bands are presented in Table~\ref{tab:Photometry8}. Our preliminary photometry was reported in \citet[][]{2024GCN.36819....1A}.

\subsubsection{EP240703c}
EP240703c was discovered by EP-WXT at 2024-07-03T18:15:00. It was detected at sky coordinates of R.A. = 289.264 deg and DEC. = -30.325 deg with an uncertainty of 3$\arcmin$ in radius \citep[][]{2024GCN.36818....1Z}. A previously known high proper motion star, LP 924-17, was found to be lying within the EP-WXT localization error circle. Although they suggested that the detected EP transient was not a stellar flare based on its X-ray light curve and spectrum. However, they did not completely rule out this possibility. We investigated the quick-look data from Swift-ToO observations of the EP-WXT localization region proposed by the EP team that constrained a relatively softer spectrum from the Swift-XRT data. Additionally, the high proper motion star is clearly detected in Swift-UVOT V-band with a magnitude of 17.76$\pm$0.21 in the AB system \citep[][]{2024GCN.36823....1Y}. Followed by EP-WXT discovery of the EP240703c, our observations from 40cm SLT started about 20.61 hours post the EP-WXT trigger in $r$-band with a total exposure of 16$\times$300s. We did not detect any new and uncatalogued optical counterpart candidate. However, the high proper motion star was also clearly detected in our observations. We noticed no significant brightness enhancement for LP 924-17 through our observations. The average magnitude was found to be 16.88$\pm$0.05 mag in $r$-band \citep[][]{2024GCN.36823....1Y}. The upper limit in $r$-band for our SLT observations is presented in Table~\ref{tab:Photometry8}.

\subsubsection{EP240708a}
The EP team reported the EP240708a detection by EP-WXT at 2024-07-08T23:28:23. The sky coordinates of this source were R.A. = 345.963 deg and DEC. = -22.840 deg with an uncertainty of 3$\arcmin$ in radius \citep[][]{2024GCN.36838....1C}. Following the EP-WXT discovery, EP-FXT also performed follow-up observations about 4 hours later. The EP-FXT clearly detected a faint X-ray source at R.A. = 345.9656 deg and DEC. = -22.8428 deg with an uncertainty of 10$\arcsec$ in radius \citep[][]{2024GCN.36840....1H}.

Further, the EP-team also obtained Swift-XRT observations about 17 hours post EP-WXT trigger, but they did not detect any evidence of a new X-ray source within the EP-WXT localization \citep[][]{2024GCN.36840....1H,2024GCN.36888....1Q}. A number of telescopes were also triggered for the search for any possible optical counterpart. None of them detected any evidence of a new and uncatalogued source within the EP-WXT localization error circle \citep[][]{2024GCN.36841....1W,2024GCN.36842....1J,2024GCN.36845....1Z,2024GCN.36852....1L}. Our observations from 40cm SLT in $r$-band started about 18.52 hours post the EP-WXT trigger with a total exposure of 36$\times$300s. We also utilized the 1m LOT in the $r$-band and $g$-band to search deeper, but we did not find any evidence of an optical counterpart. The upper limits in different bands are presented in Table~\ref{tab:Photometry9}. We reported our preliminary photometry in \citet[][]{2024GCN.36839....1L}.

\subsubsection{NVSS J004348+342626}
The EP team reported the detection of the X-ray brightening of blazar NVSS J004348+342626 by EP-WXT at 2024-07-17T03:13:01. The sky localization for the EP-WXT-detected point source was R.A. = 10.955 deg and DEC. = +34.428 with an uncertainty of 2$\farcm$2 in radius \citep[][]{2024ATel16725....1J}. There were several X-ray sources within the EP-WXT localization error circle. The Swift-XRT observations about 1.5 days post EP-WXT trigger detected an X-ray source at R.A. = 10.9534 deg and DEC. = +34.4401 with an uncertainty of only 3$\farcs$8, consistent with the spatial localization of blazar NVSS J004348+342626, confirming the EP EP-WXT trigger to be associated with this blazar \citep[][]{2024ATel16725....1J}.

It was the very first time that the flat-spectrum radio quasar (FSRQ) NVSS J004348+342626 was detected in the X-ray band. About 1.5 days post EP-WXT trigger, Swift-XRT also detected an X-ray source spatially consistent with the FSRQ position. Thus, EP-WXT and Swift-XRT confirmed the triggers to be associated with NVSS J004348+342626. The X-ray detection of this FSRQ for the very first time, along with its earlier detections in radio-band \citep[][]{1998AJ....115.1693C} and in gamma-ray \citep[][]{2020ApJS..247...33A}, makes it an interesting source. Followed by EP-WXT discovery, we triggered the 40cm SLT at the Lulin Observatory and detected the FSRQ probably in its flaring stage with an $r$-band magnitude of 18.38$\pm$0.05. An archival search in the optical bands by \citet[][]{2024ATel16729....1G} at two epochs (one at $\sim$ 3 days and another at $\sim$ 10 months) prior to EP-WXT and Swift-XRT triggers indicated to capture the quiescent phase of the FSQR with unfiltered magnitude of 18.93$\pm$0.09 mag at both the epochs. Our observations from 40cm SLT in $r$-band started about 39.55 hours post the EP-WXT trigger with a total exposure of 12$\times$300s. We reported the preliminary magnitude and corresponding upper limit from the observed field in \citet[][]{2024ATel16726....1L}. {\bf The left-hand panel of Fig.~\ref{fig:Blazar_n_stellar_flare} shows the field of view of this flaring blazar event.}

\subsubsection{EP240802a}
\label{sec:ep240802a}
EP240802a was first detected by EP-WXT at 22024-08-02T10:32:52. The sky coordinates of this source were R.A. = 287.802 deg and DEC. = -2.354 deg with an uncertainty of 1$\farcm$9 in radius \citep[][]{2024GCN.37019....1W} in radius. Following the EP-WXT discovery, EP-FXT also performed follow-up observations about 14 hours post EP-WXT discovery. The EP-FXT clearly detected an uncatalogued X-ray source at R.A. = 287.8070 deg and DEC. = -2.3125 deg with an uncertainty of 10$\arcsec$ \citep[][]{2024GCN.37019....1W} in radius.

Followed by the EP-WXT discovery and EP-FXT follow-up observation, several telescopes were triggered in search of the possible optical counterpart. About 29.22 hours post EP-WXT trigger, we observed the field of EP240802a and noticed the temporal coincidence of EP240802a with GRB 240802A \citep[][]{2024GCN.37018....1W}. This FXT triggered EP-WXT at 2024-08-02T10:32:52 and lasted for more than 500s, while the long GRB 240802A was observed to display its strongest peak at 2024-08-02T10:34:04.5, clearly overlapping with EP240802a. Till the start of our observations for EP240902a, there was no spatial position information available for GRB 240802A; however, we reported the temporal overlapping in \citet[][]{2024GCN.37021....1A}. Later, the 3-sigma error region by the IPN triangulation\footnote{http://ssl.berkeley.edu/ipn3/index.html} confirmed the positional coincidence of EP240802a with GRB 240802A \citep[][]{2024GCN.37024....1R,2024GCN.37078....1K}. Further, \citet[][]{2024GCN.37022....1S} also suggested the association of these two events. Thus, EP240802a was another FXT showing its association with a long-GRB. A number of telescopes triggered after the EP-WXT and EP-FXT trigger; however, none of them found any evidence of an optical counterpart candidate \citep[][]{2024GCN.37023....1Q,2024GCN.37029....1L,2024GCN.37094....1Z}. The earliest upper limit since the EP-WXT trigger was reported in \citet[][]{2024GCN.37094....1Z}, that too was at $\sim$0.74 days post EP-WXT trigger. Thus, there was a probability of the optical counterpart of EP240802a fading without being noticed. About 6 days post EP-WXT discovery, we again triggered the LOT in the hope of discovering the SN component (an incidence similar to EP240414a), although we did not find any evidence \citep[][]{2024GCN.37178....1Y}. The multi-epoch upper limits in $r$-band from our observations are presented in Table~\ref{tab:Photometry9}.

\subsubsection{EP240908a}
EP240908a was discovered by EP-WXT at 2024-09-08T17:28:27 at the spatial position with R.A. = 13.992 deg, DEC. = 8.089 deg, having an uncertainty of 2$\farcm$7 in radius. EP-FXT also performed a follow-up observation about 1.34 hr post EP-WXT discovery and detected an uncataloged X-ray source was detected at R.A. = 14.0031, DEC. = 8.0735 (J2000) with an uncertainty of about 10$\arcsec$ \citep[][]{2024GCN.37432....1M,2024GCN.37443....1M} in radius.

After the EP-WXT discovery, followed by EP-FXT automated observations, several ground-based telescopes were triggered to search for any associated optical counterpart of EP240908a. \citet[][]{2024GCN.37434....1F} reported an upper limit of $\sim$21.6 mag in $R$-band. Some shallow limits were also reported by \citet[][]{2024GCN.37437....1L} in $clear$- (unfiltered) band. \citet[][]{2024GCN.37438....1Q} proposed a very faint optical counterpart ($r$-band AB magnitude $\sim$24) through their observations from the Gemini-North telescope at an epoch of 0.777 days since EP-WXT discovery. They further confirmed the proposed counterpart through additional observations about 2.19 days post detection \citep[][]{2024GCN.37458....1Q}. Due to the very faint nature of the optical counter, no other telescopes could capture the optical counterpart. The observations by \citet[][]{2024GCN.37444....1P} reported an upper limit of 23.5 mag in $R$-band at an epoch of $\sim$1 day post EP-trigger. About 24.22 hours post EP-WXT trigger, our observations from 40cm SLT started in $r$-band with a total exposure of 12$\times$300s. We also only obtained an upper limit in the $r$-band as presented in Table~\ref{tab:Photometry9}. 

\subsubsection{EP240913a}
\label{sec:ep240913a}
The EP-WXT detected EP240913a at 2024-09-13T11:39:33 with R.A.= 16.681 deg, DEC. = 16.750 deg having an uncertainty of 2$\farcm$5 in radius \citep[][]{2024GCN.37492....1L}. 
Soon after its EP-WXT discovery, EP240913a was classified as a GRB event \citep[][]{2024GCN.37493....1Y} due to its positional and temporal coincidence with GRB 240913C, which was further supported by several other observations \citep[][]{2024GCN.37505....1D,2024GCN.37507....1Z,2024GCN.37510....1P,2024GCN.37538....1K}. Swift-XRT also detected an X-ray source within the EP-WXT error circle, which was proposed to be the counterpart of EP240913a \citep[][]{2024GCN.37497....1J}. No optical/NIR counterpart was reported by any of the follow-up observations \citep[][]{2024GCN.37499....1A,2024GCN.37500....1Z,2024GCN.37501....1Z,2024GCN.37511....1Z,2024GCN.37515....1S,2024GCN.37521....1L,2024GCN.37582....1Z,2024GCN.37584....1B}. Our observations from 1m LOT in $r$-band started about 26.44 hours post the EP-WXT trigger with a total exposure of 3$\times$300s. We also only obtained upper limits in $r$-band and $g$-band as mentioned in Table~\ref{tab:Photometry9}.

\subsubsection{EP240918a}
EP240918a was first detected by EP-WXT at 2024-09-18T11:24:37, and a subsequent autonomous observation started by EP-FXT two minutes later. The EP-FXT localization coordinates of this source were R.A. = 289.3937 deg and DEC. = 46.1281 deg with an uncertainty of 20$\arcsec$ in radius \citep[][]{2024GCN.37541....1Z}.

Followed by the EP-WXT discovery and autonomous EP-FXT follow-up observations, several telescopes were triggered. No optical counterpart was detected in any of the observations \citep[][]{2024GCN.37544....1W,2024GCN.37546....1S,2024GCN.37547....1F,2024GCN.37548....1L,2024GCN.37568....1Z,2024GCN.37581....1S}, even though, some observations were as early as $\sim$ two hours post EP-WXT discovery and had upper limits as deep as $\sim$\,22.8 mag. A Swift-XRT follow-up observation was conducted by \citet[][]{2024GCN.37551....1Q} about 10.7 hr post EP-WXT discovery, but no X-ray counterpart was detected. Our observations from 40cm SLT in $r$-band started about 2.30 hours post the EP-WXT trigger with a total exposure of 16$\times$300s. We also conducted a further follow-up observation using 1m LOT in $r$-band with a total exposure of 6$\times$300s. We only obtained upper limits in the $r$-band as mentioned in Table~\ref{tab:Photometry9}. We reported our preliminary photometry in \citet[][]{2024GCN.37545....1K}.

\subsubsection{EP240918b}
EP-WXT discovered the EP240918b at 2024-09-18T15:40:00 with R.A. = 258.66 deg, DEC. = 66.739 deg having an uncertainty of 2$\farcm$9 in radius \citep[][]{2024GCN.37555....1L}. About 20 hours post EP-WXT discovery, the FXT was not detected in the follow-up observations by EP-FXT \citep[][]{2024GCN.37608....1L}.

Followed by the EP-WXT discovery and EP-FXT follow-up observations, we triggered the 40cm SLT at the Lulin Observatory to search for the optical counterpart. No other telescope trigger had been reported on any platform. Our observations from 40cm SLT in $r$-band started about 22.02 hours post EP-WXT trigger with a total exposure of 7$\times$300s. We did not find evidence of any candidate counterpart down to a limit of 19 mag. We had reported our preliminary photometry in \citet[][]{2024GCN.37577....1L}.  The details of the photometry are presented in Table~\ref{tab:Photometry9}.

\subsubsection{EP240918c}
EP240918c was first detected by EP-WXT at 2024-09-18T18:06:47 with R.A. = 281.338 deg and DEC. = -13.167 deg, having an uncertainty of 2$\farcm$3 in radius \citep[][]{2024GCN.37555....1L}. Two follow-up observations by EP-FXT were performed by EP-WXT starting at 2024-09-19T12:40:28 and 2024-09-24T13:01:28, respectively. Two X-ray sources were identified within the specified EP-WXT error circle region in each observation. First, EP J184515.3-131115 was reported at R.A. = 281.3138 deg, DEC. = -13.1875 deg with an uncertainty of 10$\arcsec$ in radius. Second, EP J184516.2-130819 at R.A. = 281.3187 deg, DEC. = -13.1371 deg (J2000) with an uncertainty of 10$\arcsec$ in radius, which was consistent with the position of the star TYC 5704-6-1 \citep[][]{2024GCN.37608....1L}. Although no rapid dimming was reported for the two detected sources.

Followed by the EP-WXT discovery and EP-FXT follow-up observations, we triggered the 1m LOT at the Lulin Observatory to search for the optical counterpart. Similar to EP240918b, no other telescope trigger was reported on any platform. Our observations in $r$-band started about 19.59 hours after the EP-WXT trigger with a total exposure of 5$\times$300s. We only obtained an upper limit in the $r$-band of 18.9 mag, as presented in Table~\ref{tab:Photometry9}. We reported our preliminary photometry in \citet[][]{2024GCN.37577....1L}.

\subsubsection{EP240919a}
\label{sec:ep240919a}
EP240919a was discovered by EP-WXT at 2024-09-19T14:47:40 with a subsequent autonomous follow-up observation by EP-FXT post 10 minutes. An uncataloged source was detected by EP-FXT at the spatial position with R.A. = 334.2790 deg, DEC. = -9.7361 deg, having an uncertainty of 10$\arcsec$ in radius \citep[][]{2024GCN.37561....1L,2024GCN.37585....1L}.

Only about 80 seconds post EP-WXT discovery of EP240919a, an event was also discovered in the Fermi-GBM. The discovered transient was consistent with EP240919a in both timing and position, leading it to be also classified as a GRB event GRB 240919A \citep[][]{2024GCN.37563....1L,2024GCN.37580....1R}. The burst was also weakly detected by SVOM-GRM \citep[][]{2024GCN.37574....1S}. A high-energy counterpart detection was reported by INTEGRAL SPI-ACS \citep[][]{2024GCN.37573....1R} and interestingly, a radio counterpart of this event was also detected as reported by \citet[][]{2024GCN.37952....1G}. No plausible optical or NIR counterpart was discovered by any of the telescopes \citep[][]{2024GCN.37565....1J,2024GCN.37566....1K,2024GCN.37567....1M,2024GCN.37570....1L,2024GCN.37578....1Z,2024GCN.37579....1B,2024GCN.37583....1B,2024GCN.37610....1S}. Our observations from 40cm SLT in $r$-band started about 1.27 hours post EP-WXT trigger with a total exposure of 11$\times$300s. Although we were quick to search for the optical counterpart, we also only obtained upper limits down to 19.8 mag in as mentioned in Table~\ref{tab:Photometry9}. we reported our preliminary photometry in \citet[][]{2024GCN.37575....1A}. 

\subsubsection{EP241026b}
EP241026b was first detected by EP-WXT at 2024-10-26T18:14:30 with R.A. = 56.403 deg, DEC. = 41.031 deg, having an uncertainty of 2$\farcm$9 in radius. Followed by the EP-WXT discovery, a follow-up observation was performed using EP-FXT about 9 hr later. An uncatalogued X-ray source was reported with R.A. = 56.4058 deg, DEC. = 41.0312 deg, having an uncertainty of 10$\arcsec$ in radius \citep[][]{2024GCN.37902....1L}.

Followed by the EP-WXT discovery and EP-FXT follow-up observations, several telescopes were triggered to search for the optical counterpart of EP241026b. \citet[][]{2024GCN.37905....1L}, \citet[][]{2024GCN.37907....1M}, and \citet[][]{2024GCN.37920....1M} did not find any optical counterpart. However, about 1.48 days post EP-WXT trigger, \citet[][]{2024GCN.37938....1R} reported an obvious optical re-brightening transient at R.A. = 03$^{h}$45$^{m}$37$^{s}$.55 and DEC. = +41$^{\circ}$01$\arcmin$51$\farcs$9 that brightened by about 1.9 mag among their two epochs (first epoch reported in \citealt[][]{2024GCN.37907....1M} ) of observations. Later, \citet[][]{2024GCN.38018....1B} also claimed to have marginally detected the proposed optical counterpart about 4.31 days post EP-WXT trigger. The Keck-LRIS spectroscopic observation of this transient displayed a well-detected continuum in the range of 3400--10,200$\text{\AA}$, which put an upper limit of 1.8 for the redshift of EP241026b \citep[][]{2024GCN.38294....1Z}. Our observations from 1m LOT in $r$-band started about 142.08 hours post EP-WXT trigger with a total exposure of 12$\times$300s. We did not find any evidence of any optical counterpart candidate down to 23.5 mag. The deeper limit was favored by an excellent observing condition with seeing of $\sim$1$\arcsec$ and median airmass of $\sim$1. We reported our preliminary photometry in Aryan et al. (GCN38042).

\subsubsection{EP241101a}
EP241101a was discovered by EP-WXT at 2024-11-01T23:52:49 with R.A. = 37.763 deg, DEC. = 22.731 deg, having an uncertainty of 2$\farcm$8 in radius (GCN38039). Followed by the EP-WXT discovery, an autonomous observation by EP-FXT was also performed about 45 minutes later, but no counterpart was detected. Further analysis of the data during the EP-FXT's slew to the target before the observation started revealed an uncataloged X-ray source at R.A. = 37.7526 deg, DEC. = 22.7175 deg with an uncertainty of 10$\arcsec$ in radius (GCN38073).

Followed by the EP-WXT discovery and autonomous follow-up observations from EP-FXT, many telescopes triggered in the direction of the transient to search for the optical counterpart. Perez-Garcia et al. (GCN38047), Lipunov et al. (GCN38049), and Grossan et al. (GCN38079) reported only upper limits. However, Adami et al. (GCN38060) proposed a candidate optical counterpart at R.A. = 02$^{h}$30$^{m}$53$^{s}$.09, DEC. =  +22$^{\circ}$43$\arcmin$48$\farcs$11 with an $r$-band magnitude of 22.8$\pm$0.5. Later, Busmann et al. (GCN38064) also proposed another plausible candidate at R.A. = 02$^{h}$31$^{m}$10$^{s}$.85, DEC. =  +22$^{\circ}$44$\arcmin$54$\farcs$75 coinciding with a variable source, PSO J037.7952+22.7485 in the Pan-STARRS1 catalog. The two sources proposed above were also marked as candidates by Pankov et al. (GCN38109). However, we notice that the source proposed by Adami et al. (GCN38060) was already present in the  SDSS DR12 (SDSS J023053.10+224348.0) images, as well as in the DESI Legacy Survey images. Our observations from 1m LOT in $r$-band started about 17.08 hours after the EP-WXT trigger with a total exposure of 6$\times$300s. We reported the preliminary photometry in Aryan et al. (GCN38061). The upper limits from our observation in $r$-band and $g$-band are presented in Table~\ref{tab:Photometry9}.

\subsubsection{EP241104a}
\label{sec:ep241104a}
The EP-WXT first detected EP241104a at 2024-11-04T18:34:15 with R.A.= 32.574 deg, DEC. = 31.555 deg, having an uncertainty of 2$\farcm$7 in radius (GCN38081).

EP241104a was found spatially and temporally coincident with GRB 241104A (Fermi GBM team GCN38075) as indicated by Zhou et al. (GCN38081). Followed by the EP-WXT discovery of EP241104a, no optical follow-up observations were reported for the optical counterpart.  Our observations from 1m LOT in $r$-band started about 18.68 hours after the EP-WXT trigger with a total exposure of 6$\times$300s. We only obtained an upper limit of 22.8 mag in $r$-band as mentioned in Table~\ref{tab:Photometry9}. We reported our preliminary photometry in Wang et al. (GCN38084).

\subsubsection{EP241109a}
EP241109a was discovered by EP-WXT at 2024-11-09T06:01:55. An autonomous observation was also performed by EP-FXT, which detected an uncataloged source at the spatial position with R.A. = 18.3599 deg, DEC. =  0.0184 deg having an uncertainty of 15$\arcsec$ in radius \citep[][]{2024GCN.38140....1L}. 

\begin{figure*}
\centering
    \includegraphics[height=7.0cm,width=7cm,angle=0]{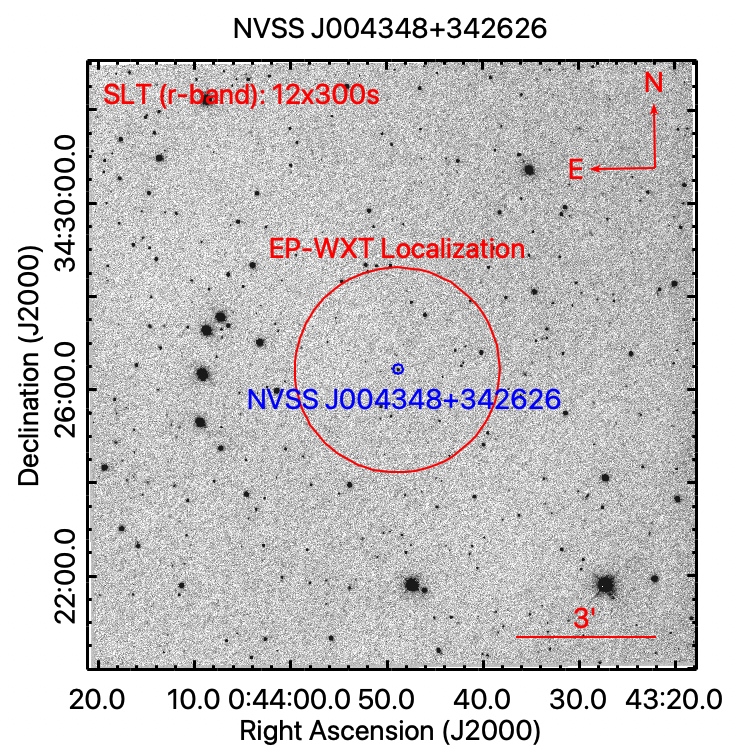}
    \includegraphics[height=7.0cm,width=7cm,angle=0]{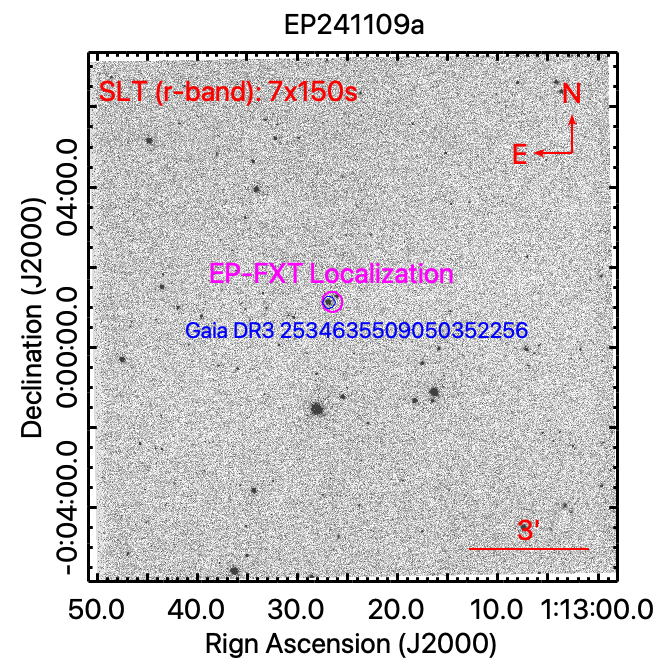}   
   \caption{{ The finding charts of the flaring blazar (left) and stellar flare (right) events from the Lulin Observatory 40cm SLT. The RA is in hh:mm:ss format while Dec is in dd:mm:ss format. The EP-WXT and EP-FXT (if available) localizations are indicated with red- and magenta-circles in the FoV, respectively.}}
    \label{fig:Blazar_n_stellar_flare}
\end{figure*}

\citet[][]{2024GCN.38140....1L} indicated the presence of a close star in the GAIA DR3 database lying within the EP-FXT error circle of EP241109a. Thus, they proposed EP241109a could be a stellar flare event. Further, the observation by \citet[][]{2024GCN.38141....1P} confirmed the event arising from a stellar flare as the proposed star diminished by 0.8 mag within 40 minutes of their observations. Later, spectroscopic observation by \citet[][]{2024GCN.38150....1Z} indicated slight excess in flux below $\sim$4000\,$\text{\AA}$. These observations further confirmed EP241109a being a stellar flare. Our observations from 40cm SLT in $r$-band started about 7.06 hours after the EP-WXT trigger with a total exposure of 7$\times$150s. We probably missed the stellar flare as our observations were late. The flared star had a magnitude of 13.99$\pm$0.03 in our observations. About 50 minutes post the first epoch of our observation, we performed further follow-up observation in $r$-band with a total exposure of 12$\times$30s. The new magnitude (13.94$\pm$0.13) was again similar to the previous one.  Besides this source, the upper limits in $r$-band from the two epochs of observations are presented in Table~\ref{tab:Photometry9}. {\bf The right-hand panel of Fig.~\ref{fig:Blazar_n_stellar_flare} shows the field of view of the star that underwent a flaring event.}

\subsubsection{EP241115a}
\label{sec:ep241115a}
EP-WXT discovered the EP241115a at 2024-11-15T05:47:20 with R.A. = 19.416 deg, DEC. = -17.954 deg having an uncertainty of 2$\farcm$6 in radius \citep[][]{2024GCN.38239....1H}.

The temporal and spatially localization of EP241115a coincided significantly with GRB 241115D \citep[][]{2024GCN.38250....1S}. A Swift-XRT ToO follow-up observation by \citet[][]{2024GCN.38251....1D} detected an uncataloged source within the EP-WXT localization error circle, and proposed that the source be associated with EP241115a. Followed by the EP-FXT detection, only a few telescopes triggered in the search for the optical counterpart of EP241115a. \citet[][]{2024GCN.38242....1L} obtained only upper limit. Our observations from 40cm SLT in $r$-band started about 30.64 hours post the EP-WXT trigger with a total exposure of 7$\times$300s. We too obtained upper limits in $r$-band and $g$-band as presented in Table~\ref{tab:Photometry9}. We reported our preliminary photometry in \citet[][]{2024GCN.38254....1F}.

\subsubsection{EP241119a}
EP241119a was first detected by EP-WXT at 2024-11-19T17:53:20 with R.A.= 84.116 deg, DEC. = 3.832 deg, having an uncertainty of 2$\farcm$3 in radius. About 9 hours post EP-WXT discovery, EP-FXT follow-up observation detected an uncataloged X-ray source at R.A. = 84.1062 deg, DEC = 3.8404 deg with an uncertainty of about 10$\arcsec$ in radius \citep[][]{2024GCN.38281....1Z}.

Followed by the EP-WXT discovery and EP-FXT follow-up observation, several telescopes were triggered to search for the optical/NIR counterpart of EP241119a. None of those observations found any evidence of optical counterpart \citep[][]{2024GCN.38282....1L,2024GCN.38286....1A,2024GCN.38288....1S,2024GCN.38610....1P}. Our observations from 40cm SLT in $r$-band started about 23.22 hours post the EP-WXT trigger with a total exposure of 24$\times$300s. We only obtained an upper limit as presented in Table~\ref{tab:Photometry9}. We reported our preliminary photometry in \citet[][]{2024GCN.38290....1A}.

\subsubsection{EP241125a}
EP-WXT discovered the EP241125a at 2024-11-25T00:06:06 with R.A.= 48.561 deg, DEC. = 37.677 deg having an uncertainty of 2$\farcm$65 in radius \citep[][]{2024GCN.38318....1W}.

Besides our observations from the Lulin Observatory, no other optical observations were reported for EP241125a.  Our observations from 1m LOT in $r$-band started about 12.09 hours post the EP-WXT trigger with a total exposure of 15$\times$120s. We did not detect any evidence of any optical counterpart candidate down to 23.0 mag as presented in Table~\ref{tab:Photometry9}. We reported our preliminary photometry in \citet[][]{2024GCN.38319....1L}.

\subsubsection{EP241126a}
EP241126a was first detected by EP-WXT at 2024-11-26T19:39:41 with R.A.= 33.744 deg, DEC. = 11.705 deg having an uncertainty of 2$\farcm$429 in radius \citep[][]{2024GCN.38335....1H}. Followed by the EP-WXT discovery, a ToO observation was performed using EP-FXT about 7.5 hr later. An uncatalogued X-ray source was reported with R.A. = 33.7444 deg, DEC. = 11.7013 deg, having an uncertainty of 10$\arcsec$ in radius \citep[][]{2024GCN.38339....1Z}.

Followed by the EP-WXT discovery and EP-FXT ToO observation, several telescopes were triggered to search for the optical counterpart of EP241126a. \citet[][]{2024GCN.38337....1F} proposed the detection of an optical counterpart with an $R$-band magnitude of $\sim$22.0 mag at R.A. = 02$^{h}$14$^{m}$57$^{s}$.96, DEC. =  +11$^{\circ}$42$\arcmin$04$\farcs$76 in their observations starting about 6.84 hr post EP-WXT discovery. This optical counterpart was also confirmed in further observations \citep[][]{2024GCN.38338....1L,2024GCN.38357....1G,2024GCN.38378....1S,2024GCN.38385....1M,2024GCN.38409....1F}. The observations by \citet[][]{2024GCN.38404....1Z} only obtain upper limits about 15.8 hr post EP-WXT discovery. Our observations from 1m LOT in $r$-band started about 16.23 hours post EP-WXT trigger with a total exposure of 6$\times$300s. We only obtained an upper limit of 22.5 mag in the $r$-band as mentioned in Table~\ref{tab:Photometry9}. Since the FXT was detected faint and our observations were late, the FXT probably faded beyond the detection limit in our images. We reported our preliminary photometry in \citet[][]{2024GCN.38344....1L}.

\subsubsection{EP241206a}
EP-WXT discovered the EP241206a at 2024-12-06T16:34:47 with R.A.= 34.702 deg, DEC. = 38.914 deg having an uncertainty of 3$\farcm$78 in radius \citep[][]{2024GCN.38457....1Y}.

Followed by the EP-WXT discovery, only a few telescopes were triggered to search for the optical counterpart. All the optical follow-up observations only obtained upper limits \citep[][]{2024GCN.38458....1L,2024GCN.38466....1X,2024GCN.38480....1S}. Our observations from 40cm SLT in $r$-band started about 22.45 hours post EP-WXT trigger with a total exposure of 8$\times$300s. We did not find any evidence of any optical counterpart in our observations down to 20.9 mag. Details of our further follow-up observations are presented in Table~\ref{tab:Photometry9}. 

\subsubsection{EP241208a}
\label{sec:ep241208a}
EP241208a was first detected by EP-WXT at 2024-12-08T16:36:13 with R.A.= 127.812 deg, DEC. = 49.082 deg having an uncertainty of 4$\arcmin$ in radius \citep[][]{2024GCN.38477....1W}. A follow-up observation was also performed using EP-FXT about 22 hr post EP-WXT discovery. An uncatalogued X-ray source was reported with R.A. = 127.8303 deg, DEC. = 49.0831 deg, having an uncertainty of 10$\arcsec$ in radius \citep[][]{2024GCN.38513....1W}.

A few telescopes were triggered in the direction of EP241208a to search for the optical counterpart. All of those observations only obtained upper limits \citep[][]{2024GCN.38514....1L,2024GCN.38560....1L}. Our observations from 40cm SLT in $r$-band started about 23.62 hours post EP-WXT trigger with a total exposure of 24$\times$300s. We also did not detect any optical counterpart candidate in our observations down to 21 mag. We reported our preliminary photometry in \citet[][]{2024GCN.38520....1W}. Later, the SVOM-ECLAIRs telescope detected a coincident long and soft transient with EP241208a \citep[][]{2024GCN.38557....1S}.

\section{Photometry logs for sources with upper limits}
Tables~\ref{tab:Photometry8} and Tables~\ref{tab:Photometry9} present the details of photometry logs for sources with upper limits.



\setcounter{table}{0}
\counterwithin{table}{section} 
\renewcommand{\thetable}{A\arabic{table}}

\input{Tables/UL_photometry}

\input{Tables/UL_photometry2}

\input{Tables/redback}

\section{Precursor searches for EP sources with optical counterparts}
\label{subsec:precursor}



\begin{figure*}
\centering
    \includegraphics[width=0.7\textwidth,angle=0]{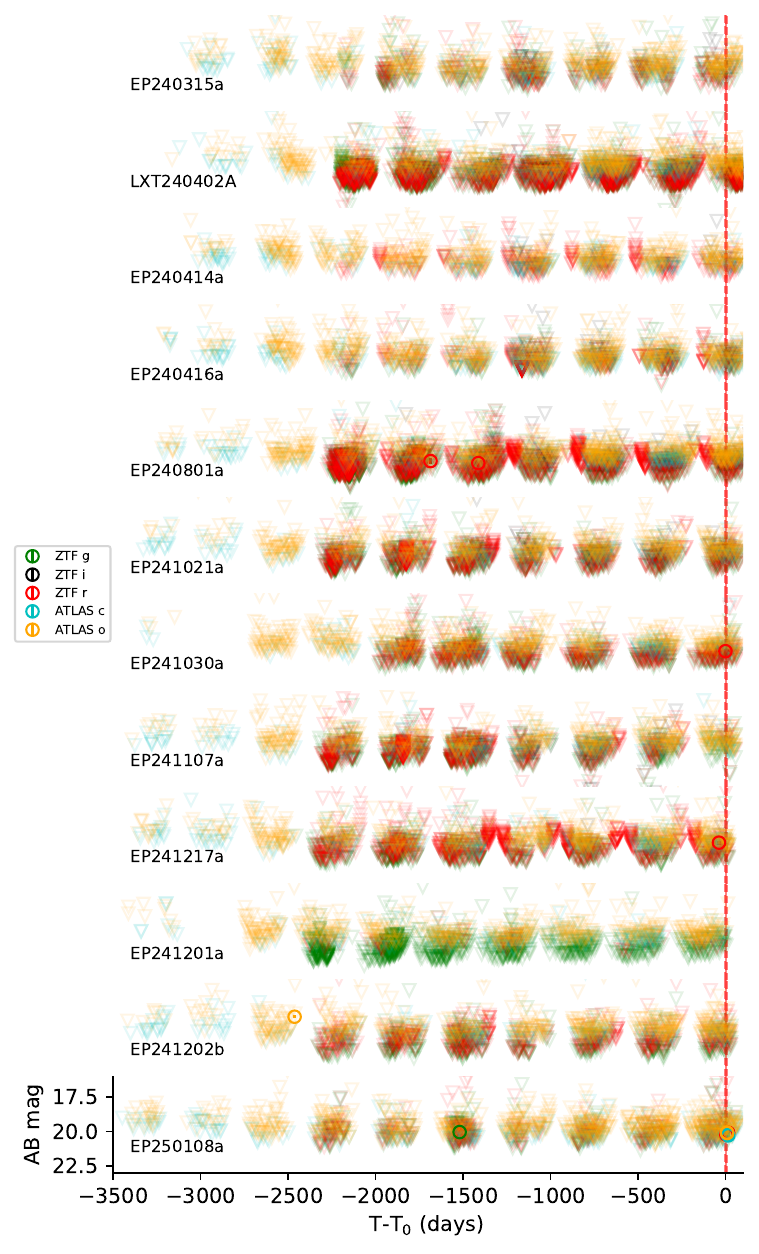}
  
   \caption{Search for precursor activity at the locations of FXTs discovered by EP-WXT, using archival data from the ZTF and ATLAS surveys over a $\sim$ 10-year baseline prior to the reported EP detection time ($T_0$, marked as day zero on the x-axis). The y-axis shows the milky way extinction-corrected AB magnitudes. Both ZTF ($g$, $r$, and $i$ bands) and ATLAS ($c$ and $o$ bands) forced photometry light curves are shown. 5$\sigma$ detection limits are used throughout. Detections are indicated by open circles, while upper limits are marked as open inverted triangles.}
    \label{fig:precursor}
\end{figure*}

As shown in Fig.~\ref{fig:precursor}, we utilized forced photometry from the ZTF and ATLAS surveys to search for potential precursor activity in archival difference images, spanning approximately 10 years prior to the reported EP detection time. This analysis reveals no significant evidence of real, astrophysical excess flux in the historical data, down to the 5$\sigma$ detection limit.
\begin{figure}
    \centering
    \includegraphics[width=0.9\linewidth]{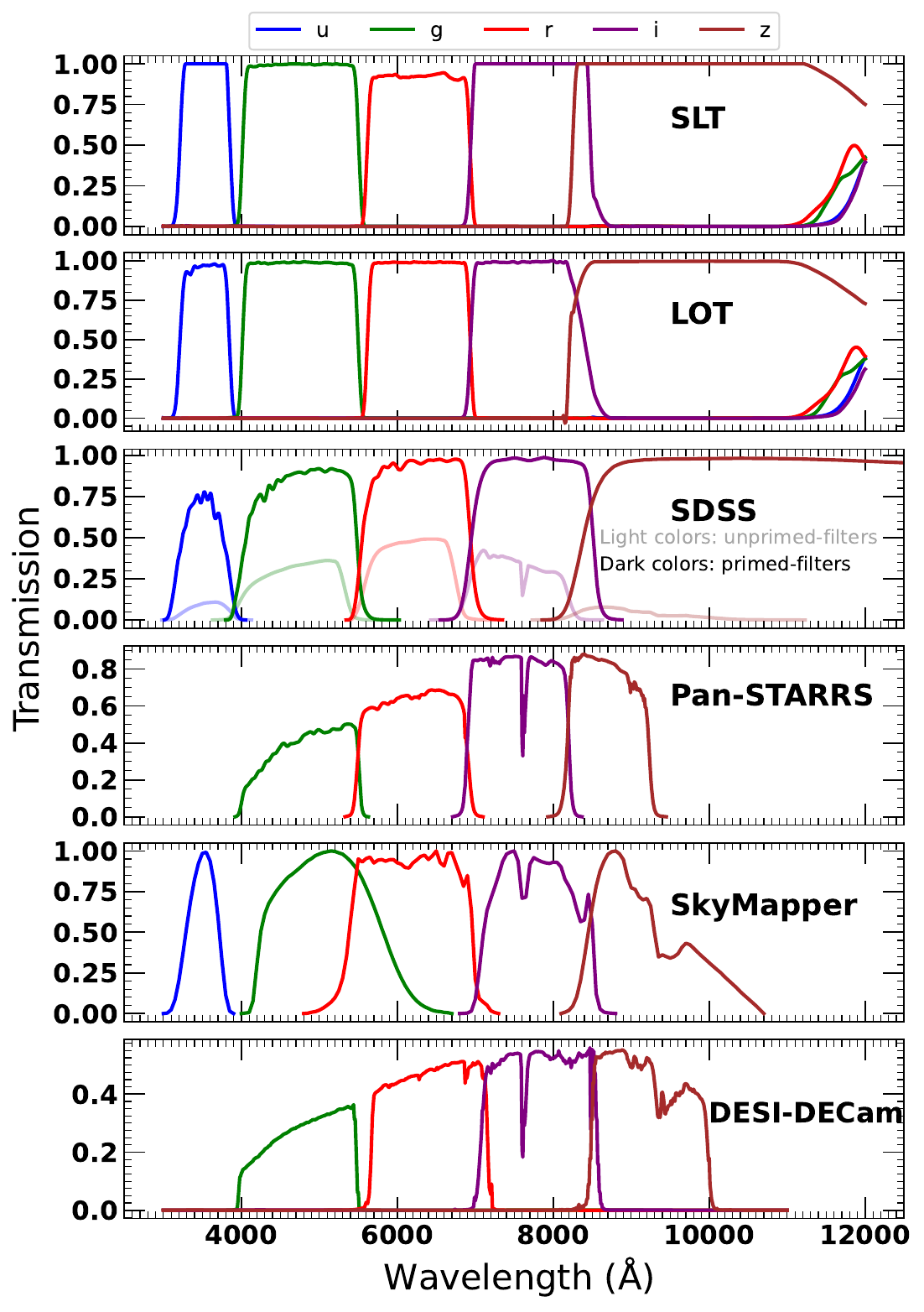}
    \caption{The comparison of the Lulin Observatory LOT and SLT filters with different sky survey filters.}
    \label{fig:filters}
\end{figure}

\section{References for the first detections and upper limits}
Here, we credit for the first optical detections or the earliest upper limits presented in Fig.~\ref{fig:gcn_summary}: EPW20240219aa: \citet[][]{2024GCN.35783....1B};
EPW20240305aa: \citet[][]{2024ATel16531....1A};
LXT240402A: \citet[][]{2024GCN.36027....1Y} and \citet[][]{2024GCN.36025....1L};
EP240315a: \citet[][]{2024ApJ...969L..14G};
EP240331a: \citet[][]{2024GCN.36011....1C};
EP240408a: \citet[][]{2024GCN.36079....1L};
EP240413a: \citet[][]{2024GCN.36098....1P};
EP240414a: \citet[][]{2024GCN.36094....1A};
EP240416a: \citet[][]{2024GCN.36139....1C};
EP240417a: \citet[][]{2024GCN.36178....1S};
EP240420a: \citet[][]{2024GCN.36202....1A};
EP240426a: \citet[][]{2024GCN.36315....1W};
EP240426b: \citet[][]{2024GCN.36335....1S};
EP240506a: \citet[][]{2024GCN.36412....1P};
EP240518a: \citet[][]{2024GCN.36529....1L};
EP240617a: This work;
EP240618a: This work;
EP240625a: \citet[][]{2024GCN.36761....1F};
EP240626a \citet[][]{2024GCN.36773....1Z};
EP240702a: \citet[][]{2024GCN.36806....1L};
EP240703a: This work;
EP240703b: \citet[][]{2024GCN.36836....1L};
EP240703c: This work;
EP240708a: \citet[][]{2024GCN.36842....1J};
EP240801a: \citet[][]{2024GCN.36998....1F};
EP240802a: \citet[][]{2024GCN.37094....1Z};
EP240804a: \citet[][]{2024GCN.37035....1A};
EP240806a: \citet[][]{2024GCN.37080....1P} ( No information of time in \citet[][]{2024GCN.37064....1L}, who reported first detection); 
EP240807a: \citet[][]{2024GCN.37105....1L};
EP240816a: \citet[][]{2024GCN.37199....1P};
EP240816b: \citet[][]{2024GCN.37191....1Z};
EP240820a: \citet[][]{2024GCN.37215....1A};
EP240904a: \citet[][]{2024ATel16807....1F};
EP240908a: \citet[][]{2024GCN.37434....1F};
EP240913a: \citet[][]{2024GCN.37500....1Z};
EP240918a: This work;
EP240918b: This work;
EP240918c: This work;
EP240919a: This work;
EP240930a: \citet[][]{2024GCN.37649....1L};
EP241021a: \citet[][]{2024GCN.37840....1F};
EP241025a: \citet[][]{2024GCN.37862....1J};
EP241026a: \citet[][]{2024GCN.37900....1W};
EP241026b: \citet[][]{2024GCN.37938....1R};
EP241030a: \citet[][]{2024GCN.37956....1K};
EP241101a: Perez-Garcia et al., GCN 38047;
EP241103a:  Izzo et al., GCN 38053;
EP241104a: Lipunov et al., GCN 38076;
EP241107a: Odeh et al.,GCN 38115;
EP241113b: \citet[][]{2024GCN.38215....1L};
EP241115a: \citet[][]{2024GCN.38242....1L};
EP241119a: \citet[][]{2024GCN.38282....1L};
EP241125a: This work;
EP241126a: \citet[][]{2024GCN.38337....1F};
EP241201a: \citet[][]{2024GCN.38418....1L};
EP241202b: \citet[][]{2024GCN.38433....1N};
EP241206a: \citet[][]{2024GCN.38466....1X};
EP241208a: \citet[][]{2024GCN.38560....1L};
EP241213a: \citet[][]{2024GCN.38556....1B};
EP241217a: \citet[][]{2024GCN.38587....1L};
EP241217b: \citet[][]{2024GCN.38597....1H};
EP241223a: \citet[][]{2024GCN.38661....1L};
EP241231b: \citet[][]{2025GCN.38788....1K};
EP250101a: \citet[][]{2025GCN.38801....1B};
EP250108a: \citet[][]{2025GCN.38878....1E};
EP250109a: \citet[][]{2025GCN.38872....1S};
EP250109b: \citet[][]{2025GCN.38921....1K};
No optical follow-ups were reported for EP240305a, EP240309a, EP240327a, and EP240709a. 


\bibliographystyle{aasjournal}




\end{document}

%% file: Tables/Detection_list.tex
\begin{rotatetable}
\begin{deluxetable*}{cccccccc}
    \renewcommand*{\arraystretch}{1.1}
    \centering

    \tablecaption{Details of the FXTs with their optical counterpart observed from Lulin Observatory.}
    \startdata
        Name   &   WXT R.A.	& WXT Dec &	WXT error & FXT R.A.	& FXT Dec	& FXT error  & Time${\rm _{FXT-WXT}}$                \\
           &   OT R.A.$^\mathsection$       & OT Dec.       &Redshift        & Optical counterpart discoverer      & OT name   &   & Time${\rm _{OT-WXT}}$          \\
        \midrule       
        EP240315a & 141.644 &	-9.547	& 3\arcmin	& 141.6483	& -9.5335	& 10\arcsec & 42~hr \\
        (GRB~240315C) & 141.64764	 & -9.53409 & 4.859$^{1}$  &  \citealt{2024GCN.35932....1S}, & AT~2024eju & & 1.1~hr
\\
    &   &   &    &  \citealt{2024ApJ...969L..14G} & & \\
        LXT240402A & -- &	-- &	--	& 245.451	& 25.763	& 10\arcsec & -- \\
        (GRB~240402B) &  245.45103  & +25.76315 & 1.551$^{2}$ & \citealt{2024GCN.36025....1L},  & -- & & 1.28~day \\
          &   &	  &	 	&  	\citealt{2024GCN.36027....1Y}$^\star \dagger$, this work  &  & \\      
        EP240414a  & 191.498 &	-9.695 &	3\arcmin	& 191.509	&-9.718	&10\arcsec &2~hr \\
        & 191.50695 & -9.71913 & 0.41$^{3}$ & \citealt{2024GCN.36094....1A}, & AT~2024gsa & & 3.13~hr \\
        &   &   &   & \citealt{2025ApJ...978L..21S}, this work &\\
        EP240416a & 203.15	& -13.612	& 3\arcmin	& --	& --	& -- &  -- \\
        &  203.14381  & -13.63029 &  -- & \citealt{2024GCN.36139....1C}$^\star$, this work & & &14.15~hr\\
        EP240801a & 345.14	& 32.61	& 2.4\arcmin	& 345.163	& 32.5927	& 10\arcsec & 180~sec\\
        & 345.16263	 & 32.59386 & 1.673$^{4}$  & \citealt{2024GCN.36998....1F} & & & 2.24~hr\\        
        EP241021a &28.852	& 5.957	& 2.4\arcmin	& 28.8483	& 5.9395	& 10\arcsec & 36.5~hr\\
        & 28.84757	 & +5.93842 & 0.75$^{5}$ & \citealt{2024GCN.37840....1F} & & & 1.77~day\\ 
        EP241030a & 343.013	& 80.449	& 2.4\arcmin	& 343.1426	& 80.4498	& 10\arcsec & 21~hr\\
        (GRB~241030A) &  343.13988  & +80.44997 & 1.411$^{6}$ & \citealt{2024GCN.37956....1K} & & & -0.73~hr$^\ddagger$ \\ 
        EP241107a & --	& --	& --	& 35.0085	& 3.3329	& 10\arcsec & 5~min \\
        &   35.01021  & +3.33394 & 0.456$^{7}$  & \citealt{2024GCN.38128....1O} & & & 1.5~hr \\
        EP241201a & 282.596	& 66.081	& 2.343\arcmin & 282.4865 & 66.0693	& 10\arcsec & 12~hr\\
        &  282.65866  & +66.05316 & -- & \citet[][]{2024GCN.38418....1L}$^\star$, this work & & & 13.4~hr \\        
        EP241202b & 45.302	& 2.441	&2.6\arcmin	& --	& --	& -- & -- \\
        & 45.33693  & +2.44084 & -- & \citet[][]{2024GCN.38433....1N}$^\star$, this work & & & 19.45~hr \\        
        EP241217a & 46.957	& 30.901	& 2.8\arcmin	& 46.9398	& 30.9299	& 20\arcsec & 7.64~hr\\
        &  46.9425 	 & +30.92942 & 4.59$^{8}$ & \citet[][]{2024GCN.38587....1L} & & & 2.5~hr \\        
        EP250108a & 55.623	&-22.509	&2.2\arcmin	& limit	& limit	& limit & 22.2~hr \\
        & 55.61829	 & -22.50591 & 0.176$^{9}$  & \citealt{2025GCN.38878....1E}, & SN~2025kg & & 31.5~hr \\
        &   &  &    & \citealt{2025arXiv250408886E} &   & &  \\ 
    \enddata
\label{tab:detection}
\footnotesize{$^\ddagger$ The OT was detected by $\it{Swift}$/UVOT shortly after the GRB trigger, and thus prior to the EP-WXT trigger time. $^\mathsection$ The coordinates (R.A. and Dec.) of the OT are adopted from the relevant discovery GCNs; while for EP240315a, EP240414a, EP241021a and EP250108a we adopted the Pan-STARRS updated coordinates. $^{1}$\citet[][]{2024GCN.35936....1S}; $^{2}$\citep[][]{2024GCN.36385....1T}; $^{3}$\citep[][]{2024GCN.36110....1J}; $^{4}$\citep[][]{2024GCN.37013....1Q}; $^{5}$\citep[][]{2024GCN.37852....1P}; $^{6}$\citep[][]{2024GCN.37959....1Z}; $^{7}$\citep[][]{2024GCN.38126}; $^{8}$\citep[][]{2024GCN.38593....1L}; $^{9}$\citep[][]{2025GCN.38908....1Z} }\\
\end{deluxetable*}
\end{rotatetable}



%% file: Tables/FXT_coordinates.tex

\begin{table*}
    \renewcommand*{\arraystretch}{1.1}
    \centering
    \caption{
        Details of FXTs having optical upper limits from Lulin observatory.}
    \begin{tabular}{ccccccccc}
        \toprule
        \midrule
        
   FXT & R.A.  & DEC.      & Error circle radius$^{*}$      &T$_{0}$       & EP-WXT Discovery   \\

      & (deg, J2000)  & (deg, J2000)     &  EP-WXT / EP-FXT        &        (UTC)        &     reference   \\
        \midrule

   EP240331a & 169.414 &  -20.042    &   20$\arcmin$ / -  & 2024-03-31T22:07:17   & \citet[][]{2024ATel16564....1P} \\          

   EP240408a & 158.840 &  -35.749    &   3$\arcmin$ / -  & 2024-04-08T17:56:30   & \citet[][]{2024GCN.36053....1H} \\ 

   EP240413a & 228.794 &  -18.800    &   20$\arcmin$ / 30\arcsec  & 2024-04-13T14:39:37   & \citet[][]{2024GCN.36086....1L} \\

   EP240506a & 213.978 &  -16.715    &   3$\arcmin$ / -  & 2024-05-06T05:01:39   & \citet[][]{2024GCN.36405....1L} \\

   EP240617a & 285.030 &  -22.561    &   3$\arcmin$ / -   & 2024-06-17T12:19:13   & \citet[][]{2024GCN.36691....1Z} \\   

   EP240618a & 281.648 &  23.833    &   3\arcmin / 30$\arcsec$   & 2024-06-18T05:43:43   & \citet[][]{2024GCN.36690....1S} \\          

   EP240625a & 310.7308 &  -15.9760    &   2\arcmin / 10$\arcsec$   & 2024-06-25T01:48:23   & \citet[][]{2024GCN.36757....1P} \\ 

   EP240626a & 263.0171 &  -13.0490    &  2\arcmin / 30$\arcsec$   & 2024-06-26T06:28:28   & \citet[][]{2024GCN.36690....1S} \\
   
   EP240702a & 328.203 &  -38.980    &   3$\arcmin$ / -  & 2024-07-02T00:50:05   & \citet[][]{2024GCN.36801....1C} \\

   EP240703a & 273.803 &  -9.681    &   3$\arcmin$ / -  & 2024-07-03T00:38:40   & \citet[][]{2024GCN.36807....1W} \\

   EP240703c & 289.264 &   -30.325    &   3$\arcmin$ / -   & 2024-07-03T18:15:00   & \citet[][]{2024GCN.36818....1Z} \\

   EP240708a & 345.9656 &  -22.8428   &   3\arcmin / 10$\arcsec$   & 2024-07-08T23:28:23   & \citet[][]{2024GCN.36818....1Z} \\   

   NVSS J004348+342626$^{\dagger}$ & 10.955 &  34.428   &   2$\farcm$2 / - & 2024-07-17T03:13:01   & \citet[][]{2024ATel16725....1J} \\ 

   EP240802a & 287.8070 &  -2.3125   &   1\arcmin.9 / 10$\arcsec$   & 2024-08-02T10:32:52   & \citet[][]{2024GCN.37019....1W} \\ 

   EP240908a & 14.0031 &  8.0735   &  - / 10$\arcsec$   & 2024-09-08T17:28:27   & \citet[][]{2024GCN.37432....1M} \\ 

   EP240913a & 16.681  &  16.750   &   2$\farcm$5 / -  & 2024-09-13T11:39:33   & \citet[][]{2024GCN.37492....1L} \\

   EP240918a & 289.3937  &  46.1281   &   - / 20$\arcsec$   & 2024-09-18T11:24:37   & \citet[][]{2024GCN.37541....1Z} \\ 

   EP240918b & 258.66  &  66.739   &   2$\farcm$9 / -  & 2024-09-18T15:40:00   & \citet[][]{2024GCN.37555....1L} \\ 

   EP240918c & 281.338  &  -13.167   &   2$\farcm$3 / 10\arcsec    & 2024-09-18T18:06:47   & \citet[][]{2024GCN.37555....1L} \\
   
   EP240919a & 334.2797  &  -9.7362   &   - \ 20$\arcsec$  & 2024-09-19T14:47:40   & \citet[][]{2024GCN.37561....1L} \\
   
   EP241026b & 56.4058  &  41.0312   &   2\arcmin.9 / 10$\arcsec$  & 2024-10-26T18:14:30   & \citet[][]{2024GCN.37902....1L} \\

   EP241101a & 37.7526  &  22.7175   &   2\arcmin.8 
 / 10$\arcsec$  & 2024-11-01T23:52:49   & GCN38039 \\   

   EP241104a & 32.574  &  31.555   &   2$\farcm$7 / - & 2024-11-04T18:34:15   & GCN38081 \\  

   EP241109a & 18.3599 &  0.0184   &  - / 15$\arcsec$  & 2024-11-09T06:01:55   & \citet[][]{2024GCN.38140....1L} \\

   EP241115a & 19.416 &  -17.954   &   2$\farcm$6 / - & 2024-11-15T05:47:20   & \citet[][]{2024GCN.38239....1H} \\

   EP241119a & 84.1062 &  3.8404   &  2\arcmin.3 / 10$\arcsec$  & 2024-11-19T17:53:20   & \citet[][]{2024GCN.38281....1Z} \\

   EP241125a & 48.561 &  37.677   &   2$\farcm$65 / -  & 2024-11-25T00:06:06   & \citet[][]{2024GCN.38318....1W} \\  

   EP241126a & 33.7444 &  11.7013   &   2\arcmin.429 / 10$\arcsec$  & 2024-11-26T19:39:41   & \citet[][]{2024GCN.38335....1H} \\  

   EP241206a & 34.702 &  38.914   &   3$\farcm$78 / - & 2024-12-06T16:34:47   & \citet[][]{2024GCN.38457....1Y} \\   

   EP241208a & 127.8303 &  49.0831   &   4\arcmin / 10$\arcsec$  & 2024-12-08T16:36:13   & \citet[][]{2024GCN.38477....1W} \\
    \midrule
    
    \end{tabular}
    \label{tab:Coordinates}
{$^*$The error circle radius is 90\% C.L. statistical and systematic.\\$\dagger$ An FSRQ blazar within the EP-WXT localization error circle.}
\end{table*}

%% file: Tables/Photometry1.tex

\begin{table*}
    \renewcommand*{\arraystretch}{1.1}
    \centering
    \caption{
        Photometric observation log for the sources with optical counterpart discovery/follow-up from Lulin observatory. The magnitudes in the table are not corrected for the expected foreground extinction following \citet{2011ApJ...737..103S}, in the direction of transients.}
    \begin{tabular}{cccccccccc}
        \toprule
        \midrule
        
  FXT & $T_{\rm start} - T_0$  & $T_{\rm mid} - T_0$       &MJD$_{\rm start}$     &Telescope       &Total exposure         &Filter        &Apparent magnitude   & $E(B-V)$  \\

      & (hr)  & (hr)     &          &                &               &              &(AB mag)       &        \\
        \midrule
  EP240315a$^{\star}$ &   19.98  &   20.20  &   60385.673  &   SLT &   6$\times$300s   &   r &   21.34$\pm$0.22    &   0.042 \\  
            &   43.22  &   43.67  &   60386.642  &   LOT &   10$\times$300s   &   i &   22.75$\pm$0.23    &    \\ 
            &   89.58  &   90.86  &   60388.573  &   SLT &   29$\times$300s   &   i &   $>$21.8    &    \\ 
            &   89.92  &   91.34  &   60388.588  &   LOT &   30$\times$300s   &   i &   22.79$\pm$0.36    &    \\ 
            &   113.21  &   114.40  &   60389.558  &   LOT &   20$\times$300s   &   i &   $>$22.1    &    \\
            
  LXT240402A &   32.31  &   32.95  &   60403.713  &   LOT &   12$\times$300s    &   r &   22.28$\pm$0.10    &   0.037 \\ 
             &   33.70  &   34.10  &   60403.771  &   LOT &   10$\times$300s    &   g &   22.86$\pm$0.09    &    \\

  EP240414a$^{\star}$ &   3.25  &   3.97  &   60414.545  &   LOT &   11$\times$300s   &   r &   21.55$\pm$0.08    &   0.033 \\ 
            &   3.70  &   3.46  &   60414.564  &   LOT &   12$\times$300s   &   i &   21.40$\pm$0.16    &    \\
            &   8.02  &   8.25  &   60414.744  &   LOT &   6$\times$300s    &   g &   21.90$\pm$0.12    &    \\
            &   28.04  &   29.58  &   60415.578  &   LOT &   19$\times$300s    &   r &   22.05$\pm$0.07    &    \\
            &   28.61  &   28.79  &   60415.602  &   LOT &   5$\times$300s    &   i &   21.69$\pm$0.23    &    \\
            &   50.71  &   51.69  &   60416.523  &   LOT &   20$\times$300s    &   r &   22.05$\pm$0.14     &    \\
            &   52.76  &   53.51  &   60416.608  &   LOT &   18$\times$300s    &   i &   22.20$\pm$0.15     &    \\
            &   73.73  &   74.37  &   60417.482  &   LOT &   11$\times$300s    &   r &   21.06$\pm$0.15     &    \\
             
  EP240416a &   14.15  &   15.40  &   60416.702  &   LOT &   7$\times$300s    &   g &   22.39$\pm$0.12    &   0.056 \\ 
            &   14.27  &   14.61  &   60416.707  &   LOT &   7$\times$300s    &   r &   22.08$\pm$0.12    &   \\
            &   14.36  &   14.94  &   60416.711  &   LOT &   7$\times$300s    &   i &   22.02$\pm$0.16    &   \\
            &   444.92  &   446.10  &   60434.651  &   SLT &   7$\times$300s    &   i &   $>$21.19    &   \\
  EP240801a$^{\star}$ &   4.66  &   5.14  &   60523.573  &   SLT &   12$\times$300s   &   r &   20.90$\pm$0.30    &   0.099 \\
            &   6.25  &   6.73  &   60523.640  &   SLT &   11$\times$300s   &   r &   21.17$\pm$0.11    & \\
            &   9.42  &   9.85  &   60523.772  &   SLT &   11$\times$300s   &   r &   21.36$\pm$0.15    & \\
  EP241021a &   31.15  &   32.17  &   60605.512  &   SLT &   24$\times$300s   &   r &   $>$19.6    &   0.048 \\        
            &   274.61  &   274.83  &   60615.656  &   LOT &   6$\times$300s   &   r &   22.06$\pm$0.33    &    \\ 
            &   325.21  &   325.69  &   60617.764  &   LOT &   12$\times$300s   &   r &   $>$22.75    &    \\ 

  EP241030a &   55.15  &   55.37  &   60615.571  &   LOT &   6$\times$300s  &   r &   21.55$\pm$0.08   &   0.121 \\  
            &   55.68  &   55.90  &   60615.593  &   LOT &   6$\times$300s  &   g &   22.27$\pm$0.12   &    \\
            &   172.64  &   172.86  &   60620.466  &   LOT &   6$\times$300s  &   r &   $>$22.06   &    \\
            
  EP241107a &   24.47  &   24.68  &   60622.610  &   LOT &   6$\times$300s   &   r &   21.99$\pm$0.16    &   0.044 \\

        \bottomrule
    \end{tabular}
    \label{tab:Photometry1}

    {$*$ The photometry for EP240315a, EP240414a, and EP240801a, were already published in \citet[][]{2024ApJ...969L..14G}, \citet[][]{2025ApJ...978L..21S}, and \citet{2025arXiv250304306J}, respectively.}
\end{table*}

%% file: Tables/Photometry2.tex

\begin{table*}
    \renewcommand*{\arraystretch}{1.1}
    \centering
    \caption{
        Table 3, continued.}
    \begin{tabular}{cccccccccc}
        \toprule
        \midrule
        
  FXT & $T_{\rm start} - T_0$  & $T_{\rm mid} - T_0$       &MJD$_{\rm start}$     &Telescope       &Total exposure         &Filter        &Apparent magnitude   & $E(B-V)$  \\

      & (hr)  & (hr)     &          &                &               &              &(AB mag)       &        \\
        \midrule
  EP241201a &   13.40  &   13.62  &   60646.434  &   LOT &   6$\times$300s   &   r &   22.69$\pm$0.41    &   0.051 \\

  EP241202b &   19.45  &   19.67  &   60647.445  &   LOT &   6$\times$300s   &   r &   22.41$\pm$0.16    &   0.085 \\
            &   20.26  &   21.34  &   60647.478  &   SLT &   24$\times$300s   &   g &   $>$21.8    &   \\
            &   21.36  &   21.58  &   60647.524  &   LOT &   6$\times$300s   &   r &   $>$22.9    &   \\
  EP241217a &   7.34  &   7.96  &   60661.539  &   LOT &   15$\times$300s    &   r &   21.18$\pm$0.10    &   0.204 \\
            &   8.68  &   8.99  &   60661.595  &   LOT &   6$\times$300s    &   g &   $>$21.88    &    \\
            &   8.77  &   9.27  &   60661.599  &   LOT &   6$\times$300s    &   i &   19.31$\pm$0.06    &    \\
            
  EP250108a$^{\star}$ &   190.53  &   191.68  &   60691.460  &   SLT &   24$\times$300s   &   i &   $>$19.8    &   0.016 \\
                      &   190.69  &   190.77  &   60691.466  &   LOT &   3$\times$300s   &   r &   20.55$\pm$0.21   &    \\
                      &   190.95  &   191.04  &   60691.478  &   LOT &   3$\times$300s   &   g &   20.83$\pm$0.11   &    \\       &   215.90  &   217.62  &   60692.517  &   SLT &   35$\times$300s   &   i &   20.55$\pm$0.21   &    \\      &   263.08  &   263.30  &   60694.483  &   LOT &   6$\times$300s   &   r &   20.24$\pm$0.06   &    \\ 
                      &   263.98  &   265.61  &   60694.520  &   SLT &   18$\times$300s   &   g &   20.92$\pm$0.24   &    \\
                      &   359.58  &   361.15  &   60698.504  &   SLT &   36$\times$300s   &   i &   20.36$\pm$0.14   &    \\
                      &   982.28  &   982.50  &   60724.450  &   LOT &   6$\times$300s   &   r &   21.43$\pm$0.10   &    \\
        \bottomrule
    \end{tabular}
    \label{tab:Photometry2}

    {$*$ The photometry for EP250108a were published in \citet{2025arXiv250408889R}.}
\end{table*}

%% file: Tables/UL_photometry.tex

\begin{table*}
    \renewcommand*{\arraystretch}{1.1}
    \centering
    \caption{
        Photometric observation log for the sources with upper limit. The magnitudes in the table are not corrected for the expected foreground extinction.}
    \begin{tabular}{ccccccccc}
        \toprule
        \midrule
        
   FXT & $T_{\rm start} - T_0$  & $T_{\rm mid} - T_0$       &MJD$_{\rm start}$      &Telescope       &Total exposure         &Filter        &Magnitude & catalog    \\

      & (hr)  & (hr)     &          &                &               &              &(AB mag)               \\
        \midrule

   EP240331a & 17.62 &  18.30   &   60401.656  & LOT   & 4$\times$4$\times$300s  & r  & $>$22.3$^*$ & Pan-STARRS1   \\
            & 17.67 &  18.24   &   60401.658  & SLT   & 3$\times$4$\times$300s+1$\times$3$\times$300s  & g  & $>$21.3$^*$ & Pan-STARRS1   \\
   
   EP240408a & 42.55 &  43.41   &   60410.520  & LOT   & 24$\times$180s  & r  & $>$22.5 & SkyMapper   \\
        & 42.74 &  43.65   &   60410.528  & SLT   & 15$\times$300s  & i  & $>$20.3 & SkyMapper   \\        
        & 65.33 &  65.85   &   60411.469  & LOT   & 20$\times$300s  & r  & $>$22.8 & SkyMapper   \\

   EP240413a & 73.55 &  73.77   &   60416.676  & LOT   & 6$\times$300s  & r  & $>$22.8 & Pan-STARRS1   \\       

   EP240506a & 32.10 &  33.56   &   60437.547  & SLT   & 30$\times$300s  & r  & $>$22.0 & SDSS   \\    

   EP240617a & 30.95 &  31.24   &   60479.803  & SLT   & 6$\times$300s  & r  & $>$19.8 & Pan-STARRS1   \\    
        & 97.94 &  98.36   &   60482.594  & SLT   & 10$\times$300s  & r  & $>$19.3 & Pan-STARRS1   \\ 
        & 98.87 &  99.30   &   60482.633  & SLT   & 10$\times$300s  & i  & $>$19.3 & Pan-STARRS1   \\    

   EP240618a & 11.81 &  12.20   &   60479.731  & SLT   & 10$\times$300s  & r  & $>$20.6 & Pan-STARRS1   \\ 
        & 12.67 &  13.06   &   60479.767  & SLT   & 10$\times$300s  & i  & $>$20.7 & Pan-STARRS1 \\
        & 35.18 &  35.57   &   60480.705  & SLT   & 10$\times$300s  & r  & $>$20.0 & Pan-STARRS1 \\   
        & 36.05 &  36.43   &   60480.741  & SLT   & 10$\times$300s  & i  & $>$20.5 & Pan-STARRS1 \\        
        & 82.43 &  82.91   &   60482.673  & SLT   & 10$\times$300s  & r  & $>$20.6 & Pan-STARRS1 \\  
        & 83.47 &  83.86   &   60482.717  & SLT   & 10$\times$300s  & i  & $>$20.6 & Pan-STARRS1 \\

   EP240625a & 40.99 &  41.37   &   60487.783  & SLT   & 10$\times$300s  & i  & $>$19.8 & SDSS\\
        & 62.39 &  62.60   &   60488.675  & SLT   & 6$\times$300s  & r  & $>$19.9 & SDSS\\        

   EP240626a & 33.16 &  33.38   &   60488.651  & SLT   & 6$\times$300s  & r  & $>$20.1 & Pan-STARRS1\\

   EP240702a & 16.05 &  17.70   &   60493.704  & SLT   & 35$\times$300s  & r  & $>$22.2 & SkyMapper\\

   EP240703a & 12.43 &  12.52   &   60494.545  & LOT   & 3$\times$300s  & r  & $>$21.8 & Pan-STARRS1\\  
         & 13.72 &  13.94   &   60494.599  & LOT   & 6$\times$300s  & i  & $>$21.6 & Pan-STARRS1\\

   EP240703c & 20.61 &  21.27   &   60495.619  & SLT   & 16$\times$300s  & r  & $>$21.5 & Pan-STARRS1\\

        \bottomrule
    \end{tabular}
    \label{tab:Photometry8}
{$^*$Deepest limiting magnitude among the four pointings. In r-band, each pointing has a total exposure time of 4$\times$300s, while in g-band, three pointings have 4$\times$300s and one pointing has 3$\times$300s of exposure.}
\end{table*}

%% file: Tables/UL_photometry2.tex

\begin{table*}
    \renewcommand*{\arraystretch}{1.1}
    \centering
    \caption{
        Table 8, continued.}
    \begin{tabular}{ccccccccc}
        \toprule
        \midrule
        
   FXT & $T_{\rm start} - T_0$  & $T_{\rm mid} - T_0$       &MJD$_{\rm start}$      &Telescope       &Total exposure        &Filter        &Magnitude & catalog    \\

      & (hr)  & (hr)     &          &                &               &              &(AB mag)               \\
        \midrule
   EP240708a & 17.02 &  18.53   &   60500.687  & SLT   & 36$\times$300s  & r  & $>$21.3 & Pan-STARRS1   \\
        & 17.16 &  17.38   &   60500.693  & LOT   & 6$\times$300s  & r  & $>$22.6 & Pan-STARRS1  \\    
        & 17.78 &  18.00   &   60500.719  & LOT   & 6$\times$300s  & g  & $>$22.4 & Pan-STARRS1  \\ 
   Blazar\footnote{Blazar NVSS J004348+342626 flaring event.} & 39.55 &  40.02   &   60509.782  & SLT   & 12$\times$300s  & r  & 18.42$\pm$0.05 & SDSS   \\
        & 39.55 &  40.02   &   60509.782  & SLT   & 12$\times$300s  & r  & $>$21.1$^{*}$ & SDSS \\
   EP240802a & 29.22 &  29.50   &   60525.657  & LOT   & 6$\times$300s  & r  & $>$21.9 & Pan-STARRS1   \\
        & 31.17 &  31.35   &   60525.738  & LOT   & 5$\times$300s  & r  & $>$21.4 & Pan-STARRS1   \\
        & 54.37 &  55.09   &   60526.705  & LOT   & 8$\times$300s  & r  & $>$21.9 & Pan-STARRS1   \\  
        & 126.93 &  127.16   &   60529.728  & LOT   & 6$\times$300s  & r  & $>$21.7 & Pan-STARRS1   \\
        & 149.04 &  149.22   &   60530.650  & LOT   & 5$\times$300s  & r  & $>$22.3 & Pan-STARRS1   \\  
        & 171.62 &  171.84   &   60531.590  & LOT   & 6$\times$300s  & r  & $>$22.0 & Pan-STARRS1   \\

   EP240908a & 24.22 &  24.69   &   60562.737  & SLT   & 12$\times$300s  & r  & $>$21.6 & SDSS   \\

   EP240913a & 26.44 &  26.53   &   60567.588  & LOT   & 3$\times$300s  & r  & $>$21.3 & Pan-STARRS1   \\
             & 26.71 &  26.93   &   60567.599  & LOT   & 6$\times$300s  & g  & $>$21.9 & Pan-STARRS1   \\

   EP240918a & 2.30 &  2.95   &   60571.571  & SLT   & 16$\times$300s  & r  & $>$20.5 & Pan-STARRS1   \\
         & 4.84 &  5.06   &   60571.677  & LOT   & 6$\times$300s  & r  & $>$21.1 & Pan-STARRS1 \\

   EP240918b & 22.02 &  22.28   &   60572.570  & SLT   & 7$\times$300s  & r  & $>$19.0 & Pan-STARRS1   \\

   EP240918c & 19.59 &  19.83   &   60572.571  & LOT   & 5$\times$300s  & r  & $>$18.9 & Pan-STARRS1   \\
   
   EP240919a & 1.17 &  1.60   &   60572.669  & SLT   & 11$\times$300s  & r  & $>$19.8 & Pan-STARRS1   \\

   EP241026b & 142.08 &  142.56   &   60615.680  & LOT   & 12$\times$300s  & r  & $>$23.5 & Pan-STARRS1   \\         

   EP241101a & 17.08 &  17.30   &   60616.708  & LOT   & 6$\times$300s  & r  & $>$22.9 & SDSS   \\ 
        & 17.61 &  17.86   &   60616.729  & LOT   & 6$\times$300s  & g  & $>$23.0 & SDSS   \\ 

   EP241104a & 18.68 &  18.93   &   60619.552  & LOT   & 6$\times$300s  & r  & $>$22.8 & SDSS   \\    

   EP241109a\footnote{Stellar flare event} & 7.06 &  7.38   &   60623.545  & SLT   & 7$\times$150s  & r  & 13.99$\pm$0.03$^{\dagger}$ & SDSS   \\  
        & 7.06 &  7.38   &   60623.545  & SLT   & 7$\times$150s  & r  & $>$19.9$^{*}$ & SDSS   \\
        & 7.95 &  8.10   &   60623.583  & SLT   & 12$\times$30s  & r  & 13.94$\pm$0.13$^{\dagger}$ & SDSS   \\ 
        & 7.95 &  8.10   &   60623.583  & SLT   & 12$\times$30s  & r  & $>$19.1$^{*}$ & SDSS   \\
        
   EP241115a & 30.64 &  30.90   &   60630.518  & SLT   & 7$\times$300s  & r  & $>$20.7 & Pan-STARRS1   \\
        &   31.27 &  31.75   &   60630.544  & SLT   & 10$\times$300s  & g  & $>$20.5 & Pan-STARRS1   \\
        &   31.31 &  31.53   &   60630.546  & LOT   & 6$\times$300s  & r  & $>$20.7 & Pan-STARRS1   \\

   EP241119a & 23.22 &  24.33   &   60634.713  & SLT   & 24$\times$300s  & r  & $>$20.8 & SDSS   \\

   EP241125a & 12.09 &  12.35   &   60639.508  & LOT   & 15$\times$120s  & r  & $>$23.0 & Pan-STARRS1   \\

   EP241126a & 16.23 &  16.45   &   60641.495  & LOT   & 6$\times$300s  & r  & $>$22.5 & Pan-STARRS1   \\

   EP241206a & 22.45 &  22.75   &   60651.626  & SLT   & 8$\times$300s  & r  & $>$20.9 & Pan-STARRS1   \\
        & 25.32 &  25.41   &   60651.746  & LOT   & 3$\times$300s  & r  & $>$22.5 & Pan-STARRS1   \\
        & 46.88 &  47.97   &   60652.644  & SLT   & 48$\times$150s  & r  & $>$21.1 & Pan-STARRS1   \\

   EP241208a & 23.62 &  24.61   &   60653.676  & SLT   & 24$\times$300s  & r  & $>$21.0 & SDSS   \\
        \bottomrule

    \end{tabular}
    \label{tab:Photometry9}
{$^*$Corresponding upper limit from the same field. $^\dagger$Magnitude of the (flaring) star. The finding charts of the Blazar NVSS J004348+342626 and the stellar flare event EP241109a from 40 cm SLT are shown in Figure~\ref{fig:Blazar_n_stellar_flare}.}
\end{table*}

%% file: Tables/redback.tex

\begin{table*}
    \renewcommand*{\arraystretch}{1.1}
    \centering
    \caption{
        Posterior distribution of the ``gaussiancore'' structured jet model from {\tt REDBACK} fittings for three of the best studies FXTs in literature, i.e., EP240315a, EP240414a, and EP241021a. We used the default range of parameters specified in {\tt REDBACK}.}
    \begin{tabular}{cccccccccc}
        \toprule
        \midrule
        
  Parameters &  Range & Nature of Priors      &EP240315a     &EP240414a       &EP241021a          \\

        \midrule
        \midrule
 $\theta_{\rm{observer}}~({\rm{rad}})$       &    [0,1.5707963267948966]   &  Sine     &   0.21$^{+0.04}_{-0.07}$    & 0.30$^{+0.03}_{-0.03}$      & 0.18$^{+0.02}_{-0.01}$      & \\
 
  $\log_{10}~E_{0}/{\rm{erg}}$       &   [44,54]    &  Uniform     &  53.57$^{+0.29}_{-0.43}$     &     53.44$^{+0.35}_{-0.40}$  & 53.72$^{+0.20}_{-0.33}$      & \\
  
  $\theta_{\rm{core}}~({\rm{rad}})$      &  [0.01,0.1]     &   Uniform    &  0.08$^{+0.01}_{-0.02}$     &  0.09$^{+0.00}_{-0.01}$     & 0.09$^{+0.01}_{-0.01}$      & \\
  
  $\theta_{\rm{truncation}}$      &   [1,8]    &   Uniform    & 5.73$^{+1.33}_{-2.10}$      &  2.47$^{+0.30}_{-0.13}$     &  1.14$^{+0.09}_{-0.08}$     & \\
  
  $\log_{10}~n_{\rm{ism}}/{\rm{cm}}^{-3}$      &   [-5,2]    &  Uniform      & 0.02$^{+1.22}_{-1.97}$      & -2.80$^{+0.40}_{-0.32}$      &  -1.61$^{+0.57}_{-0.52}$     & \\
  
  $p$      &   [2,3]    &   Uniform    & 2.49$^{+0.32}_{-0.30}$      & 2.95$^{+0.03}_{-0.06}$      &    2.59$^{+0.15}_{-0.23}$   & \\

    $\log_{10}~\epsilon_{e}$      &   [-5,0]    &   Uniform    & -0.44$^{+0.24}_{-0.38}$     &  -0.90$^{+0.30}_{-0.28}$      &  -0.82$^{+0.20}_{-0.21}$     & \\

    $\log_{10}~\epsilon_{B}$      &   [-5,0]    &  Uniform     & -1.13$^{+0.78}_{-1.17}$      &    -0.18$^{+0.13}_{-0.29}$   & -1.87$^{+0.71}_{-0.48}$      & \\

    $\xi_{N}$      &   [0.0,1.0]    &  Uniform     &  0.64$^{+0.27}_{-0.35}$     & 0.33$^{+0.26}_{-0.16}$      & 0.79$^{+0.15}_{-0.21}$      & \\    

    $\Gamma_{0}$      &  [100,2000]     &   Uniform    &  1029.39$^{+659.87}_{-640.11}$     &  1105.85$^{+608.00}_{-657.46}$     & 894.78$^{+677.39}_{-523.09}$      & \\
        \bottomrule
        \bottomrule
    \end{tabular}
    \label{tab:redback}
\end{table*}